\def\version{July 9, 2015}

\documentclass[12pt]{article}

\title  {
        Critical correlation functions for the $4$-dimensional \\
        weakly self-avoiding walk and $n$-component $|\varphi|^4$ model}

\author{
Gordon Slade\thanks{Department of Mathematics,
University of British Columbia, Vancouver, BC, Canada V6T 1Z2.
{\tt slade@math.ubc.ca}, {\tt atomberg@math.ubc.ca}}\;
and Alexandre Tomberg$^*$
}

\def\macrosPb{}
\usepackage{amsfonts}
\usepackage{amsmath,amssymb,amsthm}
\usepackage{appendix}
\usepackage{bbm} 
\usepackage{amsbsy}
\usepackage{enumerate}
\usepackage{cite}

\def\macrosHarxiv{}

\InputIfFileExists{./macros_local.tex}{}{}

\ifdefined\macrosPa
  \usepackage[textwidth=465pt,textheight=650pt,centering]{geometry} 
\else\ifdefined\macrosPb
  \usepackage[textwidth=500pt,textheight=650pt,centering]{geometry} 
\fi\fi

\ifdefined\macrosS


  \usepackage{mathptmx}
  \DeclareMathAlphabet{\mathcal}{OMS}{cmsy}{m}{n}
\fi

\ifdefined\macrosBirk
	\DeclareRobustCommand*\cal{\@fontswitch\relax\mathcal}
\else
	\usepackage[dvips]{graphicx}
\fi

\ifdefined\macrosSB

\def\UseSection{
        \numberwithin{equation}{section}
	\theoremstyle{plain}
        \newtheorem{theorem}    {Theorem}[section]
        \DefineTheorems 
}

\def\DefineTheorems{
	
	\newtheorem{lemma}      [theorem] {Lemma}
	
	\newtheorem{prop}       [theorem] {Proposition}
	
	\newtheorem{cor}        [theorem] {Corollary}

	\theoremstyle{definition}
	\newtheorem{defn}       [theorem] {Definition}

	\newtheorem{rk} 	[theorem] {Remark}
	\theoremstyle{definition}

}

\newcommand{\bt}   {\begin{theorem}}
\newcommand{\et}   {\end  {theorem}}
\newcommand{\bl}   {\begin{lemma}}
\newcommand{\el}   {\end  {lemma}}
\newcommand{\bp}   {\begin{prop}}
\newcommand{\ep}   {\end  {prop}}
\newcommand{\bc}   {\begin{cor}}
\newcommand{\ec}   {\end  {cor}}
\newcommand{\bd}   {\begin{defn}}
\newcommand{\ed}   {\end  {defn}}

\newcommand{\ba}   {\begin{array}}
\newcommand{\ea}   {\end  {array}}
\newcommand{\be}   {\begin{enumerate}}
\newcommand{\ee}   {\end  {enumerate}}
\newcommand{\bi}   {\begin{itemize}}
\newcommand{\ei}   {\end  {itemize}}

\def\eq#1\en{\begin{equation}#1\end{equation}}  
\def\eqsplit#1\ensplit{
	\begin{equation}\begin{split}#1\end{split}\end{equation}
	}
\def\eqalign#1\enalign{
	\begin{align}#1\end{align}
	}
\def\eqmul#1\enmul{
	\begin{multline}#1\end{multline}
	}
\newcommand{\eqarrstar} {\begin{eqnarray*}} 
\newcommand{\enarrstar} {\end{eqnarray*}} 
\newcommand{\eqarray}   {\begin{eqnarray}} 
\newcommand{\enarray}   {\end{eqnarray}}

\newcommand{\refeq}[1] {\eqref{e:#1}}    

\makeatletter
\newcommand{\labelcounter}[2]{{%
	\stepcounter{#1}
	\protected@write\@auxout{}%
	{\string\newlabel{#2}{{\csname the#1\endcsname}{\thepage}}}%
	{\ref{#2}}
	}}
\makeatother
\newcommand{\Nbold} {{\mathbb N}}

\newcommand{\Rbold} {{\mathbb R}}

\newcommand{\Zbold} {{\mathbb Z}}

\newcommand{\Bcal}   {\mathcal{B}} 
 
\newcommand{\Dcal}   {\mathcal{D}} 
\newcommand{\Ecal}   {\mathcal{E}}

\newcommand{\Ical}   {\mathcal{I}} 
 
\newcommand{\Kcal}   {\mathcal{K}} 
\newcommand{\Lcal}   {\mathcal{L}} 
 
\newcommand{\Ncal}   {\mathcal{N}} 
 
\newcommand{\Pcal}   {\mathcal{P}}

\newcommand{\Scal}   {\mathcal{S}}

\newcommand{\Vcal}   {\mathcal{V}} 
\newcommand{\Wcal}   {\mathcal{W}}

\newcommand{\Zcal}   {\mathcal{Z}} 

\newcommand{\Zd}    {{ {\Zbold}^d }}

\newcommand{\spose}[1] {{\hbox to 0pt{#1\hss}} }
\newcommand{\ltapprox} {\mathrel{\spose{\lower 3pt\hbox{$\mathchar"218$}}
 \raise 2.0pt\hbox{$\mathchar"13C$}}}
\newcommand{\gtapprox} {\mathrel{\spose{\lower 3pt\hbox{$\mathchar"218$}}
 \raise 2.0pt\hbox{$\mathchar"13E$}}}

\else
\fi

\UseSection   
\usepackage[usenames]{color}

\definecolor{bw}{RGB}{240, 120, 0}
\definecolor{at}{rgb}{0.0, 0.5, 0.0} 

\newcommand{\LT}{{\rm Loc}  }

\newcommand{\DV}{\Dcal}
\newcommand{\DVa}{\alpha}

\renewcommand{\to} {\rightarrow}

\newcommand{\R}{\Rbold}
\newcommand{\Z}{\Zbold}

\newcommand{\N}{\Nbold}
\newcommand{\C}{\mathbb{C}}

\newcommand{\1}{\mathbbm{1}}

\newcommand{\psib}{\bar\psi}

\newcommand{\jm}{j_\Omega}

\newcommand{\Ex}{\mathbb{E}}

\newcommand{\chicCov}{{\chi}}

\newcommand{\bubble}{{\sf B}}

\newcommand{\pair}[1]{\langle #1 \rangle}

\newcommand{\cgam}{\gamma}

\newcommand{\hldg}{h_{\rm lead}}

\newcommand{\pt}{{\rm pt}}
\newcommand{\Ipt}{I_{\rm pt}}
\newcommand{\Ipttil}{\tilde{I}_{\rm pt}}

\newcommand{\Vpt}{V_{\rm pt}}

\newcommand{\qpt}{q_{\mathrm{pt}}}

\newcommand{\Vbulk}{U}

\newcommand{\gch}{\check{g}}

\newcommand{\h}{\mathfrak{h}}
\ifdefined\macrosSB \else

\fi

\newcommand{\gbar}{\bar{g}}

\newcommand{\ggen}{\tilde{g}}
\newcommand{\sgen}{\tilde{s}}
\newcommand{\chigen}{\tilde{\chi}}
\newcommand{\mgen}{\tilde{m}}
\newcommand{\Iint}{\mathbb{I}}
\newcommand{\Igen}{\tilde{\mathbb{I}}}

\newcommand{\domRG}{\mathbb{D}}

\newcommand{\pp}{a}
\newcommand{\qq}{b}

\newcommand{\sigmab}{\bar{\sigma}}

\newcommand{\half}{\textstyle{\frac 12}}

\newcommand{\epV}{\epsilon_{V}}

\newcommand{\epdV}{\bar{\epsilon}}

\newcommand{\phib}{\bar\phi}

\newcommand{\sgn}{\mathrm{sgn}}

\newcommand{\Kspace}{\Kcal}

\newcommand{\wm}{W}

\DeclareMathOperator{\Loc}{Loc} 

\newcommand{\nl}{p}

\ifdefined\macrosH
  \usepackage{xr-hyper}
  \usepackage{hyperref}
  \hypersetup{hypertexnames=false}
  \hypersetup{colorlinks,citecolor=blue,linkcolor=blue}  

\else\ifdefined\macrosHarxiv
  \usepackage{xr-hyper}
  \usepackage{hyperref}
  \hypersetup{hypertexnames=false, hidelinks}

\else
  \newcommand{\texorpdfstring}[2]{#1}
  \usepackage{xr}
\fi\fi

\usepackage[shortlabels]{enumitem}
\setlist[itemize]{label = $\blacktriangleright$, topsep = 3pt, itemsep = -1ex}
\setlist[description]{font=\itshape}
\setlist[enumerate]{topsep = 0pt, itemsep = -1ex}

\newcommand{\htilde}{\tilde{h}}

\newcommand{\fac}[1]{(#1)^*}
\newcommand{\efac}{$h$-factorisable}
\newcommand{\Ncalefac}{\Ncal_{\text{$h$-{\rm fac}}}}

\newcommand{\inprod}[2]{\left\langle #1 ; #2 \right\rangle}
\newcommand{\varphiCx}[2]{(\varphi^{#1}_{#2})}

\newcommand{\corrfcn}{S}
\newcommand{\starpol}{S}

\newcommand{\gammap}{\gamma^{+}}
\newcommand{\gammam}{\gamma^{-}}

\newcommand{\grat}{\Gamma}

\newcommand{\mrc}{A}
\DeclareMathOperator{\Span}{span}

\newcommand{\lx}[1]{\lambda_{x, #1}} 
\newcommand{\vlx}[1]{r_{x, #1}} 

\newcommand{\qx}[1]{q_{x, #1}}
\newcommand{\vqx}[1]{R^{q_x}_{#1}}

\usepackage{graphicx}
\usepackage{tikz}
\usetikzlibrary{arrows,shapes,backgrounds,calc,through,shapes.multipart}

\date\version

\begin{document}

\maketitle


\begin{abstract}
We extend and apply
a rigorous renormalisation group method to study critical correlation
functions, on the $4$-dimensional lattice $\Z^4$,
for the weakly coupled $n$-component $|\varphi|^4$ spin model for all $n \ge 1$,
and for the continuous-time weakly self-avoiding walk.
For the $|\varphi|^4$ model, we prove that the critical
two-point function has $|x|^{-2}$ (Gaussian) decay asymptotically, for $n \ge 1$.
We also determine the asymptotic decay of the
critical correlations of the squares of
components of $\varphi$, including the logarithmic
corrections to Gaussian scaling, for $n \ge 1$.
The above extends previously known results for $n=1$ to all $n \ge 1$,
and also observes new phenomena for $n>1$, all with a new method of proof.
For the continuous-time
weakly self-avoiding walk, we determine the decay of the critical
generating function for the ``watermelon'' network consisting of $p$ weakly
mutually- and self-avoiding walks, for all $p \ge 1$, including the logarithmic
corrections.  This extends a previously known result for $p=1$, for which there
is no logarithmic correction, to a much more general setting.
In addition, for both models, we study the approach to the critical point
and prove existence of logarithmic corrections
to scaling for certain correlation functions.
Our method
gives a rigorous analysis of the weakly self-avoiding walk as the $n=0$ case
of the $|\varphi|^4$ model, and provides a unified treatment of both models, and of all the
above results.
\end{abstract}



\section{Introduction and main results}

\subsection{Introduction}

The subject of critical phenomena and phase transitions
in statistical physics has been an important source of
problems and inspiration for mathematics for over half a century,
especially in probability theory and combinatorics.
Fundamental models of statistical mechanics,
such as the Ising model, the $O(n)$ model,
the $|\varphi|^4$ model, the self-avoiding walk, percolation,
the random cluster model, dimers, and others, have become increasingly prominent in
mathematics and now form the raw material for large and diverse research
communities.

For ferromagnetic spin systems such as the Ising and $|\varphi|^4$ models,
the phase transition results from an interplay between
a local ferromagnetic interaction which causes spins to tend to align,
and thermal fluctuations which tend to destroy long-range order.
As the temperature decreases past its critical value, long-range order
emerges suddenly and in a singular manner.
Universal critical exponents and scaling limits are associated
with the phase transition.
Part of the fascination of the subject is due to the variation of
behaviour as the underlying spatial dimension $d$ changes.

The case of $d=2$ is particularly beautiful and rich, and is connected with
conformal invariance.
Magnificent advances in our rigorous understanding
of phase transitions for $d=2$ have emerged following the invention of the
Schramm--Loewner Evolution.  Nevertheless, significant challenges remain.  For example,
while we now have good understanding of the critical behaviour of site percolation
on the triangular lattice \cite{Smir01,Smir01CR,SW01},
no corresponding results have been obtained for bond
percolation on the square lattice.
Although there has been recent progress, e.g., \cite{CS12,GM14},
important issues concerning universality remain to be resolved.

The physically most interesting case of $d=3$ is the most difficult and
the least understood.  Very recently, and about 70 years after Onsager's exact
solution of the 2-dimensional Ising model, it was proved that the spontaneous
magnetisation of the 3-dimensional
Ising model vanishes at the critical temperature \cite{ADS15}.
However, the vanishing of the percolation probability for $d=3$ remains one of the most
significant open questions in probability theory, and generally the calculation of critical
exponents for $d=3$ is an enormous challenge.  An interesting exception is the
exact solution for 3-dimensional branched polymers in \cite{BI03a,KW09}.

In high dimensions, much is known.  For the Ising and
1- and 2-component $|\varphi|^4$ models,
methods involving reflection positivity have led to proofs of mean-field behaviour
for nearest-neighbour interactions
in dimensions $d>4$ \cite{Aize82,Froh82}.
Such methods have been extended to show that
in the upper critical dimension $d=4$, deviations
from mean-field behaviour are at worst logarithmic for some quantities
\cite{AG83,AF86,ACF83,BFF84}, although these references do not prove that logarithmic
corrections do exist as predicted in the physics literature.
Lace expansion methods
have been used to prove mean-field behaviour in dimensions greater than $4$,
for the self-avoiding walk \cite{HS92a}, for spread-out Ising models  \cite{Saka07},
for weakly coupled or spread-out $1$-component $\varphi^4$  \cite{Saka15},
among other models \cite{Slad06}.

Much of the attention has been devoted to models with discrete symmetry such as the Ising model,
where the group $\Z_2$ acts on the interaction by flipping all spins simultaneously.
However, from a physical perspective, continuous symmetry is highly relevant,
and from a mathematical perspective, it describes richer phenomena.
The most basic non-trivial examples of such models are the $O(n)$
and $|\varphi|^4$ models, which generalise the Ising model,  and in which
the interaction between $n$-component spins is invariant under
the orthogonal group.
It has long been understood that the $n$-component
$|\varphi|^4$ model can be obtained as a limit
of $O(n)$ models \cite{SG73,DN75}, and that the converse holds is an elementary fact.
The existence of a phase transition in nearest-neighbour models,
in which the continuous $O(n)$ symmetry is spontaneously broken,
has been proved in all dimensions $d > 2$ using the \emph{infrared bound}.
Although elegant, this method has limitations:
it is limited to reflection-positive models and does not supply detailed understanding
of the behaviour at the critical point.

Our subject in this paper is the critical behaviour of the
continuous-time weakly self-avoiding walk (or WSAW),
and of the $n$-component $|\varphi|^4$ model, for all $n \ge 1$, in the upper
critical dimension $d=4$.  Over forty years ago, de Gennes observed that
$n$-component spin models formally correspond to the self-avoiding walk in
the limit $n \to 0$ \cite{Genn72}.  Since the number of components is a natural number, the limit
$n \to 0$ is mathematically undefined, at least naively.  However, using the basis
developed in \cite{BBS-saw4-log,BBS-saw4,BBS-phi4-log}, we are able to
interpret WSAW in a mathematically rigorous manner as the $n=0$ case of
the $n$-component $|\varphi|^4$ model, and provide a unified treatment for all $n \ge 0$.
In particular, our results also apply to the case $n \ge 2$ of continuous symmetry.

The basis we build upon is a rigorous renormalisation group method.
In physics, the renormalisation group has been used simultaneously to
explain the existence of
universality and to compute the universal quantities associated
with critical phenomena.  A nonrigorous analysis of the $|\varphi|^4$ model is worked
out in \cite{WK74}, and the model is sometimes referred to as the
\emph{Landau--Ginzburg--Wilson} model.
A rigorous renormalisation group method is applied
to the $|\varphi|^4$ model in \cite{BBS-phi4-log}.  This method applies to WSAW
once the model is rewritten in terms of a supersymmetric integral representation
\cite{BIS09,McKa80,PS80}.  The supersymmetric representation we use is in terms of
a $2$-component boson field and a 2-component fermion field; the former contributes
a factor $2$ to each loop, and the latter $-2$, with the net effect that loops
do not contribute.  In \cite{BBS-phi4-log,BBS-saw4-log}, the renormalisation
group method is applied to prove that for $n\ge 0$
the susceptibility diverges at the
critical point as $\varepsilon^{-1} (\log \varepsilon^{-1})^{(n+2)/(n+8)}$,
with $n=0$ corresponding to WSAW and $n \ge 1$ corresponding to $|\varphi|^4$.
This confirms predictions of \cite{LK69,BGZ73,WR73}.
For $n=1$, it was proved much earlier in \cite{Hara87,HT87}.
Also, in \cite{BBS-saw4}, $|x|^{-2}$ decay
is proved for the critical two-point function of the $4$-dimensional WSAW.

This last result required the introduction of \emph{observables} to the
analysis, and one of our major themes is to extend the variety of
observables considered, and to apply the formalism of observables also to
the $|\varphi|^4$ model.  The latter was not done in \cite{BBS-phi4-log}.
(Somewhat related ideas were used in \cite{DH92}; different methods
were developed in \cite{Falc13}.)
Moreover, we develop
new techniques concerning reduction of the $O(n)$ symmetry of the $|\varphi|^4$
model, for $n \ge 2$.

A lesson learned from \cite{BBS-saw4-log,BBS-phi4-log} is that if we set
$n=0$ in the second-order perturbative calculations used for
the rigorous renormalisation group analysis of the $|\varphi|^4$ model,
then what results is exactly the
second-order perturbative calculations in the rigorous renormalisation group analysis
for WSAW.  This is a rigorous version of the observation of de Gennes \cite{Genn72}.
A general theory developed in \cite{BS-rg-IE,BS-rg-step} permits a unified
treatment of non-perturbative effects.  Consequently, our main task here is
to carry out appropriate perturbative calculations, with an appeal to the general
theory to bound all the error terms.  We do this in a unified way for
all $n \ge 0$, \emph{including} $n=0$.

With this approach, we derive the asymptotic decay
of several critical correlation functions, in dimension $d=4$.  For $|\varphi|^4$, we prove
$|x|^{-2}$ decay for the critical two-point function for all $n \ge 1$.  This
extends previous results for $n=1$ due to \cite{FMRS87,GK85,GK86}, to all $n \ge 1$.
In \cite{GK86}, it was also shown that for $n=1$ the critical correlation
between $\varphi_0^2$ and $\varphi_x^2$ decays as $|x|^{-4}(\log |x|)^{-2/3}$.
We extend this to general $n \ge 1$.  In addition, we
prove that for the multi-component case of
$n \ge 2$, at the critical point
there is positive correlation between same field components
at distant points, but
negative correlation between different field components.
Related results are obtained for logarithmic corrections to scaling
for correlations of fields, as the critical
point is approached.
For WSAW, we obtain the decay of the critical ``watermelon'' networks,
consisting for fixed $p \ge 1$ of $p$ weakly mutually- and self-avoiding walks
joining two distant sites, at the critical point.  This extends the result for
$p=1$ obtained in \cite{BBS-saw4}.
(An earlier related result for $p=1$ is \cite{IM94},
for a model which is neither a lattice model nor a model containing walks, but
which nevertheless shares features in common with WSAW.)
For $p \ge 2$, we also determine the logarithmic corrections to scaling for ``star networks''
consisting of $p$ weakly mutually- and self-avoiding walks which intersect at the origin,
as the critical point is approached.
Star networks are the simplest example of polymer networks of the sort
studied in \cite{Dupl89a}, and serve as building blocks for more
general networks.

We next give precise definitions of the $|\varphi|^4$
and WSAW models, followed by precise statements
of our results.

\subsection{The \texorpdfstring{$|\varphi|^4$}{phi4} model}

\subsubsection{Definition of the model}
\label{sec:phi4def}

Let $L >1$ be an integer, and let $\Lambda = \Lambda_N = \Z^d/L^N\Z^d$ be the
$d$-dimensional discrete torus of side length $L^N$.
Ultimately we are interested in the thermodynamic limit $N \to \infty$.
For convenience, we sometimes consider $\Lambda$ to be a box in $\Z^d$,
approximately centred at the origin, without opposite sides identified to create the torus.
We can then regard fixed $a,b\in \Zd$ as points in $\Lambda$ provided that $N$ is large enough,
and we make this identification throughout the paper.
In particular, we always assume that $N$ is sufficiently
large that $\Lambda$ contains the given $a,b$.

The spin field $\varphi$ is  a function $\varphi : \Lambda \to
\R^n$, or equivalently a vector $\varphi \in (\R^n)^{\Lambda}$.
We use subscripts to index $x\in \Lambda$ and superscripts for the
components $i=1,\dots, n$.
We write $|v|$ for the Euclidean norm $|v|^2 = \sum_{i=1}^n (v^i)^2$
and $v \cdot w = \sum_{i=1}^n v^i w^i$ for the Euclidean inner product on $\R^n$.
For $e\in\Z^d$ with $|e|=1$, we define the discrete gradient by
$(\nabla^e \varphi)_x =\varphi_{x+e}-\varphi_x$,
and the discrete Laplacian by
$\Delta = -\frac{1}{2}\sum_{e\in\Z^d:|e|_1=1}\nabla^{-e} \nabla^{e}$.
We write $\varphi_x \cdot (-\Delta \varphi)_x = \sum_{i=1}^n \varphi_x^i(-\Delta\varphi^i)_x$.

Given $g>0, \nu \in \R$, we define a function $U_{g,\nu,N}$ of the field by
\begin{equation} \label{e:Vdef1}
  U_{g,\nu,N}(\varphi)
  = \sum_{x\in\Lambda}
  \Big(\tfrac{1}{4} g |\varphi_x|^4 + \half \nu |\varphi_x|^2
  + \half \varphi_x\cdot (-\Delta \varphi_x)  \Big)
  .
\end{equation}
By definition, the quartic term is $|\varphi_x|^4 = (\varphi_x \cdot \varphi_x)^2$.
Then we define the expectation of a random variable $F:(\R^n)^{\Lambda} \to \R$ by
\begin{equation}
  \label{e:Pdef}
  \langle F \rangle_{g,\nu,N}
  = \frac{1}{\Zcal_{g,\nu,N}} \int F(\varphi) e^{-U_{g,\nu,N}(\varphi)} d\varphi,
\end{equation}
where $d\varphi$ is the Lebesgue measure on
$(\R^n)^{\Lambda}$,
and $\Zcal_{g,\nu,N}$ is a normalisation constant (the \emph{partition function})
defined so that $\langle 1 \rangle_{g,\nu,N} =1$.
Thus $\varphi$ is a field of classical continuous $n$-component spins
on the torus $\Lambda$, i.e., with periodic boundary conditions.

The \emph{susceptibility} is defined as the limit
\begin{equation}
  \label{e:susceptdef}
  \chi(g, \nu; n)
  = \lim_{N \to \infty} \sum_{x\in\Lambda_N}
  \pair{\varphi_a^1\varphi_x^1}_{g,\nu,N}
  = n^{-1} \lim_{N\to\infty} \sum_{x\in\Lambda_N}
  \pair{\varphi_a \cdot \varphi_x}_{g,\nu,N}
  .
\end{equation}
By translation-invariance of the measure, $\chi$ is independent of $a \in \Z^d$.
For $n=1,2$, standard correlation inequalities \cite{FFS92} imply that for
the case of free boundary conditions the limit defining the
susceptibility exists (possibly infinite) and is monotone non-increasing in $\nu$.
Proofs are lacking for $n>2$ due to a lack
of correlation inequalities in this case (as is discussed, e.g., in \cite{FFS92}),
although one expects that these facts known for $n\le 2$ should remain true also for $n>2$.
In our theorems below, we prove the existence of
the infinite volume limit with periodic boundary conditions directly
in the situations covered by the theorems, without application of any
correlation inequalities.
Our proof of existence of limits is, however, restricted to large $L$.

For $d=4$, for small $g>0$, and for all $n \ge 1$, it is proved in \cite{BBS-phi4-log}
that there is a critical value $\nu_c(g;n)$ such that, for
$\nu=\nu_c+\varepsilon$, the susceptibility diverges according to the
asymptotic formula
\begin{equation}
\label{e:chin}
    \chi(g,\nu;n) \sim
    A_{g,n}\varepsilon^{-1}(\log \varepsilon^{-1})^{(n+2)/(n+8)}
    \quad \text{as $\varepsilon\downarrow 0$},
\end{equation}
for some amplitude $A_{g,n}>0$.
Here, and throughout the paper, we write $f \sim g$ to mean $\lim f/g=1$.
In this paper, we study correlation functions
both exactly at the critical value $\nu_c(g;n)$
and in the limit as $\nu \downarrow \nu_c(g;n)$.
It is also shown in \cite{BBS-phi4-log} that
$\nu_c(g;n) = -{\sf a} g + O(g^2)$
with ${\sf a}=(n+2)G_{00} > 0$, where, for $a,b\in \Z^4$, $G_{ab}$ denotes
the massless lattice Green function.
From an analytic perspective,
\begin{equation}
    G_{ab} = (-\Delta_{\Z^4}^{-1})_{ab},
\end{equation}
where
the right-hand side is the matrix element of the inverse lattice
Laplacian acting on square-integrable scalar functions on $\Z^4$.
From a probabilistic perspective, $G_{ab}$ equals
$\frac{1}{2d}$ times
the expected number of visits to $b$ of simple random walk on $\Z^4$ started from $a$
(the extra factor $\frac{1}{2d}=\frac 18$ is due to our definition of the Laplacian).
It is a standard fact
(see, e.g., \cite{Lawl91}) that, as $|a-b|\to\infty$,
\begin{equation} \label{e:Casym}
    G_{ab} = \frac{1}{(2\pi)^2 |a-b|^2} \left(1 + O \Big(\frac{1}{|a-b|^2} \Big)\right)
    .
\end{equation}

\subsubsection{Correlation functions}

We study infinite volume correlation functions.
The existence of the infinite volume limit is not known for general $n$, and
it is part of our results that the limit does exist for $n \ge 1$, provided $g$
is sufficiently small and $L$ is sufficiently large.
We write $\pair{F}_{g,\nu} = \lim_{N\to\infty} \pair{F}_{g,\nu,N}$ when the limit exists.
We also write $\pair{F;G} = \pair{FG}-\pair{F}\pair{G}$, both in finite and infinite volume,
for the \emph{correlation} or \emph{truncated expectation} of $F,G$.
Our main results include the precise asymptotic behaviour as $|a-b|\to\infty$, for
all $n \ge 1$ and for $p=1,2$,
of the 4-dimensional infinite volume \emph{critical} truncated correlation functions
\begin{equation}
\label{e:phi4cf}
\inprod{(\varphi^i_a)^\nl}{(\varphi^j_b)^\nl}_{g,\nu_c} \quad \text{ for } 1\leq i,j \leq n.
\end{equation}
By the $O(n)$ symmetry,
it is sufficient to consider the two special cases $(i,j)=(1,1)$ and $(i,j)=(1,2)$.
The first case turns out to be positive.
Since the transformation $\varphi^1 \mapsto - \varphi^1$ does not change the measure,
the second case is zero for all odd $p$.
The second case only makes sense for $n \ge 2$, and it turns out to be \emph{negative}
for $p=2$.
In principle our methods could be used to study also $p>2$, but new
issues arise for $p>2$ and we have not pursued this case.

We define the critical correlation functions \eqref{e:phi4cf} as the limit
\begin{equation}
\label{e:cflim}
    \inprod{(\varphi^i_a)^\nl}{(\varphi^j_b)^\nl}_{g,\nu_c}
    =
    \lim_{\varepsilon \downarrow 0}\lim_{N\to\infty}
    \inprod{(\varphi^i_a)^\nl}{(\varphi^j_b)^\nl}_{g,\nu_c+\varepsilon, N}
    .
\end{equation}
Similarly, for $\nu >\nu_c$, we write
\begin{equation}
\label{e:cfsumlim}
    \sum_{x_1,x_2\in \Z^4}
    \inprod{ \varphi^i_{x_1}\varphi^j_{x_2}}{(\varphi^k_a)^2 }_{g,\nu}
    =
    \lim_{N\to\infty}
    \sum_{x_1,x_2\in \Lambda_N}
    \inprod{ \varphi^i_{x_1}\varphi^j_{x_2}}{(\varphi^k_a)^2 }_{g,\nu,N}.
\end{equation}
It is part of the statement of our results that these limits exist for small
$g>0$ and for $n \ge 1$, $p=1,2$.
However, we do require that the limit
be taken through tori $\Lambda_N = \Z^4/L^N\Z^4$ with $L$ large, as this
restriction is part of the hypotheses of results from
\cite{BS-rg-IE,BS-rg-step,BBS-saw4-log,BBS-phi4-log} upon which our analysis relies.
We therefore always tacitly assume that $L$ is large, throughout the rest of the
paper, for both the $|\varphi|^4$ and WSAW models.  When we assume that $g$ is
small in theorems, $g$ is chosen small depending on the value of $L$, and
depending also on
$n \ge 0$.

\subsection{The WSAW model}

\subsubsection{Definition of the model}

Let $X$ be the continuous-time simple random walk on the integer
lattice $\Zd$, with $d > 0$.  In more detail, $X$ is the stochastic
process with right-continuous sample paths
that takes its steps at the times of the events of a rate-$2d$ Poisson process.
Steps are taken uniformly at random to one of the $2d$ nearest neighbours of the current position,
and are independent both of the Poisson process and of all other steps.
Let $E_a$ denote the
expectation for the process
with $X(0)=a \in \Zd$.
The \emph{local time of $X$ at
$x$ up to time $T$} is the random variable $L_T(x) = \int_0^T \1_{X(t)=x} \; dt$,
and the \emph{self-intersection local time up to time $T$} is the random variable
\begin{equation} \label{e:ITdef}
  I(T) = \int_0^T \!\! \int_0^T \1_{X(t_1) = X(t_2)} \; dt_1 \, dt_2
  =
  \sum_{x\in\Z^d} \big(L_T(x)\big)^2.
\end{equation}

Given $g>0$, $\nu \in \R$, and $a,b \in \Zd$,
the continuous-time weakly self-avoiding walk \emph{two-point function} is defined by
the integral (possibly infinite)
\begin{equation}
  \label{e:Gwsaw}
  \wm^{(1)}_{ab}(g,\nu)
  =
  \int_0^\infty
  E_{a} \left(
    e^{-g I(T)}
    \1_{X(T)=b} \right)
  e^{- \nu T}
  dT.
\end{equation}
In \eqref{e:Gwsaw}, self-intersections are suppressed by the factor $e^{-gI(T)}$.
The connection between \eqref{e:Gwsaw} and the two-point function of
the usual strictly self-avoiding walk is discussed in \cite{BDS12}.
In dimension~4, \eqref{e:Gwsaw} is also known as
the two-point function of the lattice \emph{Edwards model} (with continuous time).

We define the \emph{susceptibility} by
\begin{equation}
  \label{e:suscept-def}
  \chi(g,\nu;0) =   \sum_{b\in \Z^d}  \wm^{(1)}_{ab}(g,\nu)
  =
  \int_0^\infty E_{a}(e^{-gI(T)}) e^{- \nu T} dT.
\end{equation}
By translation-invariance of the simple random walk and of \eqref{e:ITdef},
$\chi$ is independent of the point $a \in \Z^d$.
A standard subadditivity argument \cite{BBS-saw4-log} shows that for all dimensions $d > 0$
there exists a \emph{critical value} $\nu_c = \nu_c(g; 0) \in (-\infty, 0]$ (depending also on $d$)
such that
\begin{equation}
  \label{e:chi-nuc}
  \text{$\chi(g,\nu;0) < \infty$ \; if and only if \; $\nu > \nu_c$}.
\end{equation}
It is shown in \cite{BBS-saw4-log} that for $d=4$,
for small $g > 0$ and
for $\nu = \nu_c + \varepsilon$, the susceptibility diverges as
\begin{equation}
\label{e:chi0}
    \chi(g,\nu;0) \sim
    A_{g,0}\varepsilon^{-1}(\log \varepsilon^{-1})^{1/4}
    \quad \text{as $\varepsilon\downarrow 0$}.
\end{equation}
Moreover, $\nu_c(g;0) =
-{\sf a}g+O(g^2)$ with ${\sf a} = 2G_{00} >0$.

The above asymptotic formulas for the susceptibility and critical point
are both consistent with setting $n = 0$ in the corresponding statements for
the $|\varphi|^4$ model in Section~\ref{sec:phi4def}.

\subsubsection{Watermelon and star networks}

\begin{figure}
\begin{center}
\begin{tikzpicture}
\tikzstyle{vertex}=[circle, fill = black, inner sep = 1.3pt]
\tikzstyle{walk}=[semithick]
\tikzstyle{double-path}=[draw, walk, ellipse, minimum width = 1.2cm]

\begin{scope}[xshift = 0cm] 
\node[vertex, label=right:{$b$}] (b) at (1.2,0) {};
\node[vertex, label=left:{$a$}] (a) at (0,0) {};
\draw[walk] (a) -- (b);
\end{scope}

\begin{scope}[xshift = 3cm] 
\node[vertex, label=right:{$b$}] (b) at (1.2,0) {};
\node[vertex, label=left:{$a$}] (a) at (0,0) {};
\coordinate (center) at ($(a)!0.5!(b)$);

\node (ellipse) at (center) [double-path, minimum height = 1cm] {};
\end{scope}

\begin{scope}[xshift = 6cm]
\node[vertex, label=right:{$b$}] (b) at (1.2,0) {};
\node[vertex, label=left:{$a$}] (a) at (0,0) {};
\coordinate (center) at ($(a)!0.5!(b)$);

\draw[walk] (a) -- (b);
\node (ellipse) at (center) [double-path, minimum height = 1cm] {};
\end{scope}

\begin{scope}[xshift = 9cm]
\node[vertex, label=right:{$b$}] (b) at (1.2,0) {};
\node[vertex, label=left:{$a$}] (a) at (0,0) {};
\coordinate (center) at ($(a)!0.5!(b)$);

\draw[walk] (a) -- (b);
\node (ellipse) at (center) [double-path, minimum height = .6cm] {};
\node (ellipse) at (center) [double-path, minimum height = 1.1cm] {};
\end{scope}
\end{tikzpicture}
\end{center}
\caption{Watermelon networks for $p=1,2,3,5$.}
\label{fig:wat}
\end{figure}
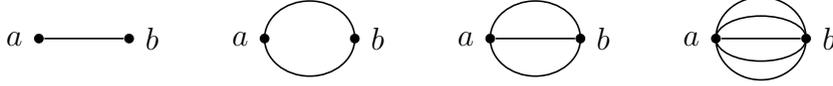

For $p \ge 1$, consider the vector of $\nl$ independent continuous-time simple random
walks on $\Z^4$:
\begin{equation}
    X(T) = \left(X^1(T_1), \dots, X^\nl(T_\nl)\right)
    \quad\text{ for } \;\; T = (T_1, \ldots, T_\nl) \in \R_+^\nl.
\end{equation}
We write $E_a$ for the expectation of $X$ with $X^k(0) = a$ for all $k$.
We define the corresponding local times, for $x \in \Z^4$, by
\begin{equation}
    L_{T_k}^k(x)
    =
    \int_0^{T_k} \1_{X^k(t) = x} dt  \quad \text{ and }
    \quad L_T(x) = L^1_{T_1}(x) + \cdots + L^\nl_{T_\nl}(x).
\end{equation}
Let $I_p(T) = \sum_{x \in \Z^4} \left(L_T(x)\right)^2$.
We write $X(T)=b$ to mean  that $X^k(T_k)=b$ for all $k =1,\ldots, p$,
and write $dT = dT_1 \cdots dT_\nl$.
The $p$-\emph{watermelon network} is then defined by
\begin{equation}
\label{e:wmdef}
\wm_{ab}^{(\nl)}(g,\nu)
=
p! \int_{\R_+^\nl} E_a \big[e^{- gI_p(T)} \1_{X(T)= b}\big] e^{-\nu \|T\|_1 } dT .
\end{equation}
By definition, $I_p(T) = \sum_{k,l=1}^p \int_0^{T_k} \int_0^{T_l} \1_{X^k(s) = X^l(t)} dsdt$,
so $I_p(T)$ measures the degree to which the $p$ walks intersect themselves, and
each other, in pairwise fashion.  The factor $e^{-gI_p(T)}$ in \eqref{e:wmdef} serves
to suppress intersections within and between the walks.
Figure~\ref{fig:wat} depicts some watermelon networks.

The $1$-watermelon network is simply the two-point function, which was
studied in \cite{BBS-saw4}.  There it was proved that the critical two-point function
obeys $\wm^{(1)}_{ab}(g,\nu_c) \sim C|a-b|^{-2}$ for small $g$.
This is as the same asymptotic behaviour \eqref{e:Casym} for the Green function.
By definition, $I_p(T) \ge \sum_{i=1}^p I_1^i(T_i)$, where the superscript $i$
indicates the self-intersection local time of  $X^i$.
This implies that
$\wm^{(p)}_{ab}(g,\nu_c) \le p!(\wm^{(1)}_{ab}(g,\nu_c))^p  \le O(|a-b|^{-2p})$.
In particular, the critical $p$-watermelon is finite for all $p \ge 1$.
Our main results provide precise asymptotics for $\wm^{(p)}_{ab}(g,\nu_c)$ for all $p \ge 1$.

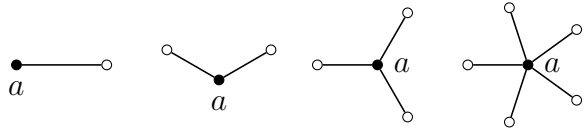
\begin{figure}
\begin{center}
\begin{tikzpicture}[scale = 0.8]

\tikzstyle{vertex}=[circle, draw = black, inner sep = 1.3pt]
\tikzstyle{vertex-pinned}=[vertex, fill = black]
\tikzstyle{walk}=[semithick]

\begin{scope}[xshift = 0cm] 
\node[vertex-pinned, label=below:{$a$}] (a) at (0,0) {};
\node[vertex] (b) at (1.5,0) {};
\draw[walk] (a) -- (b);
\end{scope}

\begin{scope}[xshift = 2.5cm, yshift=.25cm] 
\node[vertex] (b1) at (0,0) {};
\node[vertex-pinned, label=below:{$a$}] (a) at (-30:1) {};
\node[vertex] (b2) at (0:1.73) {};

\draw[walk] (b1) -- (a) -- (b2);
\end{scope}

\begin{scope}[xshift = 5cm]
\node[vertex] (b1) at (0,0) {};
\node[vertex-pinned, label=right:{$a$}] (a) at (1,0) {};
\node[vertex] (b2) at ($(a)+(60:1)$) {};
\node[vertex] (b3) at ($(a)+(-60:1)$) {};

\draw[walk] (b1) -- (a) -- (b2);
\draw[walk] (a) -- (b3);
\end{scope}

\begin{scope}[xshift = 7.5cm]
\node[vertex] (b1) at (0,0) {};
\node[vertex-pinned, label=right:{$a$}] (a) at (1,0) {};
\node[vertex] (b2) at ($(a)+(108:1)$) {};
\node[vertex] (b3) at ($(a)+(36:1)$) {};
\node[vertex] (b4) at ($(a)+(-36:1)$) {};
\node[vertex] (b5) at ($(a)+(-108:1)$) {};

\draw[walk] (b1) -- (a) -- (b2);
\draw[walk] (b3) -- (a) -- (b4);
\draw[walk] (a) -- (b5);
\end{scope}

\end{tikzpicture}

\end{center}
\caption{Star networks for $p=1,2,3,5$.}
\label{fig:starpol}
\end{figure}

For $p \ge 1$ and $a \in \Z^4$, we also define
\begin{equation}
\label{e:stardef}
\starpol^{(p)}(g,\nu)
= p! \int_{\R_+^\nl} E_a \big[ e^{- gI_p(T)} \big] e^{-\nu \|T\|_1 } dT .
\end{equation}
The right-hand side is independent of $a$ by translation invariance.
By definition, $\starpol^{(1)} (g,\nu)$ is the susceptibility $\chi(g,\nu;0)$,
while, for $p\ge 2$, $\starpol^{(p)}$ is the generating function for a \emph{star network}
of weakly self- and mutually-avoiding walks as depicted in Figure~\ref{fig:starpol}.
By a similar argument to the one employed above
for watermelon networks, $\starpol^{(p)}(g,\nu) < p! (\chi(g,\nu;0))^p$.
In particular, $\starpol^{(p)}(g,\nu) <  \infty$ for $\nu > \nu_c$.

As we will make explicit in Corollary~\ref{cor:Integral-Representation} below,
the watermelon and star networks are natural $n = 0$ analogues of the $n \ge 1$
correlation functions \eqref{e:cflim} and \eqref{e:cfsumlim}, respectively.

\subsection{Main results} \label{sec:mr}

Let $n \ge 0$ and $\nl \ge 1$ be integers.  We fix $g>0$ small and drop it
from the notation.
Exponents on logarithms turn out to be expressed in terms of
\begin{equation}
\label{e:gammadef}
\gammap_{n,\nl} = \binom{\nl}{2} \frac{n + 2}{n + 8},
\quad\quad
\gammam_{n,\nl}  = \binom{\nl}{2} \frac{2}{n + 8},
\end{equation}
with $\binom{1}{2}=0$ so that in the degenerate case
$\gammap_{n,1} = \gammam_{n,1}  =0$.  By definition, for $n=0$ we have
$\gammap_{0,\nl} = \gammam_{0,\nl}= \frac 14 \binom{p}{2}$.
We also define the constant
\begin{equation}
\label{e:bdef}
     {\sf b} = \frac{n+8}{16 \pi^2}.
\end{equation}

\begin{theorem}
\label{thm:mr-stars}
Let $d=4$.
Let $n \ge 0$ and $\nl \ge 2$ be integers.
Let $g > 0$ be sufficiently small, depending on $n,\nl$, and let
$\varepsilon = \nu - \nu_c(g; n) > 0$.
There are $g$-dependent constants $A_{n, p,\pm} >0$
such that the following hold as $\varepsilon \downarrow 0$.
\item[(i)]
For $n = 0$ and $p \ge 2$,
\begin{equation}
\label{e:starasy}
    \frac{1}{\chi(\nu;0)^p}
    \starpol^{(p)}(\nu)
    \sim
    \frac{A_{0,p,+}}{(\log \varepsilon^{-1})^{\gamma_{0,p}^+}}
    .
\end{equation}
\item[(ii)]
For $p=2$,
\begin{align}
\label{e:corrdot}
    \frac{1}{\chi^2}
    \sum_{x_1,x_2\in \Z^4}
    \inprod{ (\varphi_{x_1}\cdot \varphi_{x_2}) \,}{ |\varphi_a|^2 }_{\nu}
    &\sim
    \frac{n A_{n,2,+}}{(\log \varepsilon^{-1})^{\gamma_{n,2}^+}}
    & (n \ge 1)
    ,
\\
\label{e:corr1111}
    \frac{1}{\chi^2}
    \sum_{x_1,x_2\in \Z^4}
    \inprod{ \varphi_{x_1}^1 \varphi_{x_2}^1 }{ (\varphi_a^1)^2 }_{\nu}
    &\sim
    \frac{n-1}{n}\frac{A_{n,2,-}}{(\log \varepsilon^{-1})^{\gamma_{n,2}^-}}
    & (n \ge 2)
    ,
\\
\label{e:corr1122}
    \frac{1}{\chi^2}
    \sum_{x_1,x_2\in \Z^4}
     \inprod{ \varphi_{x_1}^1 \varphi_{x_2}^1 }{ (\varphi_a^2)^2 }_{\nu}
    &\sim
    -\frac{1}{n}\frac{A_{n,2,-}}{(\log \varepsilon^{-1})^{\gamma_{n,2}^-}}
    & (n \ge 2)
    .
\end{align}
\item[(iii)]
The amplitudes obey, as $g\downarrow 0$,
\begin{align}
\label{e:mrcasy1}
    \mrc_{n,p,\pm} &=
    p!
    ({\sf b}g)^{-\gamma_{n,p}^\pm}
    (1+O(g))
    .
\end{align}
\end{theorem}

For the case $n \ge 1$, it is part of the statement of the following theorem that
the critical correlation functions on $\Z^4$ exist
in the sense of \eqref{e:cflim}.
We write error estimates
as $|a-b|\to\infty$ in terms of
\begin{equation}
\label{e:Ecaldef}
    \Ecal_{ab}^{(\nl )} =
    \begin{cases}
    O\big((\log|a-b|)^{-1}\big) & (\nl = 1),
    \\
    O\big((\log\log|a-b|)(\log|a-b|)^{-1}\big) & (\nl \ge 2).
    \end{cases}
\end{equation}
We again drop $g$ from the notation, and, in particular, write
the critical value as $\nu_c(n)$ for all $n \ge 0$.

\begin{theorem}
\label{thm:mr-ab}
Let $d=4$.
Let $n \ge 0$ and $\nl \ge 1$ be integers.
Let $g > 0$ be sufficiently small, depending on $n,\nl$.
There are $g$-dependent constants $\mrc_{n,p,\pm}' >0$
such that the following hold as $|a-b|\to\infty$.
\item[(i)]
For $n = 0$ and $p \ge 1$,
\begin{equation}
\label{e:Wasy-0}
    \wm_{ab}^{(p)}(\nu_c(0)) =
    \frac{\mrc_{0,p,+}'}{(\log |a-b|)^{2\gammap_{0,\nl}}}
    \frac{1}{|a-b|^{2\nl}}
    \left(1+\Ecal_{ab}^{(\nl )}\right).
\end{equation}
\item[(ii)]
For $n\ge 1$ and $p=1,2$,
\begin{align}
\label{e:Wasy-1a}
    \inprod{\varphi_a^1}{\varphi_b^1}_{\nu_c(n)}
    & = \frac{\mrc_{n,1,+}'}{|a-b|^{2}}
    \left(1+\Ecal_{ab}^{(1)}\right),
\\
\label{e:Wasy-1b}
    \inprod{|\varphi_a|^2}{|\varphi_b|^2}_{\nu_c(n)}
    & = \frac{n\mrc_{n,2,+}'}{(\log |a-b|)^{2\gammap_{n,2}}}
    \frac{1}{|a-b|^{4}}
    \left(1+\Ecal_{ab}^{(2)}\right).
\end{align}
\item[(iii)]
For $n\ge 2$ and $p=2$,
\begin{align}
\label{e:Wasy-2}
    \inprod{(\varphi_a^1)^2}{(\varphi_b^1)^2}_{\nu_c(n)}
    & =
    \frac 1n
    \left( \frac{(n-1)\mrc_{n,2,-}'}{(\log |a-b|)^{2\gammam_{n,2}}}
    + \frac{\mrc_{n,2,+}'}{(\log |a-b|)^{2\gammap_{n,2}}} \right)
    \frac{1}{|a-b|^{4}}
    \left(1+\Ecal_{ab}^{(2)}\right),
\\
\label{e:phi12p-new}
     \inprod{(\varphi_a^1)^2}{(\varphi_b^2)^2}_{\nu_c(n)}
     &=
     \frac 1n
    \left( -\frac{\mrc_{n,2,-}'}{(\log |a-b|)^{2\gammam_{n,2}}}
    + \frac{\mrc_{n,2,+}'}{(\log |a-b|)^{2\gammap_{n,2}}} \right)
    \frac{1}{|a-b|^{4}}
    \left(1+\Ecal_{ab}^{(2)}\right).
\end{align}
\item[(iv)]
The amplitudes obey, as $g\downarrow 0$,
\begin{align}
\label{e:mrcasy}
    \mrc_{n,p,\pm}' &= \frac{p!}{(2\pi)^{2p}}
    \left( {\sf b}g
    \right)^{-2\gamma_{n,p}^\pm}
    (1+O(g)).
\end{align}
\end{theorem}

In Theorem~\ref{thm:mr-ab}, the interesting asymptotic behaviour as $|a-b|\to\infty$
is stressed.   However, our proof applies more generally, and gives
the following result for the case $a=b$, which provides a
natural continuity statement as $g \downarrow 0$.

\begin{theorem}
\label{thm:mr-aa}
Let $d=4$.
Let $n \ge 0$ and $\nl \ge 1$ be integers.
Let $g > 0$ be sufficiently small, depending on $n,\nl$.  Then, as $g \downarrow 0$,
\begin{align}
\label{e:Wasy-0-aa}
    \wm_{aa}^{(p)}(\nu_c(0)) & =
    G_{aa}^p
    (p! +O(g)) & (p\ge 1),
\\
\label{e:Wasy-1a-aa}
    \inprod{\varphi_a^1}{\varphi_a^1}_{\nu_c(n)}
    & = G_{aa}
    (1+O(g)) & (n\ge 1),
\\
\label{e:Wasy-1b-aa}
    \inprod{|\varphi_a|^2}{|\varphi_a|^2}_{\nu_c(n)}
    & = G_{aa}^2
    (2!n +O(g)) & (n\ge 1),
\\
\label{e:phi12p-new-aa}
     \inprod{(\varphi_a^1)^2}{(\varphi_a^2)^2}_{\nu_c(n)}
     &=
     O(g) & (n\ge 2).
\end{align}
\end{theorem}

It is worth mentioning that even to prove that the left-hand sides of
\eqref{e:Wasy-0}--\eqref{e:phi12p-new} or
\eqref{e:Wasy-0-aa}--\eqref{e:phi12p-new-aa} are \emph{finite} is a nontrivial result.
For example, it remains an open problem to
prove that the generating function for self-avoiding polygons
is finite at the
critical point in dimensions $d=2,3$ (see \cite[Section~8.1]{MS93});
this is analogous to $\wm_{aa}^{(1)}(\nu_c(0))$.

For $\nl=1$, the right-hand sides of \eqref{e:Wasy-0}--\eqref{e:Wasy-1a} give simply
$\mrc_{n,1,+}' |a-b|^{-2}(1+ O(\log|a-b|)^{-1})$.  The decay of this particular correlation function,
namely the \emph{critical two-point function},
is usually written in terms of the critical exponent $\eta$ as $|a-b|^{-(d-2+\eta)}$,
so this is a statement that $\eta$ takes its mean-field value
$\eta =0$ for all $n \ge 0$, with no logarithmic
correction to the leading behaviour.
The power $|a-b|^{-2}$ arises in our analysis via the right-hand side of \eqref{e:Casym}.

Special cases of Theorem~\ref{thm:mr-ab} have been proven previously.
For $(n,\nl)=(0,1)$, \eqref{e:Wasy-0} is the main result of \cite{BBS-saw4};
a related result for a model that involves neither a lattice nor walks appears in \cite{IM94}.
For  $(n,\nl)=(1,1)$, \eqref{e:Wasy-1a} is the main result of \cite{GK85}.
For $(n,\nl)=(1,2)$, \eqref{e:Wasy-1b}
was proved in \cite{GK86}; in this case the leading behaviour is
$|a-b|^{-4}(\log |a-b|)^{-2/3}$.
For a related model in which an ultraviolet cutoff replaces the lattice setting,
a version of \eqref{e:Wasy-1a} for
the case $(n,\nl)=(1,1)$ appears in \cite{FMRS87}.
The results:
(i)
\eqref{e:Wasy-0} for $p \ge 2$,
(ii)
\eqref{e:Wasy-1a}--\eqref{e:Wasy-1b} for $n \ge 2$ and $p =1,2$, and
(iii) \eqref{e:Wasy-2}--\eqref{e:phi12p-new} for $n \ge 2$ and $p=2$,
are new as rigorous results.

Concerning \eqref{e:mrcasy}, the factor $(2\pi)^{-2p}$ arises from the $p^{\rm th}$ power
of the Green function via \eqref{e:Casym}.
The power of $g$ in \eqref{e:mrcasy} matches the power of the logarithm in the term where
the amplitude appears.  The combination $g\log|a-b|$ is natural since there are
no logarithmic corrections for the Gaussian case $g=0$.

The exponents $\gamma_{n,\nl}^\pm$ in Theorem~\ref{thm:mr-stars} and the exponents
$2p, 2\gamma_{n,\nl}^\pm$ in Theorem~\ref{thm:mr-ab}
are predicted to be universal.
In particular, the $n=1$ exponents of \eqref{e:corrdot} and
\eqref{e:Wasy-1a}--\eqref{e:Wasy-1b} are predicted to apply to the Ising model, and the exponents
of \eqref{e:corrdot}--\eqref{e:corr1122} and \eqref{e:Wasy-1a}--\eqref{e:phi12p-new} for
$n \ge 2$ are predicted to apply to the $O(n)$ model, including the classical $XY$
(or rotor) model
for $n=2$, and the classical Heisenberg model for $n=3$.

Similarly,
the $n=0$, $\nl \ge 1$ case of \eqref{e:starasy} and \eqref{e:Wasy-0}, namely (with $\binom{1}{2}=0$),
\begin{equation}
\label{e:Wp}
    \frac{1}{\chi(\nu)^p}
    \starpol^{(p)}(\nu)
    \sim
    \frac{A_{0,p,+}}{(\log \varepsilon^{-1})^{\frac 14 \binom{\nl}{2}}}
    ,
\qquad
    \wm_{ab}^{(\nl)}(\nu_c)
    \sim \frac{\mrc_{0,p,+}'}{|a-b|^{2\nl} (\log |a-b|)^{\frac 24 \binom{\nl}{2}}},
\end{equation}
are predicted to apply
to the 4-dimensional strictly self-avoiding walk.
For $\nl \ge 2$ \emph{independent} WSAWs,
$\chi^{-p}\starpol^{(p)}$ is identically equal to $1$,
and $\wm_{ab}^{(\nl)}(\nu_c)$ is asymptotic to a multiple of $|a-b|^{-2\nl}$.
The logarithmic corrections in \eqref{e:Wp}
for $p$ weakly mutually-avoiding walks
are consistent with the interpretation that the intersection of each of the $\binom{\nl}{2}$ pairs
of walks at a vertex gives rise to a penalty
$(\log \varepsilon^{-1})^{-1/4}$ or
$(\log |a-b|)^{-\frac 14 }$
paid by each pair for joining, despite their penchant to avoid.
Related results were obtained via a non-rigorous renormalisation analysis in
\cite{Dupl86}, and a detailed non-rigorous general treatment of polymer networks,
including also dimensions below $4$, can be found in \cite{Dupl89a}.
For the case of simple random walk, the formula for star networks
in \eqref{e:Wp} is reminiscent of the fact, proved in \cite{Lawl92}, that $p$ independent
simple random walks started from the origin in $\Z^4$
do not have pairwise intersections before leaving the ball of radius $n$,
with probability asymptotic to $(\log n)^{-\frac 12 \binom{p}{2}}$
(see \cite{Dupl88} for a non-rigorous renormalisation analysis).
A number of authors have studied related matters for the case of two simple random walks
\cite{Aize85,FF85,Lawl85,Park89}.
For spread-out models of strictly SAW in dimensions $d>4$,
rigorous results for arbitrary graphical
networks were obtained in \cite{HJSS04}.  These results for $d>4$ include
a statement analogous to \eqref{e:Wp} for all $\nl \ge 1$, but
there is no logarithmic correction and the asymptotic behaviour is
simply ${\rm const} |a-b|^{-p(d-2)}$.
See also \cite[Theorem~1.5.5]{MS93} for
nearest-neighbour strictly SAW for $d \ge 6$.

For the case $n \ge 2$ and $\nl=2$, since $\gammam_{n,2} = \frac{2}{n+8} <
\frac{n+2}{n+8} = \gammap_{n,2}$,
Theorem~\ref{thm:mr-ab} gives (for $i \neq j$)
\begin{align}
\label{e:ii}
    \inprod{(\varphi^i_a)^2}{(\varphi^i_b)^2}_{\nu_c(n)}
    & \sim \frac{n-1}{n} \frac{\mrc_{n,2,-}'}{|a-b|^{4}(\log |a-b|)^{4/(n+8)}},
    \\
\label{e:ij}
    \inprod{(\varphi^i_a)^2}{(\varphi^j_b)^2}_{\nu_c(n)}
    & \sim -\frac{1}{n} \frac{\mrc_{n,2,-}'}{|a-b|^{4}(\log |a-b|)^{4/(n+8)}}.
\end{align}
On the other hand, by \eqref{e:Wasy-1b},
\begin{equation}
\label{e:norm2}
    \inprod{|\varphi_a|^2}{|\varphi_b|^2}_{\nu_c(n)}
    \sim  n\frac{\mrc_{n,2,+}'}{|a-b|^{4}(\log |a-b|)^{2(n+2)/(n+8)}}.
\end{equation}
Thus, for an individual component, $(\varphi_a^i)^2$
is more highly correlated with $(\varphi_b^i)^2$,
than is $|\varphi_a|^2$ with $|\varphi_b|^2$,
due to cancellations with the negative correlations of $(\varphi_a^i)^2$
with $(\varphi_b^j)^2$ for $i \neq j$.
Negative correlations for different components at the \emph{same} point are to be expected
since $\pair{|\varphi_a|^2}_{\nu_c(n)} <\infty$
by \eqref{e:Wasy-1a-aa},
and therefore the field
has a typical size, so making one component large must come at the cost of making
one component small.
This is similar to the fact that the squares of different components
of a uniform random variable on the sphere are negatively correlated by the
length constraint.
Our results show how this effect persists over long distances at the critical point.

In the physics literature, $|\varphi|^2$ is referred to as the \emph{energy operator},
so \eqref{e:norm2} gives the asymptotic behaviour of the energy operator correlation.
We are not aware of any reference from the physics literature where \eqref{e:ii}--\eqref{e:norm2}
are stated, though it has been
observed that the reduction of symmetry (as in
the left-hand sides of \eqref{e:ii}--\eqref{e:ij}
compared to \eqref{e:norm2}) can lead to a change in critical exponents
\cite{Ahar76,Card96}.
Reduction of symmetry plays an important role in our proof of \eqref{e:ii}--\eqref{e:norm2}:
the $O(n)$ invariant case \eqref{e:norm2} has a different renormalisation group flow
than the non-invariant cases \eqref{e:ii}--\eqref{e:ij}.

In \cite{BBS-phi4-log}, the asymptotic behaviour of the \emph{specific heat}
\begin{equation}
\label{e:cHdef}
  c_H(\nu) =  \tfrac 14 \sum_{b\in\Z^4}
  \inprod{|\varphi_a|^2}{|\varphi_b|^2}_{\nu_c + \varepsilon}
\end{equation}
(with the infinite volume limit defined similarly to \eqref{e:cfsumlim})
is studied in the limit $\epsilon \downarrow 0$.  It is proved in \cite{BBS-phi4-log},
confirming predictions of \cite{LK69,WR73},
that for $d=4$, for small $g >0$, and for $n \ge 1$, there exists
  $D(n)=D(g,n)>0$ such that, as $\varepsilon \downarrow 0$,
  \begin{equation}
  \label{e:cHasy}
    c_H(\nu_c+\varepsilon)
    \sim D(n)
    \begin{cases}
      (\log \varepsilon^{-1})^{(4-n)/(n+8)} & (n=1,2, 3)\\
      \log \log \varepsilon^{-1} & (n=4)\\
      1 & (n>4).
    \end{cases}
  \end{equation}
Interestingly, it was pointed out in \cite{LK69}, where \eqref{e:cHasy} was first
derived non-rigorously, that the
universal aspects of the phase transition for the 4-dimensional $|\varphi|^4$
model with $n=1$ should also apply
to the phase transition in a 3-dimensional uniaxial ferroelectric
substance, and the $\frac{4-1}{1+8}=\frac 13$ power in \eqref{e:cHasy}
was subsequently confirmed experimentally for the dipolar Ising ferromagnet
${\rm LiTbF}_4$ in \cite{AKG75}.
The result \eqref{e:cHasy}
is complemented by \eqref{e:norm2}, which implies that,  as $R \to \infty$,
\begin{equation}
    \sum_{b\in \Z^4: |b|\le R}
    \inprod{|\varphi_a|^2}{|\varphi_b|^2}_{\nu_c(n)}
        \sim c(n)
    \begin{cases}
      (\log R)^{(4-n)/(n+8)} & (n=1,2, 3)\\
      \log \log R & (n=4)\\
      1 & (n>4).
    \end{cases}
\end{equation}
Neither of \eqref{e:ii}-\eqref{e:ij} is summable for any $n \ge 2$, nor
are \eqref{e:Wasy-0} or \eqref{e:Wasy-1b} summable for $p=2$. The failure
of summability of \eqref{e:Wasy-0}--\eqref{e:Wasy-1a} for $p=1$ accords
with the divergence of the susceptibility at the critical point.

Notable features of our method of proof are that:
\begin{enumerate}[(i)]
\item
The case of $n=0$ is united with the
case $n \ge 1$ despite the apparent differences in the definitions of the
WSAW and $|\varphi|^4$ models.
\item
The proof proceeds via second-order perturbative
calculations \cite{BBS-rg-pt}
of the sort found in non-rigorous renormalisation group calculations in
the physics literature, but here with all error terms rigorously controlled via a
general renormalisation group method \cite{BS-rg-IE,BS-rg-step}.
\item
There is a different renormalisation group
flow due to the reduced $O(n)$ symmetry
in the proof of \eqref{e:ii}--\eqref{e:ij}, compared to
the $O(n)$ symmetric case of \eqref{e:norm2}.
This is the origin of the different powers in the logarithmic corrections.
\end{enumerate}
First steps towards the application of the method
to critical correlation functions were made
in \cite{BBS-saw4}, where the case $n=0$, $p=1$ was studied.  Here we significantly
extend the methods applied in \cite{BBS-saw4} to obtain a much more general
result, which identifies logarithmic corrections that appear when $p\ge 2$ and
reveals the new phenomena seen in \eqref{e:ii}--\eqref{e:norm2}
for the case $n \ge 2$, $p = 2$.

For the continuous-time weakly self-avoiding walk on a $4$-dimensional
\emph{hierarchical} lattice, much more has been proved \cite{BEI92,BI03c,BI03d}; in particular, the
asymptotic behaviour of the end-to-end distance is identified in \cite{BI03c}.
A rigorous analysis of the $4$-dimensional hierarchical Ising model is
given in \cite{HHW01}.
The decay of the analogue of $\langle |\varphi_a|^2; |\varphi_b|^2 \rangle_{\nu_c}$ slightly
below the critical dimension has been studied rigorously in a hierarchical
setting of quantum fields over the $p$-adics \cite{ACG13}.
A renormalisation group trajectory for the continuous-time weakly-self avoiding walk 
also slightly below the critical dimension has been constructed in \cite{MS08}.
For $n \ge 2$ and dimensions $d \ge 3$, long range order and symmetry breaking in a
related setting, in the
phase corresponding
to very large negative $\nu$, has been studied in \cite{Bala95, BO99}.

\section{Reformulation of the problem}

Initially,
the definitions of the $|\varphi|^4$ and WSAW models appear quite different.
In this section, we develop a unified formulation of the problems addressed
in our main theorems.
We begin in Sections~\ref{sec:ivl-wsaw}--\ref{sec:ir} by recalling and extending
the connection between the $|\varphi|^4$ and WSAW models,
which arises from an integral representation of WSAW.  Such integral
representations are discussed at length in \cite{BIS09}.
Using the integral representation, the WSAW star and
watermelon networks are expressed in terms
of functional integrals which involve a
complex boson field $\phi$ and a fermion field $\psi$, with quartic
self-interaction.  The renormalisation group method we apply is well suited to
the analysis of such problems with or without the fermion field, and both models
can be handled together, once we replace the Gaussian expectation for the
$|\varphi|^4$ model by a Gaussian super-expectation, as discussed in Section~\ref{sec:covGa}.
 The specific correlation functions studied in
our main theorems are obtained via the use of \emph{observable}
and \emph{external}
fields, which we
introduce in Section~\ref{sec:of}--\ref{sec:ef}. There we reformulate the
basic problem in a unified manner for both models in terms of these auxiliary fields.

\subsection{Infinite volume limit for WSAW} \label{sec:ivl-wsaw}

The integral representation for WSAW requires finite volume,
and we first
show how the watermelon and star networks on $\Z^d$ can be approximated by
networks on a torus.
Let $E_{a}^{N}$ denote the expectation corresponding to $p$ independent
continuous-time simple random walks on the torus $\Lambda_N$, started at $a \in \Lambda_N$.
Let $b \in \Lambda_N$.
For $\nl \ge 1$,
we define a finite volume version of the $\nl$-watermelon \eqref{e:wmdef} by
\begin{equation}
\label{e:wmNdef}
\wm_{ab,N}^{(\nl)}(g,\nu)
= p!
\int_{\R_+^\nl} E_a^N \big[e^{- gI_p(T)} \1_{X(T)= b}\big] e^{-\nu \|T\|_1 } dT ,
\end{equation}
and of the star network \eqref{e:stardef} by
\begin{equation}
\label{e:starNdef}
 \starpol^{(p)}_N(g,\nu)
= p! \int_{\R_+^\nl} E_a^N \big[ e^{- gI_p(T)} \big] e^{-\nu \|T\|_1 } dT
\end{equation}
(which is independent of $a$ by translation invariance).

By the argument under \eqref{e:wmdef}, $\wm_{ab,N}^{(\nl)} \le p!(\wm_{ab,N}^{(1)})^\nl$.
By the Cauchy--Schwarz inequality,
$T_1= \sum_{x\in \Lambda} L_{T_1}(x) \le (|\Lambda|I_1(T_1))^{1/2}$,
so $I_1(T_1) \ge T_1^2/|\Lambda|$, from which we conclude that
$\wm_{ab,N}^{(1)}$, and hence $\wm_{ab,N}^{(\nl)}$, is finite for all $g>0$ and $\nu\in\R$.
Similarly, for all $g>0$ and $\nu\in\R$, $\starpol^{(p)}_N(g,\nu) <\infty$.

\begin{prop}\label{prop:infinite-volume-limit}
For $ d >0$ and $g>0$,
\begin{equation}
\label{e:wmNlim}
    \wm^{(\nl)}_{ab}(g,\nu_c) = \lim_{\nu \downarrow \nu_c} \wm^{(\nl)}_{ab}(g,\nu)
    = \lim_{\nu \downarrow \nu_c} \lim_{N \to \infty} \wm^{(\nl)}_{ab,N}(g,\nu),
\end{equation}
and, for $\nu\in \R$,
\begin{equation}
\label{e:Givlc}
    \starpol^{(p)}(g,\nu)
    =
    \lim_{N \to \infty}
    \starpol_N^{(p)}(g,\nu) .
\end{equation}
\end{prop}

\begin{proof}
The first equality in \eqref{e:wmNlim} holds by monotone convergence.
An elementary proof of the second equality is given in
\cite[Proposition~\ref{saw4-prop:Glims}]{BBS-saw4} for the case of $\nl=1$.
That proof generalises directly to the case of $\nl \ge 1$, and we omit the details.

The proof of \eqref{e:Givlc} for general $p\ge 1$ is a straightforward
generalisation of the proof for $p=1$ given in
\cite[Lemma~\ref{log-lem:suscept-finvol}]{BBS-saw4-log}, and again we omit
the details.  Both sides of \eqref{e:Givlc}
are finite for $\nu>\nu_c$, but the proof gives equality also when both sides
are infinite.
\end{proof}

\subsection{Integral representation for WSAW} \label{sec:ir}

\newcommand{\carL}{M}

Let $\carL = |\Lambda_N| = L^{Nd}$.
Let $u_1,v_1,\ldots, u_\carL,v_\carL$ be standard coordinates on $\R^{2\carL}$.
Then $du_1 \wedge dv_1 \wedge \cdots \wedge du_\carL \wedge dv_\carL$ is the
standard volume form on $\R^{2\carL}$, where $\wedge$ denotes the
anticommuting wedge product.
The one-forms $du_x$, $dv_y$ generate the
Grassmann algebra of differential forms
on $\R^{2\carL}$.
We multiply differential forms using the wedge product, but
for notational simplicity we do not display the wedge explicitly, and write,
e.g., $du_xdv_y$ in place of $du_x \wedge dv_y$.  The order of differentials
in a product therefore matters.

For $p \ge 0$, a $p$-\emph{form} is a function of $u,v$ times a product of $p$ differentials,
or a sum of such.
In general, a \emph{form} $K$ is a sum of $p$-forms for $p \ge 0$,
the largest such $p$ is called the \emph{degree} of $K$ and the individual $p$-forms are called
the \emph{degree}-$p$ part of $K$.
A form which is a sum of $p$-forms for even $p$ only is called \emph{even}.
The integral of a differential form over $\R^{2\carL}$ is defined to be zero unless
the form has degree $2\carL$.
A form of degree $2\carL$ can be written as
$K = f(u,v) du_1 dv_1 \cdots du_\carL dv_\carL$, and we define
\begin{equation}
\label{e:Kintdef}
    \int K = \int_{{\mathbb R}^{2\carL}} f(u,v) du_1 dv_1 \cdots du_\carL dv_\carL,
\end{equation}
where the right-hand side is the  Lebesgue integral of $f$ over $\R^{2\carL}$.

We set $\phi_x = u_x + i v_x$, $\phib_x = u_x-iv_x$
and $d\phi_x = du_x+idv_x$, $d\phib_x = du_x-idv_x$, for $x \in \Lambda$.
Since the wedge product is
anticommutative, the following pairs all anticommute for every $x,y\in \Lambda$:
$d\phi_x$ and $d\phi_y$, $d\phib_x$ and $d\phi_y$,
$d\phib_x$ and $d\phib_y$.
Also,
\begin{equation}
\label{e:duv}
    d\phib_x d\phi_x = 2i du_x dv_x.
\end{equation}
The integral $\int f(\phi,\phib)\prod_{x\in \Lambda}d\phib_xd\phi_x$ is thus given by
$(2i)^\carL$ times the Lebesgue integral of $f(u+iv,u-iv)$ over $\R^{2\carL}$.
The product over $x$ can be taken in any order, since each factor
$d\phib_xd\phi_x$
is even.
We write
\begin{equation}
\label{e:phipsi}
    \psi_x = \frac{1}{(2\pi i)^{1/2}} d \phi_x,
    \quad
    \psib_x = \frac{1}{(2\pi i)^{1/2}} d \phib_x,
\end{equation}
with a fixed choice of the square root.
Then
\begin{equation}
\psib_x \psi_x =\frac{1}{2\pi i} d\phib_x d\phi_x = \frac{1}{\pi}du_x dv_x.
\end{equation}
We refer to $\phi,\bar\phi$ as the \emph{boson} field and to $\psi,\bar\psi$ as
the \emph{fermion} field.

Let $\Ncal^\varnothing$ denote the algebra of even differential forms.
An element $K\in\Ncal^\varnothing$
can be written as
\begin{equation} \label{e:psipsib}
  K=
  \sum_{k=0}^{2\carL}
  \sum_{s,t: s + t=2k}
  \sum_{x_1,\ldots,x_s\in \Lambda}
  \sum_{y_1,\ldots, y_t \in \Lambda} K_{x,y}
  \psi^x \bar\psi^{y}
  ,
\end{equation}
where $x=(x_1,\ldots,x_s)$, $y=(y_1,\ldots,y_t)$,
$\psi^x = \psi_{x_1}\cdots\psi_{x_s}$,
$\psib^y = \psib_{y_1}\cdots\psib_{y_t}$, and where each $K_{x,y}$
(including the degenerate case $s=t=0$) is a function of
$(\phi,\phib)$.
We fix a positive integer $p_\Ncal \ge \max\{10,2p+4\}$ and impose the smoothness condition
that elements of $\Ncal^\varnothing$ are such that the coefficients $K_{x,y}$
are in $C^{p_\Ncal}$
(the reason for this particular choice of $p_\Ncal$ is discussed
in Section~\ref{sec:pNcal}.)

Given a finite index set $J$, let $K=(K_j)_{j\in J}$ with each $K_j \in \Ncal^\varnothing$.
Let $K_j^{0}$ denote the degree-zero part
of $K_j$.
Given a $C^\infty$ function $F : \R^{J} \to\mathbb C$,
we define $F(K)$ by
its power series about the degree-zero part of $K$
(which we assume to be real), i.e.,
\begin{equation}
\label{e:Fdef}
    F(K) = \sum_{\alpha} \frac{1}{\alpha !}
    F^{(\alpha)}(K^{0})
    (K - K^{0})^{\alpha}.
\end{equation}
Here $\alpha$ is a multi-index, with $\alpha ! = \prod_{j\in J}\alpha_j !$,
and
$(K - K^{0})^{\alpha}
=\prod_{j\in J} (K_{j} - K_{j}^{0})^{\alpha_{j}}$.
The summation terminates as soon as
$\sum_{j\in J}\alpha_j=\carL$ since higher-order forms must vanish,
and the order of the product on the right-hand side does not matter since each $K_j$
is assumed to be even.

For $x \in \Lambda$,
we define the differential forms
\begin{gather}
\label{e:taudef}
\tau_x = \phi_x \bar{\phi_x} + \psi_x \bar{\psi_x},
\\
\label{e:addDelta}
\tau_{\Delta,x} = \frac{1}{2}
\Big( \phi_{x} (- \Delta \bar\phi)_{x} + (- \Delta \phi)_{x} \bar\phi_{x} +
\psi_{x}  (- \Delta \bar\psi)_{x} + (- \Delta \psi)_{x}  \bar\psi_{x} \Big),
\end{gather}
where $\Delta = \Delta_\Lambda$ is the lattice Laplacian defined above \eqref{e:Vdef1}.
The forms $\tau_x$ and $\tau_{\Delta,x}$ both have real degree-zero parts.
The following proposition is a minor extension of \cite[Theorem~5.1]{BIS09};
we provide a self-contained proof in Appendix~\ref{app:intrep}.
The integrand on the left-hand side
of \eqref{e:intrep1az} is defined as in \eqref{e:Fdef}, e.g.,
$e^{-\tau_x} = e^{-|\phi_x|^2}(1+\psi_x\bar\psi_x)$,
and the integral is as in \eqref{e:Kintdef}.
On the right-hand side, $S_p$ denotes the set of permutations of $1,\ldots,p$.
\begin{prop}
\label{prop:Integral-Representation}
For $d>0$, $g>0$, $\nu \in \R$, $p \ge 1$, and $A=(a_1,\ldots,a_p)$, $B=(b_1,\ldots,b_p)$
with each $a_i,b_j \in \Lambda_N$,
\begin{equation}
\label{e:intrep1az}
\begin{aligned}
    \int e^{-\sum_{x\in\Lambda} \big(\tau_{\Delta ,x} + g\tau_x^2 + \nu\tau_x \big)}
    \bar{\phi}_{a_1}\cdots &\phib_{a_p} \phi_{b_1} \cdots \phi_{b_p} \\
    &=
    \sum_{\sigma\in S_p}
    \int_{\R_+^\nl} E_A^N \big[e^{- gI_p(T)} \1_{X(T)= \sigma(B)}\big] e^{-\nu \|T\|_1 } dT
    ,
\end{aligned}
\end{equation}
where on the right-hand side $X^i(0)=a_i$ and $X^i(T_i)=\sigma(b_i)$.
\end{prop}

\begin{cor}
\label{cor:Integral-Representation}
For $d>0$, $g>0$, $\nu \in \R$, $p \ge 1$, and $a,b,b_1, \ldots, b_p \in \Lambda_N$,
\begin{align}
\label{e:int-representation-stars}
    \starpol^{(p)}_N(g,\nu)
    &=
    \sum_{b_1,\ldots,b_p\in\Lambda_N}
    \int e^{-\sum_{x\in\Lambda_N} \big(\tau_{\Delta ,x} + g\tau_x^2 + \nu\tau_x \big)}
     \bar{\phi}_a^{\nl} \phi_{b_1}\cdots\phi_{b_p}
     ,
    \\
\label{e:int-representation-wm}
\wm_{ab,N}^{(\nl)}(g,\nu)
&=
\int
e^{-\sum_{x\in\Lambda_N} \big(\tau_{\Delta ,x} + g\tau_x^2 + \nu\tau_x \big)} \bar{\phi}_a^{\nl} \phi_b^\nl.
\end{align}
\end{cor}

\begin{proof}
This is an immediate consequence of Proposition~\ref{prop:Integral-Representation}
and the definitions of $\starpol^{(p)}_N,\wm_{ab,N}^{(\nl)}$.
\end{proof}

\subsection{Change of variables and Gaussian approximation} \label{sec:covGa}

To unify the treatment of the $|\varphi|^4$ and WSAW models,
for the $|\varphi|^4$ model instead of \eqref{e:taudef}--\eqref{e:addDelta}
we define
\begin{equation}
\label{e:tauphi}
\tau_x = \half |\varphi_x|^2,
\quad \tau_x^2 = \tfrac14 |\varphi_x|^4,
\quad \tau_{\Delta,x} = \half \varphi_x \cdot (-\Delta \varphi)_x.
\end{equation}
For either model, given $g,\nu,z\in \R$,
we write
\begin{equation}
U_{g,\nu,z;x} = g \tau_x^2 + \nu \tau_x + z\tau_{\Delta ,x}.
\end{equation}
The polynomial $U_{g,\nu,1;x}$ appears in \eqref{e:Pdef} with $\tau$ and $\tau_\Delta$
interpreted as in \eqref{e:tauphi}, and it appears in the right-hand sides
of \eqref{e:int-representation-stars} and \eqref{e:int-representation-wm} with the interpretation \eqref{e:taudef}--\eqref{e:addDelta}.
Given $X \subset \Lambda$ and $g_0,\nu_0,z_0\in \R$, we define
\begin{equation}
\label{e:Vtil0def}
U_{0} (X)
= \sum_{x\in X} U_{g_0,\nu_0,z_0;x}
= \sum_{x\in X} \big(g_{0} \tau_x^2 + \nu_{0} \tau_x + z_{0}\tau_{\Delta ,x}\big).
\end{equation}

To write our principal quantities as perturbations of a Gaussian, we make
an appropriate change of variables.
For $|\varphi|^4$, given $z_0 > -1$
and $m^2 > 0$, by definition,
\begin{equation} \label{e:Vsplit}
  U_{g,\nu,1;x}(\varphi)
  = U_{0,m^2,1;x}((1+z_0)^{-1/2}\varphi)
  + U_{g_0,\nu_0,z_0;x}((1+z_0)^{-1/2}\varphi)
  ,
\end{equation}
with
\begin{equation} \label{e:gg0}
  g_0 = g(1+z_0)^2, \quad \nu_0 = (1+z_0)\nu-m^2
  .
\end{equation}
The equations \eqref{e:gg0} can equivalently be written as
\begin{equation}
\label{e:g0g}
    g = \frac{g_0}{(1+z_0)^2}, \quad
    \nu = \frac{\nu_0+m^2}{1+z_0}
  .
\end{equation}
For the moment, we regard $m^2,z_0$ as parameters that can be chosen arbitrarily.
In Section~\ref{sec:npflow},
we make careful choices of these, corresponding to ``physical mass'' and
``wave function renormalisation'' in the physics literature.
Let $C=(-\Delta+m^2)^{-1}$, with $\Delta$ the discrete Laplacian on $\Lambda_N$
(acting on scalar functions).
For $|\varphi|^4$, the Gaussian expectation with covariance $C$ is defined by
\begin{equation}
\label{e:Gex}
    \Ex_C F = \pair{F}_{0,m^2,N} .
\end{equation}
Given a function $F(\varphi)$ we write
$F'(\varphi)=F((1+z_0)^{1/2}\varphi)$.
Using \eqref{e:Vsplit} and the change of variables
$\varphi_x \mapsto \varphi'=(1+z_0)^{1/2}\varphi_x$,
we obtain
\begin{equation}
\label{e:ExF}
    \pair{F}_{g,\nu,N}
    = \frac{\Ex_C  F' e^{-U_{0}(\Lambda)}}{\Ex_C e^{-U_{0}(\Lambda)}} .
\end{equation}

For WSAW, we use the Gaussian \emph{super-expectation}
\begin{equation}
\label{e:Gsex}
    \Ex_C F = \int F e^{-\sum_{x\in\Lambda}( \tau_{\Delta,x} + m^2 \tau_x)},
\end{equation}
defined for
$F \in \Ncal^\varnothing$ such that the integral exists.
In \eqref{e:Gsex}, $\tau_x$ and $\tau_{\Delta,x}$ are now the differential
forms defined in \eqref{e:taudef}--\eqref{e:addDelta}, which
incorporate both boson and fermion fields.
Such integrals are discussed at length for our context in \cite{BIS09, BS-rg-norm}.
By Corollary~\ref{cor:Integral-Representation} and an analogue of \eqref{e:Vsplit},
\begin{equation}
\label{e:wmEC}
    \wm^{(\nl)}_{ab,N}(g,\nu)
    =
    (1+z_0)^p \Ex_C\left(e^{-U_0(\Lambda)} \bar\phi_a^\nl \phi_b^\nl\right)
    .
\end{equation}
Unlike in \eqref{e:ExF}, there is no division by a partition function.
In fact, as a result of \emph{supersymmetry} (see \cite{BIS09}), here
$\Ex_C e^{-U_0(\Lambda)}= 1$.
In addition, since $\Ex_C (e^{-U_0(\Lambda)}\phib_a^p)= \Ex_C (e^{-U_0(\Lambda)}\phi_b^p)= 0$,
there is no subtracted term in \eqref{e:wmEC},
like there is in the truncated correlation  \eqref{e:phi4cf} for the $|\varphi|^4$ model,

\subsection{Observable field} \label{sec:of}

As is often the case in statistical mechanics, we compute
correlation functions as derivatives with respect to an external field, which we
refer to as an \emph{observable} field.  We do this in a manner similar to what
is done in \cite{BBS-saw4} for the case $(n,p)=(0,1)$.

\subsubsection{Observable field for \texorpdfstring{$|\varphi|^4$}{phi4}}
\label{sec:phi4observables_representation}

Given $n \ge 1$,
let $\corrfcn_{ij} = \pair{ (\varphi_a^i)^p  ; (\varphi_b^j)^p}_{g,\nu,N}$,
which is what we wish to compute.
This defines a symmetric $n \times n$ matrix whose diagonal elements are the same,
and whose off-diagonal elements are also the same.

We use the notation
$\varphi_x^p$, which is equal to $\varphi_x$ when $p=1$, and to
the vector whose components are $(\varphi_x^i)^2$ for $p=2$.
Recall the definition of $U_0$ in \eqref{e:Vtil0def}.  Given a vector
$h \in \R^n$, and given \emph{observable fields}
$\sigma_a,\sigma_b \in \R$, we define $V_0$ (which depends on $h,n,p$) by
\begin{equation}
\label{e:V0n1}
    V_{0;x}
    =
    U_{0;x}  - \sigma_a (\varphi_a^p \cdot h) \1_{x=a}
    - \sigma_b (\varphi_b^p \cdot h) \1_{x=b}.
\end{equation}
Although the observable fields carry subscripts $a,b$, they represent two
constant fields which are independent of spatial location; the indicator functions
on the right-hand side of \eqref{e:V0n1} serve to localise the observables at $a,b$.
Let $D_{\sigma_a}$ denote the operator
$\frac{\partial}{\partial \sigma_a}$ at $\sigma_a = \sigma_b = 0$, and similarly
for higher derivatives.
By \eqref{e:ExF} and calculation of the derivative,
\begin{equation}
\label{e:corrdiff}
    h \cdot \corrfcn h =
    \inprod{ \varphi_a^p \cdot h\,}{\varphi_b^p \cdot h}_{g,\nu,N}
    =
    (1+z_0)^p
    D_{\sigma_a\sigma_b}^2
\log \Ex_C e^{- V_{0} (\Lambda)}.
\end{equation}
Given the values of $h \cdot \corrfcn h$ for two choices of $h$, the matrix
elements of $\corrfcn$ can be computed easily.

We define
\begin{equation}
\label{e:Ncaldef}
    \Ncal^\varnothing = \Ncal^\varnothing(\Lambda) = C^{p_\Ncal}((\R^n)^\Lambda,\R)
\end{equation}
to be the space of real-valued functions of the fields having at least
$p_\Ncal$ continuous derivatives, where $p_\Ncal$  is fixed
as in Section~\ref{sec:ir}.  This is the space of random variables of initial
interest, but because of the introduction of the observable fields, we are
interested in functions not only of $\varphi \in (\R^n)^\Lambda$ but also of
$\sigma_a,\sigma_b$.  On the other hand, our ultimate interest in the dependence
on the observable fields is the computation of the derivative appearing in \eqref{e:corrdiff}.
For this, we have no need to examine any
dependence on $\sigma_a,\sigma_b$ beyond terms of the form $1, \sigma_a,\sigma_b,
\sigma_a\sigma_b$.  We formalise this via the introduction of a quotient space,
in which two functions of $\varphi,\sigma_a,\sigma_b$ become equivalent if
their formal power series in the observable fields agree to
order $1,\sigma_a, \sigma_b,\sigma_a\sigma_b$, as follows.

Let $\widetilde\Ncal$ be the space of real-valued functions of $\varphi,\sigma_a,\sigma_b$
which are $C^{p_\Ncal}$ in $\varphi$ and $C^\infty$ in $\sigma_a,\sigma_b$.
Consider the elements of $\widetilde\Ncal$ whose formal power series
expansion to order $1,\sigma_a, \sigma_b,\sigma_a\sigma_b$ is zero.
These elements form an ideal $\Ical$ in $\widetilde\Ncal$,
and the quotient algebra $\Ncal =\widetilde\Ncal/\Ical$ has a direct sum decomposition
\begin{equation}
\label{e:Ncaldecomp4}
\Ncal = \widetilde\Ncal/\Ical = \Ncal^\varnothing \oplus \Ncal^a \oplus \Ncal^b \oplus \Ncal^{ab}.
\end{equation}
The elements of $\Ncal^a, \Ncal^b , \Ncal^{ab}$
are given by elements of $\Ncal^\varnothing$ multiplied
by $\sigma_a$, by $\sigma_b$, and by $\sigma_a\sigma_b$ respectively.
As functions of the observable field, elements of $\Ncal$ are then
identified with polynomials in the external field with terms only of
order $1,\sigma_a, \sigma_b,\sigma_a\sigma_b$.
For example, we identify
$e^{(\varphi_a\cdot h)\sigma_a + (\varphi_b\cdot h)\sigma_b}$ and
$1+(\varphi_a\cdot h)\sigma_a + (\varphi_b\cdot h)\sigma_b
+ (\varphi_a\cdot h)(\varphi_b\cdot h)\sigma_a\sigma_b$, as
both are elements of the same equivalence class in the quotient space.
An element $F \in \Ncal$ can be written as
\begin{equation}
\label{e:Fdecomp}
    F = F_\varnothing + \sigma_a F_a + \sigma_b F_b +\sigma_a \sigma_b F_{ab},
\end{equation}
where $F_\alpha \in \Ncal^\varnothing$ for each $\alpha \in \{\varnothing, a,b,ab\}$.
We define projections $\pi_\alpha : \Ncal \to \Ncal^\alpha$ by
$\pi_\varnothing F = F_\varnothing$,
$\pi_a F = \sigma_a F_a$, $\pi_b F = \sigma_b F_b$,
and $\pi_{ab} F = \sigma_a\sigma_b F_{ab}$.

\subsubsection{Observable field for WSAW}

For WSAW, we introduce
\emph{observable fields} $\sigma_a,\sigma_b \in \mathbb{C}$, and we
extend \eqref{e:psipsib} by now allowing the coefficients $K_{x,y}$ to be functions
of $\sigma_a,\sigma_b$ as well as of the boson field $\phi,\bar\phi$.
Let $\widetilde\Ncal$ be the resulting algebra of differential forms, and let
$\Ical$ denote the ideal in $\widetilde\Ncal$ consisting of those elements of
$\widetilde\Ncal$ whose formal power series expansion in the external field
to order $1,\sigma_a, \sigma_b,\sigma_a\sigma_b$ is zero.
The quotient algebra $\Ncal =\widetilde\Ncal/\Ical$ again has the direct sum decomposition
\begin{equation}
\label{e:Ncaldecomp}
\Ncal = \widetilde\Ncal/\Ical = \Ncal^\varnothing \oplus \Ncal^a \oplus \Ncal^b \oplus \Ncal^{ab},
\end{equation}
where elements of $\Ncal^a, \Ncal^b , \Ncal^{ab}$
are respectively given by elements of $\Ncal^\varnothing$ multiplied
by $\sigma_a$, by $\sigma_b$, and by $\sigma_a\sigma_b$.
For example, $\phi_x \bar\phi_y \psi_x \bar\psi_x
\in \Ncal^\varnothing$, and $\sigma_a \bar\phi_x \in \Ncal^a$.
As functions of the external field, elements of $\Ncal$ are
again identified with polynomials in the external fields with terms of order
$1,\sigma_a,\sigma_b,
\sigma_a\sigma_b$.
We use canonical projections $\pi_\alpha$ also for WSAW,
as defined below \eqref{e:Fdecomp}.

Let
\begin{equation}
\label{e:V0n0}
V_{0;x}  =
\Vbulk_{0;x}   - \sigma_a \phib_{a}^\nl \1_{x=a} - \sigma_b \phi_{b}^\nl \1_{x=b}.
\end{equation}
Then the expectation $\Ex_C e^{-V_0(\Lambda)}$ is well-defined for any $p \ge 1$,
including large $p$, since the superficially dangerous factor
$\exp[\sigma_a \phib_{a}^\nl + \sigma_b \phi_{b}^\nl ]$ is equivalent
to a polynomial in the fields, which is integrable.
With this interpretation, for all $p \ge 1$,
\begin{equation}
\label{e:intobs}
\wm^{(\nl)}_{ab,N}(g,\nu)
= (1+z_0)^p
D_{\sigma_a\sigma_b}^2
\Ex_C e^{- V_{0} (\Lambda)}.
\end{equation}
In view of the observations below \eqref{e:wmEC}, we may equivalently write
\begin{equation}
\label{e:intobslog}
\wm^{(\nl)}_{ab,N}(g,\nu)
= (1+z_0)^p
D_{\sigma_a\sigma_b}^2
\log \Ex_C e^{- V_{0} (\Lambda)},
\end{equation}
which has the same form as \eqref{e:corrdiff}.

\subsection{External field} \label{sec:ef}

The observable field is a \emph{constant} real external field which
couples to the $p^{\rm th}$ power of the field only at the points $a,b$,
due to the indicators in \eqref{e:V0n1} and \eqref{e:V0n0}.
For the proof of Theorem~\ref{thm:mr-stars},
we introduce a different kind of external field, taking values in $\R^n$, varying in space,
and coupled everywhere to the field $\varphi$ (or to $\phi$ and $\bar\phi$ for $n = 0$), as follows.

\subsubsection{External field for \texorpdfstring{$|\varphi|^4$}{phi4}}

Let $n \ge 1$.
We refer to a function $J : \Lambda \to \R^n$ as an \emph{external field}.
We define the inner product $(\cdot\, , \cdot)$ of two fields as
$(\varphi,J) = \sum_{x \in \Lambda} \varphi_x \cdot J_x$, where
$\varphi_x \cdot J_x$ is the standard dot product on $\R^n$.
We typically write  $H : \Lambda \to \R^n$ for a \emph{constant} field, with
$H_x = H_0$ for every $x \in \Lambda$.
For $k$ a non-negative integer, we write $D^k_J(H)$ for the operation of
$k^{\rm th}$ directional derivative with respect to $J$ at $J=0$, with each derivative
taken in direction $H$.
Then by \eqref{e:ExF} and direct computation of the derivative, for $p=1,2$,
(using symmetry for $p=2$)
\begin{equation}
\label{e:stardiff}
    \langle (\varphi, H)^p ; \varphi_a^p \cdot h \rangle_{g,\nu,N}
    = (1 + z_0)^p
    D_J^p(H) D_{\sigma_a}
    \log \Ex_C e^{- V_{0} (\Lambda) + (\varphi, J)}.
\end{equation}
Finite volume correlations as in \eqref{e:corrdot}--\eqref{e:corr1122}
can be written
in the form \eqref{e:stardiff} with
appropriate choices of $H, h \in \R^n$.

\subsubsection{External field for WSAW}

For WSAW, we use conjugate external fields $J,\bar J: \Lambda \to \C$.
Let $1$ denote the constant test function $1_x=1$ for all $x \in \Lambda$.
We define $D_{\bar J}^k$ to be the operator of $k$ directional derivatives
with respect to $\bar J$ in the direction 1 at $(J, \bar J) = (0,0)$, i.e.,
$D_{\bar J}^kF(J,\bar J)
= \frac{\partial}{\partial s_1}|_0 \cdots \frac{\partial}{\partial s_k}|_0
F(0,0+s_11+\cdots s_k1)$.
Direct computation gives
\begin{equation}\label{e:star-to-integral}
\starpol^{(p)}_N (g,\nu)
= (1+z_0)^p
D_{\bar J}^\nl D_{\sigma_a}
\,\Ex_C e^{- V_{0} (\Lambda) + (J,\phib)+(\bar J, \phi)}.
\end{equation}

\section{Perturbative renormalisation group flow} \label{sec:approach}

In Section~\ref{sec:pGi}, we recall how a Gaussian expectation (or super-expectation)
can be evaluated progressively in an iterative fashion.
This provides the basis for the renormalisation group approach.
In Section~\ref{sec:fp}, we identify a class of local field polynomials
that is important for our
analysis, and recall the projection operator $\LT$ from \cite{BS-rg-loc},
which projects $\Ncal$ onto local polynomials.
In Section~\ref{sec:Idef} we recall from \cite{BS-rg-IE} the definition of a replacement
$I(V,\Lambda)$ for $e^{-V(\Lambda)}$, which is better
suited to the renormalisation group iteration.
We also define the perturbative flow of coupling
constants used in the proof of Theorem~\ref{thm:mr-ab}.
Finally, in Section~\ref{sec:Vptdef},
we perform an explicit computation of the perturbative flow for observables, in
Proposition~\ref{prop:pt}, which provides the basis for the computation of the
logarithmic powers in our main results.

\subsection{Progressive Gaussian integration} \label{sec:pGi}

We call $C=(-\Delta_{\Lambda_N} + m^2)^{-1}$ the \emph{covariance}.
According to \eqref{e:corrdiff}, \eqref{e:intobslog},
one of our goals is the computation of the expectation
\begin{equation}
\label{e:goal}
    \Ex_C e^{-V_0(\Lambda)}.
\end{equation}
For $n\ge 1$, $V_0$ is given by \eqref{e:V0n1} and the expectation is the standard Gaussian
expectation \eqref{e:Gex}.  For $n=0$,
$V_0$ is given by \eqref{e:V0n0} and the expectation is
the Gaussian super-expectation \eqref{e:Gsex}.
We compute these expectations \emph{progressively}, using covariance decomposition.

We use decompositions of the two covariances
$(-\Delta_\Zd + m^2)^{-1}$ and $(-\Delta_{\Lambda_N} +m^2)^{-1}$.
For $\Zd$, the covariance exists for $d>2$ for all
$m^2 \ge 0$, but for $\Lambda_N$ we must restrict to $m^2>0$ since the
finite-volume Laplacian is not invertible.
In \cite[Section~\ref{pt-sec:Cdecomp}]{BBS-rg-pt},
results from \cite{Baue13a,BGM04} are applied to
define a sequence
$(C_j)_{1 \le j < \infty}$ (depending on $m^2 \ge 0$)
of positive definite covariances on $\Zd$
such that
\begin{equation}
\label{e:ZdCj}
    (\Delta_\Zd + m^2)^{-1} = \sum_{j=1}^\infty C_j
    \quad
    \quad
    (m^2 \ge 0).
\end{equation}
For $j \ge 0$, we define the partial sums
\begin{equation}
\label{e:wjdef}
    w_j = \sum_{i=1}^j C_i, \qquad w_0=0.
\end{equation}
The covariances $C_j$ are translation invariant,
and have the \emph{finite-range} property
\begin{equation}
  \label{e:frp}
  C_{j;xy} = 0 \quad \text{if \; $|x-y| \geq \frac{1}{2} L^j$}
  .
\end{equation}
For $j<N$, the covariances $C_j$ can therefore be identified with
covariances on $\Lambda=\Lambda_N$, and we use both interpretations.
For $m^2>0$, there is also a covariance $C_{N,N}$ on $\Lambda$ such that
\begin{equation}
\label{e:NCj}
    (-\Delta_{\Lambda_N} + m^2)^{-1} = \sum_{j=1}^{N-1} C_j + C_{N,N}
    .
\end{equation}
Good estimates on $C_j$ and $C_{N,N}$ are given in
\cite[Proposition~\ref{pt-prop:Cdecomp}]{BBS-rg-pt}.

For $n \ge 1$, we write $\Ex_C\theta F$  for the convolution of $F$ with the
Gaussian measure,
i.e., given an integrable $F \in \Ncal$, we define
\begin{equation}
\label{e:thetadef}
  (\Ex_C\theta F)(\varphi) = \Ex_C F(\varphi+\zeta),
\end{equation}
where the expectation $\Ex_C$ acts on $\zeta$ and leaves $\varphi$ fixed.
It is thus a conditional expectation.

For $n = 0$, we use a copy $\Lambda'$ of $\Lambda$, and
in addition to the fields $\phi,\bar\phi,\psi,\bar\psi$ on $\Lambda$
we introduce a boson field $\xi,\bar\xi$ and
fermion field $\eta, \bar\eta$ on $\Lambda'$, with
$\eta = \frac{1}{\sqrt{2\pi i}}d\xi$,
$\bar\eta = \frac{1}{\sqrt{2\pi i}}d\bar\xi$.  Then
we consider the ``doubled'' algebra $\Ncal(\Lambda \sqcup
\Lambda ')$ containing the original
fields and also these
additional fields.
We define a map $\theta : \Ncal(\Lambda ) \to \Ncal(\Lambda \sqcup \Lambda')$
by making the replacement in an element of $\Ncal$ of $\phi$ by $\phi+\xi$,
$\bar\phi$ by $\bar\phi+\bar\xi$, $\psi$ by $\psi+\eta$, and $\bar\psi$ by $\bar\psi+\bar\xi$.
Then for $F \in \Ncal(\Lambda)$, $\Ex_C\theta F$ is obtained by regarding the expectation
as an integral over the variables $\xi,\bar\xi,\eta,\bar\eta$
which leaves the variables $\phi,\bar\phi,\psi,\bar\psi$ fixed.

According to
\cite[Proposition~\ref{norm-prop:conv}]{BS-rg-norm}, for both WSAW and $n \ge 1$ we have
\begin{equation}
    \label{e:progressive}
    \Ex_{C}\theta F
    =
    \big( \Ex_{C_{N,N}}\theta \circ \Ex_{C_{N-1}}\theta \circ \cdots
    \circ \Ex_{C_{1}}\theta\big) F
    .
\end{equation}
This expresses the expectation on the left-hand side as a progressive
integration.
To compute the expectation $\Ex_C e^{-V_0(\Lambda)}$ of \eqref{e:goal}, we use
\eqref{e:progressive}
to evaluate the more general quantity $\Ex_C \theta e^{-V_0(\Lambda)}$ progressively.
Namely, we define
\begin{equation}
\label{e:Z0def}
  Z_{j+1} = \Ex_{C_{j+1}}\theta Z_j \quad\quad
  (0 \le j<N),
\end{equation}
with $Z_0 = e^{-V_0(\Lambda)}$, and with an abuse of notation in that we interpret $C_N$
as $C_{N,N}$.  By \eqref{e:progressive}, we can evaluate $\Ex_CF$ by
setting the fields equal to zero in
\begin{equation}
\label{e:ZN}
    Z_N = \Ex_C \theta Z_0.
\end{equation}
Thus we are led to study the recursion $Z_j \mapsto Z_{j+1}$.
We write $\Ex_{j} = \Ex_{C_j}$,
and leave implicit the dependence of the covariance $C_j$
on the mass $m$.

Given a \emph{scale} $j \in \{0,1,\ldots,N\}$,
we partition $\Lambda = \Z^d/L^N\Z^d$ into a
disjoint union of $L^{d(N-j)}$ scale-$j$ \emph{blocks} of side length $L^j$,
and denote the
set of all such blocks by $\Bcal_j$.
One block contains the origin at its corner and is of the form
$\{x\in \Lambda:  |x|_{\infty} < L^{j}\}$,
and all other blocks are translates of this one by vectors in
$L^{j} \Zd$.
A scale-$j$ \emph{polymer} is a union of
scale-$j$ blocks,
and we write
$\Pcal_j=\Pcal_j(\Lambda)$ for the set of scale-$j$ polymers.
Given $a,b\in \Lambda$,
an important scale is the \emph{coalescence scale} $j_{\pp \qq}$, defined by
\begin{equation}
   \label{e:Phi-def-jc}
    j_{\pp \qq}
    =
    \big\lfloor
   \log_{L} (2 |\pp - \qq|)
   \big\rfloor
   .
\end{equation}
Thus $j_{ab}$ is the unique integer such that
\begin{equation}
\label{e:jabbds}
    \tfrac 12 L^{j_{ab}} \le |a-b| < \tfrac 12 L^{j_{ab}+1}.
\end{equation}
By \eqref{e:frp}, the smallest $j$ for which $C_{j;ab}\neq 0$ is possible is $j=j_{ab}+1$.

\subsection{Field polynomials and the operator \texorpdfstring{$\LT$}{Loc}} \label{sec:fp}

\subsubsection{Approximation via cumulant expansion}

To illustrate the ideas involved in the study of the recursion $Z_j \mapsto Z_{j+1}$,
we consider the computation of $Z_1 = \Ex_{1}\theta e^{-V_0(\Lambda)}$,
at the level of formal power series accurate to second order in the coupling constants
of $V_0$.
This can be done by expansion of $e^{-V_0(\Lambda)}$ to second order, and the result
can be written
\begin{equation}
Z_1 = \Ex_{C_1} \theta e^{-V_0}
\approx
\exp \left(-\Ex_{C_1} \theta V_0 + \frac 12 \Ex_{C_1} \theta \left(V_0;V_0\right)
\right),
\end{equation}
where
$\Ex\theta(V_0;V_0) = \Ex\theta V_0^2 - (\Ex \theta V_0)^2$, and
$\approx$ denotes approximation accurate to second order in the sense of formal power
series.
This is an instance of the \emph{cumulant expansion}.
Then
\begin{equation}\label{e:Z1}
Z_1
\approx e^{-H_1} \quad\text{with}\quad H_1
= \Ex_{C_1} \theta V_0 - \frac 12 \Ex_{C_1} \theta \left(V_0;V_0\right).
\end{equation}

The polynomial $H_1$ can be computed explicitly, as follows.
We define operators
\begin{equation}
\label{e:LapC}
    \Lcal_C = \frac 12
    \sum_{u,v \in \Lambda}
    C_{u,v}
    \sum_{i=1}^n
    \frac{\partial}{\partial \varphi_{u}^i}
    \frac{\partial}{\partial \varphi_{v}^i}
    ,\quad
    \Lcal_C
    =
    \sum_{u,v \in \Lambda}
    C_{u,v}
    \left(
    \frac{\partial}{\partial \phi_{u}}
    \frac{\partial}{\partial \bar\phi_{v}}
    +
    \frac{\partial}{\partial \psi_{u}}
    \frac{\partial}{\partial \bar\psi_{v}}
    \right)
    ,
\end{equation}
for the $|\varphi|^4$ and WSAW models,  respectively.
Then,
for a polynomial  $A$ in the fields,
\begin{equation}
\label{e:eLap}
  \Ex_C \theta A = e^{\Lcal_C}A,
\end{equation}
where the exponential on the right-hand side is defined by its power series expansion
(a finite series when applied to a polynomial);
see \cite[Lemma~\ref{norm-lem:*heat-eq}]{BS-rg-norm} for a proof.
The computation of the first
term $\Ex_{C_1} \theta V_0$ of $H_1$ in \eqref{e:Z1} is elementary; details for observables
are given in Section~\ref{sec:Vptdef} below.
The second term $\Ex_{C_1} \theta \left(V_0;V_0\right)$ is bilinear and
can be computed as a sum of terms arising from the monomials in $V_0$.
For the case $(n,p)=(1,1)$, using \eqref{e:eLap} we find that one of these terms is
\begin{equation}
\Ex_{C_1} \theta \left(\sigma_a \varphi_a;\sigma_b \varphi_b\right)
= \sigma_a\sigma_b \left(\varphi_a\varphi_b + C_{1;ab}\right)
 - \sigma_a\sigma_b \varphi_a\varphi_b
= \sigma_a\sigma_b C_{1;ab}.
\end{equation}
There is no $\sigma_a\sigma_b$ term in $V_0$, and the creation of
such a term in $H_1$ is welcome, as the second derivatives on
the right-hand sides of \eqref{e:corrdiff} and \eqref{e:intobs} would produce
a non-zero result when applied to $e^{-H_1}$ but not to $e^{-V_0}$.

One lesson learned from the above computation is that expectation can create new terms
that did not appear in $V_0$, such as $\sigma_a\sigma_b C_{1;ab}$.
To accommodate this,
we will define an $n$-dependent class of polynomials $\Vcal_h$ that
is stable under the action of the progressive integration,
to second-order approximation as above.

A second lesson from the above computation is that not all terms in $H_1$
are \emph{local},
due to the nonlocal nature of the
 operator $\Lcal_C$ in \eqref{e:LapC}.
To deal with this issue, we use the projection operator $\Loc$
of \cite{BS-rg-loc},
which we discuss in Section~\ref{sec:loc}.

\subsubsection{Local field polynomials} \label{sec:Vcal}

Given $h \in \R^n$, $h \neq 0$, we define a class of local polynomials $\Vcal_h$
that can be used to parametrise the result of progressive expectations.
It is necessary to keep track of the dependence on the vector $h$ for $n \geq 2$,
whereas for $n = 0$ and $n = 1$ we simply set $h = 1$.
We define
\begin{align}
\label{e:rhoabdef}
\rho_x^a (h) &=
\begin{cases}
\bar{\phi}_x^\nl & (n=0) \\
\left(\varphi_x^\nl \cdot h\right) / |h| & (n\ge 1),
\end{cases}
&
\rho_x^b (h) &=
\begin{cases}
\phi_x^\nl & (n=0) \\
\left(\varphi_x^\nl \cdot h\right) / |h| & (n\ge 1).
\end{cases}
\end{align}
In \eqref{e:rhoabdef}, the superscripts $a, b$ on the left-hand sides have significance only
for the case $n = 0$ where they indicate whether or not there is a bar on $\phi$.
For $n \ge 1$, $\rho_x^a = \rho_x^b$ and there is no dependence on $a,b$.
Also, note that $\rho_x^\alpha$ depends on the vector $h$ only through its direction.
We also need the monomial
\begin{equation}
    \tau_{\nabla\nabla,x} =
    \begin{cases}
   \frac 12
   \sum_{e\in\Z^d:|e|_1=1}
   \left(
   (\nabla^e \phi)_x (\nabla^e \bar\phi)_x +
   (\nabla^e \psi)_x (\nabla^e \bar\psi)_x
   \right)
   & (n=0)
   \\
  \tfrac14 \sum_{e\in\Z^d:|e|_1=1} \nabla^e\varphi_x \cdot \nabla^e \varphi_x
  &
  (n\ge 1).
  \end{cases}
\end{equation}
Then we define the polynomials (for $\alpha =a,b$)
\begin{equation}
V_{\varnothing ,x} = g\tau_x^2 + \nu\tau_x + z\tau_{\Delta,x} + y \tau_{\nabla\nabla,x} + u,
\quad
V_{\alpha,x} =
\begin{cases}
    \lambda_\alpha \rho_x^\alpha(h)
    & (n=0)
    \\
    \lambda_\alpha \rho_x^\alpha(h) + t_\alpha
    & (n\ge 1),
\end{cases}
\end{equation}
and define a set of functions $V \mapsto V_x$ by
\begin{equation}
\label{e:Vmulticomponent}
    \Vcal_h = \{
    V: V_x =
    V_{\varnothing,x} - \sigma_a V_{a,x}\1_{x=a}  - \sigma_b V_{b,x}\1_{x=b}
    - \sigma_a\sigma_b \textstyle{\frac 12} ( q_a\1_{x=a} + q_b\1_{x=b})
    \}.
\end{equation}
Given $X \subset \Lambda$, we also define
\begin{equation}
\label{e:Vcalesig}
    \Vcal_h(X) = \{V(X) = \textstyle{\sum_{x\in X}} V_x : V \in \Vcal_h \}.
\end{equation}
The scalar coefficients in the above polynomials are all real numbers for the $|\varphi|^4$ model.
For the WSAW, all are real except $\lambda_a, \lambda_b, q_a, q_b$
which are permitted to be complex (this is discussed further in Section~\ref{sec:realcc} below).

Two useful subspaces of $\Vcal_h$ are the subspace
$\Vcal_h^{(0)}$
consisting of elements of $\Vcal_h$ with
$u = y = t_a = t_b = q_a = q_b = 0$,
and the subspace $\Vcal_h^{(1)}$
consisting of elements with $y = 0$.
The polynomial $V_0$ of \eqref{e:goal} lies in the subset of $\Vcal_h^{(0)}$ with
$\lambda_a = \lambda_b = 1$.

\subsubsection{Localisation}
\label{sec:loc}

Let $X\subset\Lambda$.
We now recall some basics about the localisation operator $\Loc_X :
\Ncal \to \Vcal(X)$,  which projects $\Ncal$ onto a vector space $\Vcal(X)$
of local polynomials that in general contains more monomials than $\Vcal_h(X)$.
The definition and general theory of this operator is given in
\cite{BS-rg-loc}, and we adapt the theory here to incorporate the observables.

By definition, the operator $\Loc_X$
respects the direct sum decomposition \eqref{e:Ncaldecomp4}--\eqref{e:Ncaldecomp} of $\Ncal$,
in the sense that
$\LT_X \pi_\alpha = \pi_\alpha \LT_X$ for each $\alpha = \{\varnothing, a,b,ab\}$.
We omit discussion of a detail that limits the domain of $\Ncal$ to avoid issues
associated with ``wrapping around'' the torus $\Lambda$, this point is discussed
carefully in \cite{BS-rg-loc}.
The restrictions $\Loc_X|_{\Ncal^\alpha}$ are defined individually for each
$\alpha$.  As discussed in detail in \cite{BS-rg-loc}, their definitions require:
(i) specification of the dimensions of the fields,
(ii) choice of a maximal monomial dimension $d_+(\alpha)$ for each $\alpha$, and
(iii) choice of covariant field polynomials $\hat P$ which form the basis for the vector
space $\Loc_X(\Ncal^{\alpha})$ (see \cite[Definition~\ref{loc-def:Vcal}]{BS-rg-loc}).
Item~(iii) is done exactly as in \cite[\eqref{loc-e:Pm-def}]{BS-rg-loc}; this item
does not play any significant role in the present paper and we will not discuss it further.
Also, since we do not make explicit use of $\Loc_X|_{\Ncal^\varnothing}$
in this paper, we do not specify its definition
in detail, which is identical to what is used in \cite{BBS-phi4-log,BBS-saw4-log}.
For the observable components of $\LT_X$, we use the following.
\begin{enumerate}[(i)]
\item
The \emph{dimensions} of the fields are simply
\begin{equation}
\label{e:phidim}
    [\varphi]= 1,
    \qquad
    [\phi] = [\bar\phi] = [\psi] = [\bar\psi] = 1.
\end{equation}
By definition, the dimension of a monomial
$\nabla^\eta \zeta$ is
$|\eta|_1+[\zeta]=|\eta|_1+1$, where $\eta$ is a multi-index and
$\zeta$ may be any of $\varphi$  or $\phi,\bar\phi,\psi,\bar\psi$.
Here
$\nabla^\eta=\nabla_{1}^{\eta_1} \dotsb \nabla_{d}^{\eta_d}$ for a
multi-index $\eta=(\eta_1,\dotsc,\eta_d)$, where
$\nabla_{k}$ denotes the
finite-difference operator $(\nabla_{k}f)_x=f_{x+e_k}-f_x$, with $e_k$ the $k^{\rm th}$
standard unit vector.
The dimension of a product
of such monomials is the sum of the dimensions of the factors in the product.
This is the same as in \cite{BBS-phi4-log,BBS-saw4-log}.

\item
For $\alpha = ab$, we set $d_+(ab) = 0$.
For $\alpha = a$ and $\alpha = b$,
we make the scale dependent choice $d_+^{(j)}(a) = d_+^{(j)}(b) = p \1_{j < j_{ab}}$.
The reduction of $d_+^{(j)}(a)$ at the coalescence scale $j_{ab}$
(defined in \eqref{e:Phi-def-jc})
is a decision that simplifies some aspects
in Section~\ref{sec:pfmr1} below; this decision was also taken in \cite{BBS-saw4}.
\end{enumerate}
We use a superscript to emphasise the scale dependence in the above choices,
and write $\LT_{X}^{(j)}$ for the operator $\LT_{X}$ with the above $j$-dependent choice of
$d_+$.

\subsection{Definition of \texorpdfstring{$I_j$}{I}}
\label{sec:Idef}

We now recall several definitions which are made and explained in \cite{BBS-rg-pt,BS-rg-IE}.
We use the direct sum decompositions $\Ncal = \Ncal^\varnothing \oplus \Ncal^a \oplus
\Ncal^b \oplus \Ncal^{ab}$ of \eqref{e:Ncaldecomp4}--\eqref{e:Ncaldecomp},  the
canonical projections $\pi_\alpha$ as defined under \eqref{e:Fdecomp},
and the abbreviation $\pi_*=1-\pi_\varnothing = \pi_a+\pi_b+\pi_{ab}$.

For polynomials $A,B$ in the fields, and with $\Lcal_C$ given by \eqref{e:LapC}, we define
\begin{align}
  \label{e:FCAB}
    F_{C}(A,B)
    &= e^{\Lcal_C}
    \big(e^{-\Lcal_C}A\big)
    \big(e^{-\Lcal_C}B\big) - AB
    ,
    \\
\label{e:Fpi}
    F_{\pi,C}(A,B)
    &=
    F_{C}(A,\pi_\varnothing B)
    + F_{C}(\pi_* A,B).
\end{align}
Recall that the covariance $w_j$ is defined by \eqref{e:wjdef}.
For a local polynomial $V$ in the fields, and for a polymer $X \in \Pcal_j$, we  set
\begin{equation}
  \label{e:WLTF}
  W_j(V,X) = \frac 12 \sum_{x\in X} (1-\LT_{x}^{(j)}) F_{\pi,w_j}(V_x,V(\Lambda)).
\end{equation}
(The definition \eqref{e:WLTF} is inapplicable for the final scale $j=N$;
this special case is discussed in \cite[Section~\ref{IE-sec:finalscale}]{BS-rg-IE}.)
Then, for $X \in \Pcal_j$,
we define the \emph{interaction} functional
\begin{equation}
  \label{e:Idef}
  I_j(V,X) = e^{-V(X)}\prod_{B \in \Bcal_j(X)}(1+W_j(V,B)).
\end{equation}
For $j=0$, where $w_0=0$, we interpret the above
as $I_0(V,X)=e^{-V(X)}$.

Let $\Lcal_{j+1}=\Lcal_{C_{j+1}}$.  Given $V$, we  define
\begin{equation} \label{e:Palt}
P_{j,x}
= \frac 12 \sum_{y \in \Lambda}
\big(\Loc_{x}^{(j+1)} F_{\pi,w_{j+1}} (e^{\Lcal_{j+1}} V_x,e^{\Lcal_{j+1}} V_y)
- e^{\Lcal_{j+1}} \Loc_{x}^{(j+1)}  F_{\pi,w_j} (V_x,V_y) \big),
\end{equation}
and set
\begin{equation}
\label{e:Vptdef}
V_{\pt,j+1,x}(V)
= e^{\Lcal_{j+1}} V_x -  P_{j,x}  .
\end{equation}
The definition \eqref{e:Vptdef} is equivalent to the
definition in \cite[\eqref{pt-e:Vptdef}]{BBS-rg-pt}, by
\cite[Lemmas~\ref{pt-lem:Palt}--\ref{pt-lem:Fexpand}]{BBS-rg-pt}.
The definition is motivated in
\cite{BBS-rg-pt}, where it is shown that
\begin{equation} \label{e:Iex}
  \Ex_{j+1}\theta I_j(V,\Lambda) \approx I_{j+1}(\Vpt, \Lambda) ,
\end{equation}
with ``$\approx$'' as in Section~\ref{sec:fp} above.
Equation~\eqref{e:Iex} shows that, to second order, $I$ enjoys a form of stability under
expectation when $V$ is advanced to $\Vpt$.
However, at this point there is no uniformity in scale $j$ or volume $\Lambda$
in the error estimate.

\subsection{Perturbative flow of coupling constants}
\label{sec:Vptdef}

The perturbative flow of the bulk coupling constants $g,\nu,z,y$
is given in \cite{BBS-rg-pt}
for WSAW and in \cite{BBS-phi4-log} for $|\varphi|^4$.
In \cite{BBS-rg-pt} it is also given for the observable
coupling constants $\lambda_\alpha,q_\alpha$ for WSAW, for
the specific case $p=1$.
In the following proposition, we extend the perturbative computation to the
observables needed for our main results,
and compute  $\pi_* \Vpt$ for all $p \ge 1$ for WSAW, and for $p=1,2$ for
$|\varphi|^4$.
For this, we need some preliminaries.

\begin{defn}
\label{def:M2}
For $n \ge 2$, we write $M_2(n)$
for the set of $n\times n$ matrices of the form $rI+sJ$, with $r,s\in \R$,
$I$ the identity matrix, and $J$ having all
entries equal to $1$.
\end{defn}

The vector
\begin{equation}\label{e:epm}
    e^{+} = (1,1,\ldots,1)
    \in \R^n
\end{equation}
appears frequently in our analysis.
Every matrix in $M_2(n)$ has eigenspaces $E^\pm$, where $E^+={\rm span}(e^+)$ with
eigenvalue $r+ns$, and $E^-$ is the orthogonal complement
$E^-=(E^+)^\perp$ with eigenvalue $r$.

Let $I$ denote $1 \in \R$ when $n=0$ and the $n \times n$ identity matrix for $n \ge 1$.
We define a matrix $T$, which is in $M_2(n)$
for $n \ge 2$,   by
\begin{equation}
    T  =
    \begin{cases}
    \binom{p}{2} \frac 14 I & (n=0)
    \\
    \binom{p}{2} \frac 13 I & (n=1)
    \\
    \binom{p}{2}\frac{2}{n+8} I + \binom{p}{2}\frac{1}{n+8}J
    & (n \ge 2).
    \end{cases}
\end{equation}
The matrix $T$ is the zero matrix for $p=1$ (as $\binom{1}{2}=0$), and otherwise
has eigenspace $E^+$ with
eigenvalue $\gammap_{n,p}=\binom{p}{2}\frac{n+2}{n+8}$, and for $n \ge 2$ also
has eigenspace $E^-=(E^+)^\perp$ with eigenvalue
$\gammam_{n,p}=\binom{p}{2}\frac{2}{n+8}$.
The correspondence between the matrix $T$ for $n \ge 1$ and the
value we have assigned to $n=0$ should
be understood via the eigenvalues, as $\frac{0+2}{0+8}=\frac{2}{0+8} = \frac 14$.
For $n=0,1$ there is only $\gamma^+$ and $E^+$.
For
$n \ge 2$ and $p=2$, we have $\gamma^- = \frac{2}{n+8} < \frac{n+2}{n+8}= \gamma^+ <1$,
and this is the only setting where both eigenvalues play a role in our analysis.

For $q:\Lambda \to \R$, let $q^{(n)}=\sum_{x \in \Lambda}q_{0,x}^n$.
Let $C=C_{j+1}$, $w=w_j$, and, for $g,\nu\in\R$, let
\begin{equation}
\label{e:deltadef}
    \nu^+=\nu + g(n+2)C_{00},
    \quad\quad
    \delta_j[f(\nu,w)] = f(\nu^+, w+C) - f(\nu,w),
\end{equation}
\begin{equation}
\label{e:betadef}
    \beta_j = (n+8) \delta_j[w^{(2)}].
\end{equation}
We define the matrix
\begin{equation}
\label{e:Ajdef}
    A_j =
    \begin{cases}
    (1-p\delta_j[\nu w^{(1)}])I - \beta_jg T & (j + 1 < j_{ab})
    \\
    I & ( j + 1 \ge j_{ab}).
    \end{cases}
\end{equation}
Thus $A_j$ is
$n\times n$ for $n \ge 1$ and $1 \times 1$ for $n=0$.  For $n \ge 2$, $A_j \in M_2(n)$.
The eigenvalues and eigenvectors of $A_j$ play an important role in identifying
the logarithmic corrections in Theorem~\ref{thm:mr-ab}.
The eigenspaces  are $E^\pm$, with eigenvalues
\begin{equation}
\label{e:fdef}
    f_j^\pm =
    \begin{cases}
    1- p\delta_j [\nu w^{(1)}] - \beta_j g \gamma^\pm_{n,p} & (j + 1 < j_{ab})
    \\
    1 & (j + 1 \ge j_{ab}).
    \end{cases}
\end{equation}
In our applications, $g$ and $\nu$ are small enough that $f_j^\pm >0$.

The following proposition computes $\Vpt$ as a function of $V$.
For its statement, we define
\begin{equation}
\label{e:varsigdef}
    \varsigma_j
    =
    C_{0,0}
    ( 1 - \1_{j + 1 < j_{ab}}2\nu w^{(1)})
    + \1_{j + 1 < j_{ab}} \nu^+\delta_j[ w^{(2)}]
    + \1_{j + 1 \ge j_{ab}}\delta_j[\nu w^{(2)}]
    .
\end{equation}

\begin{prop}
\label{prop:pt}
Let $d=4$.
Let $p \ge 1$ for WSAW, and $p = 1, 2$ for $|\varphi|^4$.
Let $V \in \Vcal_h$ with $|h| = 1$.
Then $V_{\pt,j+1}(V) \in \Vcal_{h_\pt}$, and for $x = a, b$,
$h_\pt$ and $\pi_*V_{\pt,j+1}$ are given by
\begin{align}
\label{e:hpt}
h_\pt
    & =
\left(A_j h\right) / |A_j h|,
\\
\label{e:lampt}
\lambda_{\pt,x} &= |A_j h| \lambda_x,
 \\
\label{e:qpt}
q_{\pt,x} & = q_x +  \nl! \lambda_a \lambda_b \delta_j[w^\nl_{ab}],
\\
\label{e:tpt}
    t_{\pt,x}
    & = t_x
    + \1_{n \ge 1}\1_{p=2} \lambda_x( e^+\cdot h)
    \varsigma_j
    .
\end{align}
In particular, if $h \in E^\pm$, then
$h_\pt =  h$ and $\lambda_{\pt,x}h_\pt  = f^\pm_j \lambda_x h$.
\end{prop}

It is clear from \eqref{e:hpt} that for $V \in \Vcal_h$ it is  in general not  the
case that $\Vpt$ lies in $\Vcal_h$
when $n \ge 2$.
Instead, $\Vpt \in \Vcal_{h_\pt}$ for a new direction $h_\pt$.
However, if $h$ is in one of the eigenspaces $E^\pm$, then $h_\pt=h$.
To have  $h_\pt=h$ is a desirable simplification, and this gives the
eigenspaces $E^\pm$ a special significance.

The proof of Proposition~\ref{prop:pt}
involves similar but not identical calculations for $n=0$ and $n \ge 1$.
However, once Proposition~\ref{prop:pt} is proved, the remaining analysis for
the proof of our main results is unified for all $n \ge 0$.

As noted below \eqref{e:jabbds},
$j = j_{ab}$ is the smallest scale $j$ for which $C_{j+1,ab} \neq 0$ is possible,
and so $\delta_i[w^\nl_{ab}]$ can be nonzero for the first time also when $i = j_{ab}$.
Therefore the first scale for which $\qpt-q$ can be nonzero
is $q_{\pt,j_{ab}+1}$.

For the rest of this section, we write
$w=w_j$, $C=C_{j+1}$ and $\Lcal = \Lcal_{C_{j+1}}$.
The first step in the proof of
Proposition~\ref{prop:pt} is
the computation of the first term in $\Vpt = e^{\Lcal}V -P$ of \eqref{e:Vptdef},
provided by the following lemma.

\begin{lemma}
\label{lem:EV}
Let $n=0$ and $p \ge 1$, or let $n \ge 1$ and $p=1,2$.  For $V \in \Vcal_h$,
\begin{equation}
\label{e:EV}
    e^{\Lcal} V_x
    =
    V_x + g(n+2)C_{00}\tau_x
    + \1_{n \ge 1} \left(\delta u_{\pt} - \1_{p=2}\lambda_x(e^+ \cdot h) C_{00}\sigma_x \right),
\end{equation}
where $\delta u_{\pt}$ is an explicit quadratic function of $g,\nu,y+z$.
\end{lemma}

\begin{proof}
The computation of $e^{\Lcal} \pi_{\varnothing} V_{x}$ is carried out in
\cite{BBS-phi4-log,BBS-saw4-log} and agrees with the above formula.
In particular, $\delta u_\pt = 0$ for $n=0$, and $\delta u_\pt$ is given by
\cite[\eqref{phi4-log-e:uptlong}]{BBS-phi4-log} for $n \ge 1$.
For the observable part, for $n=0$ we have $\Lcal \pi_*V=0$ and hence
$\pi_* e^{\Lcal} V_x = \pi_* V_x$, as in \eqref{e:EV}.
For $n \ge 1$ and $\nl = 1,2$, we have
$\Lcal^2\pi_* V=0$, so $\pi_* e^{\Lcal_{C}} V_x
= \pi_* V_x + \Lcal\pi_* V_x$.  Direct calculation of $\Lcal\pi_* V_x$
gives the final term of \eqref{e:EV}.
\end{proof}

To compute $\pi_*P_x$ we use \eqref{e:Palt}, i.e.,
\begin{equation}
\label{e:pistarP}
\pi_* P_x
= \frac 12 \sum_{y \in \Lambda} \bigg(\Loc_x \pi_* F_{\pi,w + C}(e^{\Lcal} V_x, e^{\Lcal} V_y)
- e^{\Lcal} \Loc_x \pi_* F_{\pi,w}(V_x,V_y)\bigg),
\end{equation}
in conjunction with \eqref{e:Fpi} which implies
\begin{equation}
\label{e:LocpistarF}
\pi_* F_{\pi,w}(V_x,V_y)
= 2  F_{w}( \pi_* V_x, \pi_\varnothing V_y)
+  F_{w}(\pi_* V_x, \pi_* V_y).
\end{equation}
For the following lemma, for each pair $x,y\in \Lambda$, we define an
$n \times n$ matrix ($1\times 1$ if $n=0$)
\begin{equation}
    M_{xy} = \1_{j+1<j_{ab}} \left(\nu p w_{xy}I +  g(n+8)w_{xy}^2 T \right).
\end{equation}

\begin{lemma}
\label{lem:LocF}
Let $n=0$, $p \ge 1$, or $n \ge 1$, $p=1,2$.  For $V \in \Vcal_h$,
\begin{align}
\label{e:LocF1}
    \LT_x F_w (\pi_*V_x, \pi_\varnothing V_y)
    &=
    -\sigma_x \lambda_x \left( (  M_{xy} \varphi_x^p\cdot h) +
    \1_{n \ge 1} \1_{p=2} \nu w_{xy}^2 (e^+ \cdot h) \right),
    \\
\label{e:LocF2}
    \LT_x F_w (\pi_*V_x, \pi_* V_y)
    &=
    -\sigma_a \sigma_b p! \lambda_a\lambda_b |h|^2 w_{xy}^p (\1_{x=a}\1_{y=b}+ \1_{x=b}\1_{y=a}),
\end{align}
where for $n=0$ we interpret $\varphi_x$ on the right-hand side of \eqref{e:LocF1}
as $\phib_a$ for $x=a$ and $\phi_b$ for $x=b$.
\end{lemma}

\begin{proof}
We evaluate $F$ using \cite[Lemma~\ref{pt-lem:Fexpand}]{BBS-rg-pt}, which implies
that for $n\ge 1$,
\begin{equation}
\label{e:Fsum}
    F_{C}(A_x,B_y) = \sum_{k=1}^D  \frac{1}{k!}
    \sum_{i_1,\ldots,i_k=1}^n
    \sum_{\substack{u_l,v_l \in \Lambda\\(l=1,\dots,k)}}  \left(\prod_{l=1}^k C_{u_l,v_l}\right)
    \frac{\partial^k A_x}{\partial \varphi_{u_1}^{i_1}\cdots \partial \varphi_{u_k}^{i_k}}
    \frac{\partial^k B_y}{\partial \varphi_{v_1}^{i_1}\cdots \partial \varphi_{v_k}^{i_k}},
\end{equation}
with $D = \deg A \wedge \deg B$.
For $n=0$, there is a related formula that also involves the fermions.

For \eqref{e:LocF1}, we first note that there is no contribution from
the terms involving $t_\alpha$ or $q_\alpha$ in $V_x$,
since the sum in \eqref{e:Fsum} starts at $k=1$ and hence always
involves differentiation with respect to $\varphi$, which is absent in these terms.
The cases $\pi_a$ and $\pi_b$ are symmetric, and we therefore only consider $\pi_a$.
It can be argued on the basis of dimensional considerations
that there is no contribution due to the terms
$y\tau_{\nabla\nabla}+z\tau_\Delta$ in $\pi_\varnothing V$.
For the remaining calculation,
we use the notation appropriate for $n \ge 1$ and comment on what is different for
$n=0$.
To prove \eqref{e:LocF1}, we therefore compute
\begin{equation}
    \LT_a F_w (\pi_aV_a, \pi_\varnothing V_y)
    =
    - \lambda_a \sum_{i=1}^n h^i
    \left(
    g\LT_x F_w (\sigma_a (\varphi_a^i)^p , \tau_y^2  )
    +
    \nu \LT_x F_w (\sigma_a (\varphi_a^i)^p , \tau_y)
    \right).
\end{equation}

For $n \ge 1$ and  $p =1, 2$,
\begin{equation}
F_{w}\left(\varphiCx{i}{x}^\nl, |\varphi_y|^2 \right)
= F_{w}\left(\varphiCx{i}{x}^\nl, \varphiCx{i}{y}^2 \right)
= 2\nl w_{xy} \varphiCx{i}{x}^{\nl-1} \varphiCx{i}{y}
+ 2 \binom{\nl}{2} w^2_{xy}
,
\end{equation}
while for $n=0$ and $p \ge 1$,
\begin{equation}
F_{w}\left(\phib_x^\nl, |\phi_y|^2 \right)
= \nl w_{xy} \phib_{x}^{\nl-1} \phib_{y}
.
\end{equation}
Thus, for all $(n,p)$ under consideration,
\begin{equation}
    \LT_a F_{w}\left(\sigma_a \varphiCx{i}{a}^\nl, \tau_y \right)
    = \sigma_a \left(\1_{j+1<j_{ab}} \nl w_{xy} \varphiCx{i}{a}^{\nl}
    + \1_{n \ge 1}\1_{p=2}  w^2_{xy} \right)
    ,
\end{equation}
with the modification noted below \eqref{e:LocF2} for $n=0$.
For $n \ge 1$ and  $p =1, 2$,
\begin{align}
\label{e:Fp4}
&F_{w}\left(\varphiCx{i}{x}^\nl, |\varphi_y|^4 \right)
= F_{w}\left(\varphiCx{i}{x}^\nl, \varphiCx{i}{y}^4 \right)
+
2
F_{w}  \left(\varphiCx{i}{x}^\nl, \varphiCx{i}{y}^2 \right)
\sum_{j:j\neq i}\varphiCx{j}{y}^2 ,
\\
&F_{w}\left(\varphiCx{i}{x}^\nl, \varphiCx{i}{y}^4 \right)
= 4 \nl  w_{xy} \varphiCx{i}{x}^{\nl-1} \varphiCx{i}{y}^{3}
+ 12 \binom{\nl}{2} w^2_{xy}
\varphiCx{i}{y}^{2},
\end{align}
while for $n=0$ and $p \ge 1$,
\begin{equation}
    F_{w}\left(\phib_x^\nl, |\phi_y|^4 \right)
    = 2 \nl  w_{xy} \phib_x^{\nl-1} \phi_y \phib_y^2
    + 2 \binom{\nl}{2} w^2_{xy} \phib_x^{\nl-2} \phib_y^{2}.
\end{equation}
The terms of total degree above $p$ are annihilated by $\LT$, and
\begin{equation}
    \LT_a \left[ F_{w} \left(\sigma_a \varphiCx{i}{x}^\nl, \varphiCx{i}{y}^2 \right)
    \sum_{j\neq i}\varphiCx{j}{y}^2 \right]
    =
    \sigma_a\1_{n \ge 1} 2 \binom{p}{2} w_{xy}^2 \sum_{j:j \ne i}(\varphi_a^j)^2.
\end{equation}
Thus, for all $(n,p)$ under consideration, we have
\begin{equation}
    \LT_a F_{w}\left(\sigma_a \varphiCx{i}{x}^\nl, \tau_y^2 \right)
    =
    \sigma_a\1_{j+1<j_{ab}}
     (n+8)w_{xy}^2
    \left(T \varphi^p_a
    \right)^i
    .
\end{equation}
Assembly of the above completes the proof of \eqref{e:LocF1}.
We omit the simpler proof of \eqref{e:LocF2}.
\end{proof}

\begin{proof}[Proof of Proposition~\ref{prop:pt}]
Equation~\eqref{e:qpt} states that
\begin{equation}
\label{e:qpt-pf}
q_{\pt,x} = q_x + \nl! \lambda_a\lambda_b
\delta[w^\nl_{ab}],
\end{equation}
and this
is an immediate consequence of \eqref{e:EV}, \eqref{e:pistarP} and \eqref{e:LocF2}.
To prove \eqref{e:hpt}--\eqref{e:lampt}, we use
\begin{equation}
\label{e:lampt-pf}
\lambda_{\pt,x} h_\pt
 =
\lambda_x h   -
    \1_{j + 1 < j_{ab}}\left(\nl \delta[\nu w^{(1)}]I -  \beta gT\right) \lambda_x h
    .
\end{equation}
The first term on the right-hand side arises from \eqref{e:EV},
and the rest of the right-hand side of
\eqref{e:lampt-pf} arises from $P_a$, via \eqref{e:pistarP} and the first term on
the right-hand side of \eqref{e:LocF1} (using also Lemma~\ref{lem:EV}).

Finally, we prove \eqref{e:tpt}.
The $t_x$ term in $e^\Lcal V$ is equal to
$-t_x-\1_{n \ge 1}\1_{p=2} C_{00}\lambda_x (e^+ \cdot h)$,
by Lemma~\ref{lem:EV}.  The contribution due to $-P$ arises only from the
first term on the right-hand side of \eqref{e:LocpistarF},
and only for $n\ge 1$ and $p = 2$, by Lemma~\ref{lem:LocF}.
Thus we seek the contribution to $t_{\pt,x}$ due to
\begin{equation}
    -\sum_{y\in\Lambda} \left(
    \LT_x F_{w+C}(e^\Lcal V_x,e^\Lcal V_y) - e^\Lcal \LT_x F_w(V_x,V_y)
    \right).
\end{equation}
We apply Lemma~\ref{lem:EV} and \eqref{e:LocF1}
to see that the first term contributes a $t_x$-term which is
equal to
$\1_{n \ge 1}\1_{p=2} \nu^+ w_{j+1}^{(2)}\lambda_x( e^+ \cdot h)$,
and the second contributes
\begin{equation}
    -\sum_{y \in \Lambda} \lambda_x ( M_{xy} e^\Lcal \varphi_x^p \cdot h)
    - \1_{n \ge 1}\1_{p=2} \nu w^{(2)}
    .
\end{equation}
The latter produces a $t_x$-term
\begin{equation}
    - \1_{n \ge 1}\1_{p=2}
    \left(
    \1_{j + 1 < j_{ab}}[p\nu w^{(1)} + g(n+2)w^{(2)}]C_{00}
    +
    \nu w^{(2)}
    \right)
    \lambda_x (e^+ \cdot h)
    ,
\end{equation}
where we have used $Te^+= \frac{n+2}{n+8}e^+$ for $p=2$.
This leads to
\begin{equation}\label{e:tpt-pf}
\begin{aligned}
t_{\pt,x}
= t_x + \1_{n \ge 1}&\1_{p=2} \lambda_x (e^+ \cdot h)
\Big(C_{0,0} + \delta[\nu w^{(2)}] \\
- &\1_{j + 1 < j_{ab}}C_{0,0}[2\nu w^{(1)} + g(n+2)w^{(2)}] \Big),
\end{aligned}
\end{equation}
which is equivalent to \eqref{e:tpt} by definition of $\nu^+$
and $\varsigma$.  This completes the proof.
\end{proof}

\section{Non-perturbative renormalisation group coordinate}
\label{sec:npflow}

Proposition~\ref{prop:pt} gives the evolution of the observable coupling
constants, as defined by the map $V\mapsto \Vpt$.  As discussed around
\eqref{e:Iex},
this map describes the effect of taking the expectation at a single
scale, but only at a perturbative level.
In this section, we
present aspects of the formalism of \cite{BS-rg-step,BBS-saw4-log},
which introduces and employs a non-perturbative renormalisation group
coordinate $K$.  With this coordinate,
Proposition~\ref{prop:pt} can be supplemented so as
to obtain a rigorous non-perturbative analysis, including observables.
A new ingredient is required here to deal with observables when $n \ge 2$,
namely the notion of $h$-factorisability which is defined in
Section~\ref{sec:sym} and developed further in Section~\ref{sec:hfac}.

\subsection{Circle product}

Recall that the sets $\Bcal_j$ and $\Pcal_j$ of scale-$j$ blocks and polymers in
$\Lambda$ are defined in Section~\ref{sec:pGi}.
For maps $F,G: \Pcal_j \to \Ncal$, we define the \emph{circle product}
$F\circ G: \Pcal_j \to \Ncal$ by
\begin{equation}
\label{e:circdef}
  (F \circ G)(X) = \sum_{Y \in \Pcal_j(X)} F(X \setminus Y) G(Y)
  \qquad
  (X \in \Pcal_j).
\end{equation}
The empty set $\varnothing$ is a polymer, as is $\Lambda$, so the sum over $Y$
always includes $Y=\varnothing$, and includes $Y=\Lambda$ when $X=\Lambda$.
Every map $F:\Pcal_j \to \Ncal$ that we encounter obeys $F(\varnothing)=1$.
The circle product is commutative and associative,
and has unit element $\1_\varnothing$
defined by $\1_\varnothing(X) = 1$ if $X=\varnothing$ and otherwise $\1_\varnothing(X)=0$.

We define
\begin{equation} \label{e:I0K0}
  I_0(X) = e^{-V_0(X)}, \quad K_0(X) = \1_{\varnothing}(X).
\end{equation}
Then
\begin{equation} \label{e:Z0I0K0}
  Z_0 =e^{-V_0(\Lambda)} = I_0(\Lambda) = (I_0 \circ K_0)(\Lambda).
\end{equation}
We wish to maintain the form
of \eqref{e:Z0I0K0} after each expectation in the progressive expectation
\eqref{e:progressive}.
Namely, we seek to define polynomials $U_j \in \Vcal_h^{(0)}$,
constants $u_j$, $t_{a,j}$, $t_{b,j}$, $q_{a,j}$, $q_{b,j}$,
and a \emph{non-perturbative coordinate}  $K_j : \Pcal_j \to \Ncal_j$,
such that $Z_j$ of \eqref{e:Z0def} is given by
\begin{equation}
\label{e:IcircKnew}
    Z_j = e^{\zeta_j}(I_j\circ K_j)(\Lambda),
    \qquad
    \zeta_j= - u_j|\Lambda| +(t_{a,j} \sigma_a+ t_{b,j}\sigma_b)
    + \textstyle{\frac 12} (q_{a,j} + q_{b,j}) \sigma_a\sigma_b
    ,
\end{equation}
with $I_j=I_j(U_j)$ given by \eqref{e:Idef}.
We systematically use the symbol
$U$ for elements of $\Vcal_h^{(0)}$
and $V$ for other polynomials.
Let  $\delta\zeta_{j+1} = \zeta_{j+1}-\zeta_j$.
Then \eqref{e:Z0def} can equivalently be written as
\begin{equation} \label{e:IcircKdu}
  \Ex_{j+1}\theta(I_j \circ K_j)(\Lambda)
  =
  e^{-\delta \zeta_{j+1}}(I_{j+1} \circ K_{j+1})(\Lambda)
  .
\end{equation}

By the definition in \eqref{e:ZN}, $Z_N = Z_N(\varphi) = (\Ex_C \theta Z_0)(\varphi)$,
and we seek to write this as $Z_N=e^{\zeta_N}(I_N \circ K_N)(\Lambda)$.
At the final scale, $\Pcal_N = \{\varnothing, \Lambda_N\}$, and
$I_N = e^{ - U_N} (1 + W_N)$ by \eqref{e:Idef}, so
\begin{equation}
\label{e:ZNfinal}
Z_N = e^{\zeta_N}\left(I_N  + K_N \right) = e^{\zeta_N}\left(e^{ -U_N} (1 + W_N)  + K_N \right).
\end{equation}
To prove Theorems~\ref{thm:mr-stars} and~\ref{thm:mr-ab},
our goal is to achieve \eqref{e:ZNfinal} with $W_N$ and $K_N$ as error terms, so that
the partition function $Z_N(\varphi)$ is to leading order equal to $e^{\zeta_N - U_N(\varphi)}$.
Assuming this, we can evaluate the derivatives from Section~\ref{sec:of} easily.
For example, as $U_N(0) = 0$,
the correlation function \eqref{e:corrdiff} or
the watermelon network \eqref{e:intobslog}
are given by $D^2_{\sigma_a\sigma_b} \zeta_N = \tfrac{1}{2} (q_{a,N} + q_{b,N})$.
We discuss this in more detail in Section~\ref{sec:rg-flow-analysis},
where we show that for both models, the important information is ultimately encoded in
the observable coupling constants $\qx{N}$ and $\lambda_{x,N}$.

\subsection{Symmetries and symmetry reduction}
\label{sec:sym}

New considerations concerning symmetry, not present in \cite{BS-rg-IE,BS-rg-step},
are needed for our analysis of observables when $n \ge 2$.
We present the relevant definitions here.

\begin{defn}
\label{def:latticesym}
\emph{Lattice symmetry.} Let $\mathcal{A}$ denote the set of graph automorphisms of $\Lambda$,
i.e., bijections that preserve nearest neighbours.
An automorphism $A\in\mathcal{A}$ acts on $\Ncal$ via
$AF(\varphi) = F(A\varphi)$, where $(A\varphi)_x = \varphi_{Ax}$.
We say that a local monomial $M_x \in \Ncal$ is \emph{Euclidean invariant}
if $A M_x = M_x$ for all $A \in \mathcal{A}$ that fix $x$.
We say that a function $F : \Pcal_j \to \Ncal$ is \emph{Euclidean covariant} if $A(F(X))=F(AX)$
for all automorphisms $A$ of $\Lambda$ and all $X\in \Pcal_j$.
\end{defn}

\begin{defn}
\label{def:fieldsym}\emph{Field symmetry.}
For $n \ge 1$, an $n \times n$ real matrix $m$ acts on $F \in \Ncal$ via
$(mF)(\varphi) = F(m \varphi)$.  There is no action of $m$ on $\sigma_a$ or $\sigma_b$.
Given a group $G$ of $n \times n$ matrices, we say that $F \in \Ncal$ is $G$-invariant if $mF = F$ for all $m \in G$.
\\
For $n = 0$, let $G=U(1)$ be the group
$\{z \in \C : |z|=1\}$
with complex multiplication.
We set $\sigma_a = \sigma^p$ and $\sigma_b = \bar\sigma^p$,
with $\sigma \in \C$.
Then $m \in U(1)$ acts on $F \in \Ncal$ by $(mF)(\sigma,\bar\sigma,\phi,\phib,\psi,\psib)
= F(m\sigma,\bar{m}\bar\sigma, m \phi, \bar{m}\phib, m \psi, \bar{m}\psib)$.
We say that $F$ is \emph{$U(1)$-invariant}, or \emph{gauge invariant},
if $mF=F$ for all $m\in U(1)$.
\\
The supersymmetry operator $Q$ is defined, e.g., in \cite[Section~\ref{pt-sec:Vpt}]{BBS-rg-pt}
or \cite[Section~6]{BIS09}, and
we say that $F$ is \emph{supersymmetric} if $QF=0$.
Supersymmetry is special to the $n = 0$ case and does not play a role for observables;
the rest of this paper can be read without delving into its precise meaning.
\end{defn}

By definition, for $n=0$, elements of $\Vcal_h$ of \eqref{e:Vmulticomponent}
are $U(1)$-invariant.
For $n \ge 1$, we use the following matrix groups:
\begin{itemize}
\item
$G = O(n)$, the group of $n\times n$ orthogonal matrices.
\item
$G = S(n)$, the permutation subgroup of $O(n)$, consisting of the $n!$ matrices obtained
by permutations of the columns of the identity matrix.
\item
$G = R(n)$,
the reflection subgroup of $O(n)$ consisting of the $2^n$ diagonal matrices with diagonal
elements in $\{-1,+1\}$.
\end{itemize}

Although $O(n)$-invariance will hold for the \emph{bulk}
space $\Ncal^\varnothing$ for all $n \ge 1$, for $n \ge 2$ the $O(n)$ symmetry
can be reduced by choice of $h$.  This can be seen already from the $\varphi_a^p \cdot h$
term in $V_{0,x}$,
which is not $R(n)$ invariant when $p=1$, and which is not $S(n)$-invariant for $p=2$ unless
$h$ is in the eigenspace $E^+$ spanned by
$e^+=(1,\ldots,1)$.
We now define a weaker property that replaces
$O(n)$-invariance for the observable terms
when $n \geq 2$, and that plays a role in the definition of the Banach space in
which the non-perturbative coordinate $K$ lies.

\begin{defn}
\label{def:efac}
Let $n \ge 1$ and fix $h \in \R^n$.  We say that
$F \in \Ncal$  is $h$-\emph{factorisable} if  for $\alpha =a,b$:
\begin{enumerate}[(i)]
\item
there exists
$F_\alpha^* \in (\pi_\varnothing\Ncal)^n$ (depending on $h$, not unique) such that
$\pi_\alpha F = \sigma_\alpha (F_\alpha^*\cdot h)$,
and
\item
$(P F_\alpha^*)(\varphi) = F_\alpha^* (P\varphi)$ for all $P \in S(n)$, where by
definition $P F_\alpha^*$ is the result of permuting the components of $F^*_\alpha$
with the permutation $P$.
\end{enumerate}
We write $\Ncalefac = \{F \in \Ncal : F \; \text{is $h$-factorisable}\}$ for
the vector space of $h$-factorisable elements of $\Ncal$.
\end{defn}

In the following definition, $h$ does not play a direct role
as a vector when $n=0,1$ but we nevertheless use it as a notational device
to write $\Ncal_h$ as the vector subspace of $\Ncal$ that obeys the conditions
listed in the definition.
For $n=0$, we say that $F$ \emph{has no constant part} if its degree-zero part (as a form)
is equal to zero when evaluated at $\phi=\bar\phi=0$.

\begin{defn}
\label{def:Ncalh}
For $n \ge 0$, let $\Ncal_{h}$ denote the subspace of all $F \in \Ncal$ such that
\begin{enumerate}[(i)]
\item
If $n = 0$, $\pi_\varnothing F$ is supersymmetric,
$F$ is $U(1)$-invariant, and $F$ has no constant part.
\item
If $n \geq 1$, $F\in \Ncalefac$,
$\pi_\varnothing F$ is $O(n)$-invariant, and if in addition $p = 2$, then $F$ is $R(n)$-invariant.
\end{enumerate}
\end{defn}

By Proposition~\ref{prop:pt}, if we choose $h \in E^\pm$, then
$V_\pt : \Vcal_h \to \Vcal_h$.
The symmetry restrictions of $\Ncal_h$, particularly $h$-factorisation, are
used to carry this perturbative fact over to the non-perturbative renormalisation group coordinate
and show that $V_j\in \Vcal_h$ for all $j$.
The two powers $\gamma_{n,p}^+$ and $\gamma_{n,p}^-$
for the logarithmic corrections in Theorems~\ref{thm:mr-stars}--\ref{thm:mr-ab}
will arise from the distinction between $h \in E^+$ and $h \in E^-$.

\subsection{The non-perturbative coordinate \texorpdfstring{$K$}{K}}

We now discuss the definition of the non-perturbative coordinate $K$.
This requires several definitions, as preparation.

A polymer $X\in\Pcal_j$ is \emph{connected} if for any $x,y\in X$ there exists a path
of the form
$x_0=x,x_1,\ldots,x_{n-1},x_n=y$ with $\|x_{i+1}-x_i\|_\infty = 1$ for all $i$.
Every polymer can be partitioned into connected components, and we denote the
set of connected components of $X$ by ${\rm Comp}_j(X)$.
Let $\Scal_j\subset\Pcal_j$ denote the set of connected polymers
consisting of at most $2^d=16$
blocks;  elements of $\Scal_j$ are called \emph{small sets}.
(The specific number 16 plays a special role in \cite{BS-rg-step}, but not here.)
The \emph{small set neighbourhood} of $X$ is
\begin{equation}
    X^\Box = \bigcup_{Y \in\Scal_j: X\cap Y \not =\varnothing } Y.
\end{equation}

For $n \ge 0$,
we define
$\Ncal(X)$ to consist of those elements of $\Ncal$ in \eqref{e:Ncaldecomp} which
depend on the boson, fermion (for $n=0$), and external fields only at points in $X$, where
we regard the external field
$\sigma_x$ as located at $x$ for $x=\pp,\qq$.
At scale $j$, $K$ lies in the space $\Kcal_j$ of maps from $\Pcal_j$ to
$\Ncal$, given in the following definition.

\begin{defn}
\label{def:Kspace}
Let $h = 1$ for $n = 0, 1$ and $h\in \R^n$ for $n \ge 1$.
Let $\Kspace_{j} = \Kspace_{j} (h,\Lambda_N)$ be the vector space of functions $K :
\Pcal_j  \to \Ncal$ with the properties:
\begin{itemize}
\item Field locality: $K(X) \in \Ncal(X^{\Box})$ for each connected $X\in\Pcal_j$.
Also, (i) $\pi_a K(X) =0$ unless $a \in X$, (ii) $\pi_b K(X) =0$ unless $b \in X$,
and (iii) $\pi_{ab}K(X)=0$ unless $a \in X$ and $b \in X^\Box$ or vice versa, and
$\pi_{ab}K(X)=0$ if $X\in\Scal_j$ and $j<j_{ab}$.

\item Symmetry:
$\pi_\varnothing K$ is Euclidean covariant, and $K(X) \in \Ncal_h$ for all $X\in\Pcal_j$.

\item Component factorisation: $K (X) = \prod_{Y
\in {\rm Comp}_j( X)}K (Y)$ for all $X\in\Pcal_j$.
\end{itemize}
\end{defn}

In \cite[Section~\ref{step-sec:mr}]{BS-rg-step}, the
scale dependent \emph{renormalisation
group map} from a domain
in $\Vcal_h^{(0)} \times \Kcal_j$ to $\Vcal_h^{(1)} \times \Kcal_{j+1}$
is defined, which we write as
\begin{equation}
\label{e:RGmap}
    (U,K) \mapsto (V_+,K_+).
\end{equation}
In \eqref{e:RGmap} and elsewhere, to simplify the notation we systematically drop
labels $j$ for scale, and indicate scale $j+1$ simply by $+$.
We use the map \eqref{e:RGmap}, which satisfies \eqref{e:IcircKdu}.
The discussion in \cite{BS-rg-IE,BS-rg-step} is written explicitly
for the WSAW with the observable having power
$p=1$, but it applies in our present more general setting with the modifications
discussed
in Section~\ref{sec:step} below.

The map $(U,K) \mapsto V_+$ is explicit and relatively simple, and is defined as follows.
Let $\LT_{Y,B}$ denote the operator defined by
$\LT_{Y,B} F = P_Y(B)$, where $P_Y$ is the polynomial determined by $P_Y(Y) = \Loc_Y F$.
We define a map $V \mapsto V^{(1)}$ from
$\Vcal_h$ to $\Vcal_h^{(1)}$ by replacing
$z\tau_{\Delta}+y\tau_{\nabla\nabla}$ in $V\in \Vcal_h$ by
$(z+y)\tau_{\Delta}$ in $V^{(1)}$.
Let $h \in E^\pm$.
We also define a map $V \mapsto V^{(0)}$ from
$\Vcal_h$ to $\Vcal_h^{(0)}$ by replacing
$z\tau_{\Delta}+y\tau_{\nabla\nabla}$ in $V$ by
$(z+y)\tau_{\Delta}$ and replacing
$u,t_a,t_b,q_a,q_b$ in $V$ by zero.
As in \cite[Section~\ref{step-sec:Rconstruction}]{BS-rg-step}, the map $(U,K) \mapsto V_{+}$
is given by
\begin{equation}
\label{e:Vplusdef}
  V_{+}(U,K) = \Vpt^{(1)}(U-Q) \quad\text{with}\quad
  Q(B)
  =
  \sum_{Y \in \Scal:Y \supset B}
  \LT_{Y,B} \left( \frac{K (Y)}{I(Y,V)} \right),
\end{equation}
where $\Vpt$ is the explicit quadratic polynomial
map $V \mapsto \Vpt$ discussed in Section~\ref{sec:Vptdef}.
When $K=0$,  $V_{+}(U,0)$ is simply $\Vpt^{\smash{(1)}}(U)$.
We write $V_+=(\delta \zeta_+,U_+)$, and in particular
$\delta \zeta_+(U,0) = \delta \zeta_\pt(U)$ and $U_+(U,0) = \Vpt^{\smash{(0)}}(U)$.
We express estimates on $V_+$
in terms of $R_+$ defined by
\begin{equation} \label{e:Rdef}
    R_+(U,K) = V_+(U,K)-V_+(U,0)
    = V_+(U,K)-\Vpt^{\smash{(1)}}(U)
    \in \Vcal_h^{(1)}
    .
\end{equation}

As in \cite[\eqref{step-e:piVKplus}]{BS-rg-step},
the renormalisation group map has the property
\begin{equation} \label{e:bulk}
  \pi_\varnothing V_+(U,K) = V_+(\pi_\varnothing U,\pi_\varnothing K), \quad
  \pi_\varnothing K_+(U,K) = K_+(\pi_\varnothing U,\pi_\varnothing K).
\end{equation}
Thus, under the map \eqref{e:RGmap},
the \emph{bulk coordinates} $(\pi_\varnothing V_j, \pi_\varnothing K_j)$ satisfy
a closed evolution independent of the observables. We denote this evolution map by
$(V_+^\varnothing,K_+^\varnothing)$.  Then the bulk part of \eqref{e:RGmap} becomes
\begin{equation} \label{e:RGmapbulk}
  (\pi_\varnothing V_{+}, \pi_\varnothing K_{+})
    = (V_+^\varnothing(\pi_\varnothing U,\pi_\varnothing K),
       K_+^\varnothing(\pi_\varnothing U,\pi_\varnothing K)).
\end{equation}

\subsection{Existence of bulk flow}

A critical global renormalisation group flow of the bulk coordinates
is constructed in \cite{BBS-saw4-log} for WSAW and in \cite{BBS-phi4-log} for $|\varphi|^4$.
In particular, there is a construction of $(\pi_\varnothing V_{j},\pi_\varnothing K_j)$, obeying
\eqref{e:RGmapbulk}  for all $j$,
such that \eqref{e:IcircKdu} holds
if $\sigma_a = \sigma_b = 0$.
The bulk flow provides detailed information about the
sequence $\pi_\varnothing V_{j}$, and estimates on $\pi_\varnothing K_j$
sufficient for studying the infinite volume limit at the critical point.

For the bulk flow, we change perspective on which variables are independent.
Both $|\varphi|^4$ and WSAW have parameters $g,\nu$.
In \eqref{e:Vtil0def}, additional parameters $m^2$, $g_0$, $\nu_0$, $z_0$  are introduced.
For the moment we consider $m^2,g_0,\nu_0,z_0$
as four independent variables and do not
work with
$g,\nu$ directly.
We relate $m^2,g_0,\nu_0,z_0$ to the original parameters $g,\nu$
in Section~\ref{sec:changevariables} below.

To state the result about the bulk flow,
let $\gbar_j$ be the $(m^2,g_0)$-dependent sequence determined by
$\gbar_{j+1}=\gbar_j - \beta_j\gbar_j^2$, with $\gbar_0=g_0$, and
with $\beta_j=\beta_j(m^2) = (n+8)\delta[w^{(2)}_j]$ as in
\eqref{e:betadef}.
For $m^2>0$, we define the \emph{mass scale} $j_m$
to be the largest integer $j$ such that $mL^j \le 1$, and we set
$j_0=\infty$.
By definition, $\lim_{m \downarrow 0}j_m=\infty$.
Given  $\Omega >1$ ($\Omega=2$ is a good choice),
we define
\begin{equation}
\label{e:chicCovdef}
    \chicCov_j = \Omega^{-(j-j_m)_+},
\end{equation}
where $x_+=\max\{x,0\}$.
By \cite[Lemma~\ref{pt-lem:wlims}]{BBS-rg-pt}, $\beta_j = O(\chi_j)$
(\cite[Lemma~\ref{pt-lem:wlims}]{BBS-rg-pt} actually shows that
$\beta_j =O(\Omega^{-(j-\jm)_+})$ for another scale $j_\Omega$
used in \cite{BBS-rg-pt,BS-rg-IE,BS-rg-step},
but $\Omega^{-(j-\jm)_+}$ and $\chi_j$ are comparable
by \cite[Proposition~\ref{pt-prop:rg-pt-flow}]{BBS-rg-pt}.)
By
\cite[Proposition~\ref{log-prop:approximate-flow}]{BBS-saw4-log}
and
\cite[\eqref{log-e:chigbd-bis}]{BBS-saw4-log} respectively,
the bounds
\begin{equation}
\chi_j \gbar_j^p \leq O\left(\frac{g_0}{1+g_0j}\right)^p \quad (p \geq 0),
\qquad
\label{e:chisum}
\sum_{k=j}^\infty \chi_k \gbar_k^p = O(\chi_j \gbar_j^{p-1}) \quad (p > 1),
\end{equation}
hold uniformly in $(m^2,g_0) \in [0,\delta)^2$, for a small $\delta >0$.
The sequence $\gbar_j$ converges to $0$ when $m^2=0$ but not when $m^2>0$.

For WSAW, the following theorem is a consequence of
\cite[Proposition~\ref{log-prop:KjNbd}]{BBS-saw4-log}.
For $|\varphi|^4$, it is \cite[Theorem~\ref{phi4-log-thm:flow-flow}]{BBS-phi4-log}.
The latter also controls the flow of the coupling constant $u_j$, which is used
for the analysis of the pressure in \cite{BBS-phi4-log} but is not needed here.
The domains $\domRG_j^\varnothing$, and the $\Wcal_j$-norms  on the space $\Kcal_j$,
which appear in the theorem are discussed following its statement.

\begin{theorem} \label{thm:bulkflow}
  Let $d=4$, $n \ge 0$, and let $\delta>0$ be sufficiently small.
  Let $N \ge 1$.
  Let $(m^2,g_0)\in [0,\delta)^2$ and $\sigma_a=\sigma_b=0$.
  There exist $M>0$ and an infinite sequence
  of continuous functions $U_j=(g_j^c,\nu_j^c,z_j^c)$ of $(m^2,g_0)$,
  independent of the volume parameter $N$, such that
  for initial conditions $U_0 = (g_0,\nu_0^c,z_0^c)$ and
  $K_0=\1_{\varnothing}$,
  a flow  $(U_j,K_j) \in \domRG_j^\varnothing$
  exists such that \eqref{e:RGmapbulk} holds for all $j+1 < N$,
  and, if $m^2 \in [\delta L^{-2(N-1)}, \delta)$, also for $j+1=N$.
  Moreover,
  $g_j^c = O(\gbar_j)$, $z_j^c = O(\chi_j \gbar_j)$, $\nu_j = O(\chi_j L^{-2j} \gbar_j)$,
  and
  \begin{equation}
    \label{e:VVbar1-bulkflow}
    \|K_j\|_{\Wcal_j}
    =     \|\pi_\varnothing K_j\|_{\Wcal_j}
    \leq M\chicCov_j \gbar_j^3
    \quad ( j \le N).
  \end{equation}
\end{theorem}

In the remainder of the paper, we often drop the superscripts and write simply
\begin{equation}
    U_j=(g_j,\nu_j,z_j)
\end{equation}
for the sequence provided by Theorem~\ref{thm:bulkflow}.
The stated continuity of $U_j$ is not part of the statements of
\cite[Proposition~\ref{log-prop:KjNbd}]{BBS-saw4-log}
or \cite[Theorem~\ref{phi4-log-thm:flow-flow}]{BBS-phi4-log},
but it is established in \cite[Section~\ref{log-sec:Fcont}]{BBS-saw4-log}.

The definition of the $\Wcal_j$ norm on $\Kcal_j$
in \eqref{e:VVbar1-bulkflow} is discussed
at length in \cite{BS-rg-step},
and we do not repeat the details here.
The inequality \eqref{e:VVbar1-bulkflow} provides various estimates on $K_j(X)$
and on its derivatives with respect to fields, in terms of the size of the polymer $X$.
Some examples of its use are given in Lemma~\ref{lem:WK} below.  For example,
as noted explicitly in
\cite[\eqref{step-e:Kg1}]{BS-rg-step}, \eqref{e:VVbar1-bulkflow}
with $j = N$ implies that
\begin{equation}
\label{e:Kgnull}
    |\pi_\varnothing K_N(\Lambda)|
    \le M \chi_N \gbar_N^3
\end{equation}
(with fields set equal to zero on the left-hand side),
uniformly in $m^2 \in [\delta L^{-2(N-1)}, \delta)$.

The $\Wcal_j=\Wcal_j(\sgen)$ norm depends on a parameter
$\sgen = (\mgen^2,\ggen) \in [0,\delta)^2$,
whose significance is discussed in \cite[Section~\ref{log-sec:flow-norms}]{BBS-saw4-log}.
Useful choices of this parameter depend on the scale $j$, as well as
on approximate values of the mass parameter $m^2$ of the covariance and the coupling constant $g_j$.
We use the convention that when the parameter $\sgen$ is omitted,
it is given by $\sgen = s_j = (m^2,\ggen_j(m^2,g_0))$, where
$\ggen = \ggen_j$ is defined in terms of the initial condition $g_0$ by
\begin{equation} \label{e:ggendef}
  \ggen_j = \ggen_j(m^2,g_0)
  = \gbar_j(0,g_0) \1_{j \le j_m} + \gbar_{j_m}(0,g_0) \1_{j > j_m}.
\end{equation}
By
\cite[Lemma~\ref{log-lem:gbarmcomp}]{BBS-saw4-log},
\begin{equation}
\label{e:gbarggen}
    \ggen_j=\gbar_j+O(\gbar_j^2),
\end{equation}
so the sequences $(\ggen_j)$ and $(\gbar_j)$ are the same to
leading order.
Moreover,
\begin{equation}
\label{e:ggbar}
    g_j=\gbar_j(1+O(\gbar_j |\log \gbar_j|));
\end{equation}
this follows from \cite[\eqref{log-e:TVV2}, \eqref{log-e:VVbar2app}]{BBS-saw4-log}
for WSAW and the same result holds for $n \ge 1$ according to \cite{BBS-phi4-log}.
Thus the sequences $\ggen_j$, $\gbar_j$ and $g_j$ are essentially interchangeable,
and in particular error bounds expressed in terms of any one of them are equivalent.

The domain $\domRG_j^\varnothing = \domRG_j^\varnothing(\sgen)
\subset \Vcal_h^\varnothing \times \Kcal_j^\varnothing$ also depends on $\sgen$
(with the convention mentioned above when $\sgen$ is omitted),
is independent of $h$ as we deal only with the bulk here,
and is defined as follows.
For the universal constant $C_\DV \ge 2$ determined in \cite{BBS-saw4-log}, for $j<N$,
\begin{equation} \label{e:domRGbulk}
  \domRG_j^\varnothing(\sgen)
  = \{(g,\nu,z)\in \R^3 :
    C_{\DV}^{-1} \ggen <  g < C_{\DV} \ggen,
    \; L^{2j}|\nu|,|z| \le C_\DV \ggen \}
    \times B_{\Wcal_j^\varnothing}(\DVa\chigen_j\ggen^3).
\end{equation}
The first factor is the stability
domain defined in \cite[\eqref{step-e:DV1}]{BS-rg-IE}, restricted to
the bulk coordinates and real scalars.
In the second factor,
$B_X(a)$ denotes the open ball of radius $a$
centred at the origin of the Banach space $X$, and $\DVa$ is as in
\cite[Theorem~\ref{log-thm:step-mr-fv}]{BBS-saw4-log}
and
\cite[Theorem~\ref{phi4-log-thm:step-mr}]{BBS-phi4-log};
for concreteness we use $\DVa = 4 M$ where $M$ is the constant of Theorem~\ref{thm:bulkflow}
(the same choice was made above \cite[Proposition~\ref{log-prop:flow-flow}]{BBS-saw4-log}).
The space $\Kcal^\varnothing$ is the restriction of $\Kcal$ to elements $K$ with $\pi_*K(X) = 0$
for all polymers $X$. Since, by \eqref{e:bulk}, the renormalisation group acts triangularly,
the distinction between $\Wcal$ and $\Wcal^\varnothing$ is unimportant for the bulk flow, and $\Wcal^\varnothing$
is denoted by $\Wcal$ in \cite{BBS-saw4-log}.

\subsection{Properties of the bulk flow}

We provide some details about the flow of bulk coupling constants, for later use.

The \emph{bubble diagram} is defined by
\begin{equation}
\label{e:bubbledef}
  \bubble_{m^2} = (n+8) \int_0^\infty \!\!\!
  \int_0^\infty P(X(T) = Y(S)) e^{-m^2T} e^{-m^2 S} \; dT \, dS
  ,
\end{equation}
where $X,Y$ are independent continuous-time simple random walks (taking
steps at the events of a rate-$(2d)$ Poisson process).
For $d=4$, it is an exercise in calculus (see \cite[\eqref{log-e:freebubble}]{BBS-saw4-log})
to see that
\begin{equation}
\label{e:newbubble}
    \bubble_{m^2} \sim  {\sf b} \log m^{-2}
    \;\;\;
    \text{as $m^2 \downarrow 0$, with ${\sf b} = \frac{n+8}{16 \pi^2}$.}
\end{equation}
We recall from \cite[Lemma~\ref{pt-lem:betalim}]{BBS-rg-pt} that
\begin{equation}
\label{e:betalim}
    \beta_j = {\sf b} \log L +O(L^{-j})
    \quad\text{for $m^2=0$} .
\end{equation}

\begin{lemma} \label{lem:ginfty}
  For $(m^2,g_0) \in (0,\delta)^2$,
  the limit $g_\infty = \lim_{j \to \infty} g_j$  exists,
  is continuous in $(m^2,g_0)$, and extends continuously to $[0,\delta)^2$.
  For $g_0 \in (0,\delta)$,
  \begin{equation} \label{e:ginfty}
    g_\infty \sim \frac{1}{\bubble_{m^2}}
    \qquad \text{as $m^2 \downarrow 0$}
    .
  \end{equation}
\end{lemma}

\begin{proof}
For $n=0$, this is \cite[Lemma~\ref{log-lem:ginfty}]{BBS-saw4-log}, adapted
from its statement for the sequence $\gch_j$ to the sequence $g_j$.
That this adaptation is possible is discussed at the end of
\cite[Section~\ref{log-sec:finscale}]{BBS-saw4-log}.
For $n \ge 1$, \eqref{e:ginfty} also holds, as indicated in \cite[\eqref{phi4-log-e:ginfty}]{BBS-phi4-log}.
\end{proof}

For the next lemma, recall that $\Ecal_{ab}^{(p)}$ is defined in \eqref{e:Ecaldef}.

\begin{lemma}
\label{lem:ggenasy}
As $|a - b| \to \infty$, $L^{j_{ab}} = 2 |a - b| + O(1)$.  If $j_{ab} < j_m$ then
$g_{j_{ab}}^{-1} ={\sf b} (\log |a - b|) ( 1 + \Ecal_{ab}^{(2)} )$.
\end{lemma}

\begin{proof}
It is an immediate consequence of \eqref{e:jabbds} that
$L^{j_{ab}}=2|a-b| + O(1)$.

For $j_{ab}<j_m$, we have $\ggen_{j_{ab}}=\gbar_{j_{ab}}$ with $\gbar_{j_{ab}}$
defined by the
sequence $\beta_j$ given by $m^2=0$.
By \eqref{e:gbarggen}--\eqref{e:ggbar} it suffices to prove that
\begin{align}
\label{e:gjabnew}
    \gbar_{j_{ab}}(0)^{-1}
    &= {\sf b} (\log |a-b|) ( 1+\Ecal_{ab}^{(2)})
    .
\end{align}
It is shown in the proof of \cite[Lemma~\ref{flow-lem:elementary-recursion}]{BBS-rg-flow} that
if $\psi: \R_+ \to \R$ is absolutely continuous then
\begin{equation}
\label{e:gbarsumbis}
  \sum_{l=j}^{k} \beta_l \psi(\bar g_l) \bar g_l^2
  = \int_{\bar g_{k+1}}^{\bar g_{j}} \psi(t) \; dt
  + O\left(\int_{\bar g_{k+1}}^{\bar g_{j}} t^2 |\psi'(t)| \; dt
  \right)
  .
\end{equation}
Let $\beta_\infty = {\sf b}\log L$.
We set $\psi(t) = t^{-2}$ in \eqref{e:gbarsumbis},
and apply \eqref{e:betalim},
to obtain
  \begin{equation} \label{e:gjinv}
    \bar g_{k}^{-1}
    = \bar g_0^{-1} + \sum_{j=0}^{k-1} \beta_j + O(|\log \bar g_{k}|)
    = \bar g_0^{-1} + \beta_\infty k + O(1) + O(|\log \bar g_{k}|)
    .
  \end{equation}
In particular,
$\gbar_k^{-1} = O(\gbar_0^{-1} + \beta_\infty k) = O(k)$
(with $g_0$-dependent constant).
Therefore,
\begin{equation}
    \gbar_k^{-1} = \beta_\infty k + O(\log k).
\end{equation}
This gives \eqref{e:gjabnew} and completes the proof.
\end{proof}

\begin{lemma}
\label{lem:nu}
Let $\delta_j=\delta_j[\nu w^{(1)}]$ and
$\delta_j'= \nu_{j+1}w_{j+1}^{(1)} - \nu_j w_j^{(1)}$.
Then $\delta_j=O(\chi_j \gbar_j)$ and $|\delta_j - \delta_j'|=O(\chi_j \gbar_j^2)$.
Also, the sequence $g_j$ obeys
\begin{equation}
\label{e:gjflow}
    g_{j+1}=(1-e_j)g_j \quad\text{with}\quad
    e_j = \beta_j g_j +4\delta_j'  + \tilde r_j,
    \quad \tilde r_j = O(\chi_j \gbar_j^2).
\end{equation}
\end{lemma}

\begin{proof}
By \cite[Lemma~\ref{pt-lem:wlims}]{BBS-rg-pt}, $w_j^{(1)}=O(L^{2j})$
and by \cite[Proposition~\ref{pt-prop:Cdecomp}]{BBS-rg-pt}, $C_{j+1;ab} =O(\chi_j L^{-2j})$.
With \eqref{e:deltadef} and \eqref{e:domRGbulk}, we therefore have
\begin{equation}
\begin{aligned}
\delta_j
&= (\nu_j + (2+n) g_j C_{j+1;00})(w_j^{(1)} +C_{j+1}^{(1)}) - \nu_j w_j^{(1)}\\
&= \nu_jC_{j+1}^{(1)} + (2+n) g_j C_{j+1;00}w_{j+1}^{(1)}\\
&= O(\ggen_j L^{-2j})O(\chi_j) + O(g_j)O(\chi_jL^{-2j})O(L^{2j}) = O(\chi_j \gbar_j).
\end{aligned}
\end{equation}
For the second statement, by definition
\begin{equation}
    \delta_j' =
    \left(\nu_{j+1} - (\nu_j + (2+n) g_j C_{j+1;00}) \right) w_{j+1}^{(1)}.
\end{equation}
The subtracted terms in the difference on the right-hand side cancel the first-order
part of $\nu_{j+1}$ (see \cite[\eqref{pt-e:nupta}]{BBS-rg-pt}),
leaving only the higher-order terms which are bounded by $O(\chi_j L^{-2j}\gbar_j^2)$
according to \cite[\eqref{step-e:factnu}]{BS-rg-step}.
This leads to the desired bound on $\delta_j'$.

Finally, to prove \eqref{e:gjflow}, we recall from
\cite[\eqref{log-e:gchrec}]{BBS-saw4-log} that
\begin{equation} \label{e:gchrec}
  \gch_{j+1}
  = \gch_j - \beta_j \gch_j^2 + r_j
  \quad\text{with}\quad r_j=O(\chicCov_j \gch_j^3)
  = O(\gbar_j^3),
\end{equation}
where, by \cite[\eqref{pt-e:gch-def1}]{BBS-rg-pt},
\begin{align}
  \label{e:gch-def1}
  \gch_j  & = g_j +4g_j\nu_j w_j^{(1)}
  .
\end{align}
Then \eqref{e:gjflow} follows from substitution of \eqref{e:gch-def1} into \eqref{e:gchrec}.
\end{proof}

Recall from \eqref{e:fdef} that
the eigenvalues of the matrix $A_j$ defined in \eqref{e:Ajdef}
are $f_j = 1- p\delta_j [\nu w^{(1)}] - \beta_j g_j \gamma$ for $j+1<j_{ab}$,
and otherwise $f_j=1$.
Now $g_j,\nu_j$ (and also $z_j$) are given by the flow of the bulk
coupling constants determined in Theorem~\ref{thm:bulkflow}.
The constant $\gamma$ is given by $\gamma = \gamma_{n,p}^\pm$, depending on
the values of $(n,p)$ and the choice of $h \in E^\pm$.
For $j \le J$, we write
\begin{equation}
\label{e:Pidef}
    \Pi_{j,J} = \prod_{i=j}^J f_i,
    \quad\quad
    \Pi_j=\Pi_{0,j}.
\end{equation}
The value of $\Pi_{j,J}$ depends on $\gamma$, and we write $\Pi_{j,J}^\pm$ for its
values when $\gamma=\gamma^\pm$.
The matrix product $A_J A_{J-1} \cdots A_j$ has eigenvalues
$\Pi_{j,J}^\pm$, with the eigenvalue $\Pi_{j,J}^-$ only occurring for $n \ge 2$ and $p=2$.
Error estimates in the following lemma depend on $\gamma$, but this is unimportant
since $\gamma$ may be regarded as fixed.

\begin{lemma}
\label{lem:Pi}
Let $(m^2,g_0) \in [0,\delta]$.
Let $0 \le j \le j_{ab}$, $j \le J  <\infty$,  $J_{ab}=\min\{J , j_{ab}\}$, and $\gamma \in \R$.
There exists $\alpha_j = 1 + O(\gbar_j)$ such that
\begin{equation}
    \Pi_{j,J}
    = \alpha_j \left( \frac{g_{J_{ab}+1}}{g_j} \right)^\gamma
    \big( 1 + O(\chi_{J_{ab}} \gbar_{J_{ab}}) \big).
\end{equation}
\end{lemma}

\begin{proof}
Since $f_i=1$ for $i+1 \ge j_{ab}$, it suffices to restrict attention to
$J$ with $J+1<j_{ab}$.
Let $\delta_i=\delta_i[\nu w^{(1)}]$.
By Lemma~\ref{lem:nu}, $g_{i+1}=(1-e_i)g_i$ with $e_i$ given by \eqref{e:gjflow}.
As noted below \eqref{e:chicCovdef}, $\beta_j = O(\chi_j)$.
By Lemma~\ref{lem:nu},
$\delta_i=O(\chi_i \gbar_i)$.
Therefore, $e_i=O(\chi_i \gbar_i)$.
Let $\delta_i'= \nu_{i+1}w_{i+1}^{(1)} - \nu_i w_i^{(1)}$.
By Lemma~\ref{lem:nu},
$|\delta_i - \delta_i'|=O(\chi_i \gbar_i^2)$.
By \eqref{e:fdef},
\begin{equation}
    f_i = (1-\gamma e_i)(1+d_i) ,
\end{equation}
with
\begin{equation}
d_i = (1-\gamma e_i)^{-1}((4\gamma -p)\delta_i + \gamma \tilde r_i)
    = (4\gamma -p)\delta_i' +O(\chi_j \gbar_i^2)
    .
\end{equation}
By Taylor's theorem,
for small $t$,
\begin{equation}
    1-\gamma t = (1-t)^\gamma (1+O(t^2)).
\end{equation}
Therefore,
\begin{equation}
f_i = (1-e_i)^\gamma (1+O(\chi_i \gbar_i^2))(1+d_i)
    = \left(\frac{g_{i+1}}{g_i} \right)^\gamma (1+E_i),
\end{equation}
with
\begin{equation}
E_i = d_i+ O(\chi_i \gbar_i^2)= (4\gamma -p)\delta_i' +O(\chi_i \gbar_i^2)
    = O(\chi_i \gbar_i^2).
\end{equation}

Let
\begin{equation}
    \alpha_j = \prod_{i=j}^\infty (1+E_i).
\end{equation}
Since $\sum_i E_i$ is finite by \eqref{e:chisum}, the
infinite product converges, and moreover \eqref{e:chisum} implies that
$\alpha_j = 1+O(\sum_{i=j}^\infty E_i) = 1+O(\chi_j \gbar_j)$.
With \eqref{e:Pidef}, we obtain
\begin{equation}\label{e:Pi-jJ-def}
    \Pi_{j,J} = \alpha_j \left( \frac{g_{J+1}}{g_j} \right)^\gamma
    \alpha_J^{-1}
    = \alpha_j \left( \frac{g_{J+1}}{g_j} \right)^\gamma (1+O(\chi_J \gbar_J)),
\end{equation}
and the proof is complete.
\end{proof}

For $j \ge 0$, in view of Lemma~\ref{lem:Pi} it is natural to define
\begin{equation}
\label{e:gratdef}
    \grat_j = (g_j/g_0)^\gamma.
\end{equation}

\begin{lemma}
\label{lem:grat}
As $|a-b| \to \infty$,
if $j_{ab} < j_m$ then
\begin{align}
\label{e:gratasy}
    \grat_{j_{ab}} &=
    \left( \frac{1}{{\sf b} g_0 \log|a-b|} \right)^\gamma
     ( 1+\Ecal_{ab}^{(p)})
     .
\end{align}
\end{lemma}

\begin{proof}
Since
$L^{j_{ab}}=2|a-b| + O(1)$ by Lemma~\ref{lem:ggenasy},
\eqref{e:gratasy} follows from \eqref{e:gratdef} and Lemma~\ref{lem:ggenasy}.
The error estimate improves for $p=1$ because in this case $\gamma=0$;
in fact $\grat_j=1$ for all $j$ when $p=1$ so the error in fact vanishes.
\end{proof}

\subsection{Change of variables}
\label{sec:changevariables}

Theorem~\ref{thm:bulkflow} is stated in terms of the parameters $m^2,g_0$, rather than
the parameters $g,\nu$ that define the WSAW and $|\varphi|^4$ models.
The following proposition,
proved in \cite[Proposition~\ref{log-prop:changevariables}(ii)]{BBS-saw4-log}
for WSAW and \cite[\eqref{phi4-log-e:tildemap}]{BBS-phi4-log}
for $|\varphi|^4$,
relates these sets of parameters via the functions $z_0^c,\nu_0^c$ of
Theorem~\ref{thm:bulkflow} and \eqref{e:gg0}.  The critical value $\nu_c$
enters the analysis here, for the first time.

\begin{prop} \label{prop:changevariables}
Let $d = 4$, $n \ge 0$, and $\delta_1 > 0$ be small enough.
There exists a function $[0,\delta_1)^2 \to [0,\delta)^2$, that we denote by
$(g,\varepsilon) \mapsto (\tilde m^2(g,\varepsilon),\tilde g_0(g,\varepsilon))$,
such that \eqref{e:gg0} holds with $\nu = \nu_c(g)+\varepsilon$,
if $z_0 = z_0^c(\tilde m^2,\tilde g_0)$ and $\nu_0 = \nu_0^c(\tilde m^2,\tilde g_0)$.
The functions $\tilde m, \tilde g_0$ are right-continuous as $\varepsilon \downarrow 0$,
and satisfy $\tilde m^2(g,0) = 0$, and $\tilde m^2(g,\varepsilon) > 0$ if $\varepsilon > 0$.
\end{prop}

We also define the right-continuous functions (as $\varepsilon \downarrow 0$)
\begin{equation}
    \tilde z_0(g,\varepsilon) =
    z_0^c(\tilde m^2(g,\varepsilon),\tilde g_0(g,\varepsilon)),
    \quad\quad
    \tilde \nu_0(g,\varepsilon) =
    \nu_0^c(\tilde m^2(g,\varepsilon),\tilde g_0(g,\varepsilon)).
\end{equation}
Starting from $(g,\nu)$, Proposition~\ref{prop:changevariables}
provides $(\mgen^2,\ggen_0)$, and then Theorem~\ref{thm:bulkflow} provides
an initial condition $U_0=(\ggen_0,\tilde z_0,\tilde \nu_0)$ for which there
exists a global bulk flow of the renormalisation group map.
This needs to be supplemented by the observable flow, whose perturbative part
is given by Proposition~\ref{prop:pt}.
In the next section, we analyse the complete renormalisation group
flow, including the non-perturbative corrections for the observable flow.

\section{Complete renormalisation group flow}
\label{sec:pfmr1}

We now augment the bulk flow provided by Theorem~\ref{thm:bulkflow}
to obtain a complete renormalisation group flow, including observables.
In Section~\ref{sec:par}, we introduce the  domain for the complete
renormalisation group flow.  The main result concerning a single renormalisation
group step, Theorem~\ref{thm:step-mr-fv}, is stated in Section~\ref{sec:1rg} with proof
deferred to Section~\ref{sec:step}.  In Sections~\ref{sec:crg}--\ref{sec:indlim}, we apply
Theorem~\ref{thm:step-mr-fv} to conclude that
the renormalisation group step can be iterated indefinitely.
This is used in Sections~\ref{sec:pf-stars}--\ref{sec:pfmr} to prove our
main results Theorems~\ref{thm:mr-stars}--\ref{thm:mr-aa}.

\subsection{Parameters, norms and domains}
\label{sec:par}

We use several norms, and domains defined via these norms.
The norms extend those in
\cite[Section~\ref{step-sec:norms}]{BS-rg-step}
where only the two-point function was considered,
to handle the new observables present here.

The following sequences $\h_j$ and $\h_{\sigma,j}$ each
have distinct values in two distinct cases, which we
identify as either the $\h=\ell$ or $\h=\htilde$ cases.
This $\htilde$, which is called $h$ in \cite{BS-rg-IE,BS-rg-step,BBS-saw4-log,BBS-phi4-log},
is not related to and
should not be confused with the vector $h\in\R^n$ used to define the space $\Vcal_h$.
The two options for $\h_j,\h_{\sigma,j}$ are used to construct the $T_{\phi,j}(\h_j)$ norm
in \cite{BS-rg-step}.

For $\ell_0,k_0>0$ as in \cite[Section~\ref{step-sec:2np}]{BS-rg-step},
and for $j \ge 0$, let
\begin{equation}
\label{e:elldef}
    \h_j =
    \begin{cases}
    \ell_j = \ell_0 L^{-j} & (\h=\ell)
    \\
    \htilde_{j} = k_0 \ggen_j^{-1/4}L^{-j} & (\h=\htilde).
    \end{cases}
\end{equation}
With the notation $x \wedge y = \min\{x,y\}$ and $x_+=\max\{x,0\}$, we also define
\begin{equation}
\label{e:hsigdef}
\h_{\sigma,j} =
\grat_{j\wedge j_{ab}}^{-1} \ell_{j \wedge j_{ab}}^{-p}
2^{\nl(j - j_{ab})_+}
\times
\begin{cases}\ggen_j & (\h = \ell)
\\
\ggen_j^{\nl/4} & (\h = \htilde).
\end{cases}
\end{equation}
The occurrence of $\grat$ in \eqref{e:hsigdef} is a feature that is not visible
in \cite{BBS-saw4}, since if $p=1$ then $\gamma=0$ and $\grat=1$.  The definition here
is more subtle, as it anticipates the ultimate appearance of logarithmic corrections for $p\ge 2$.
It plays an important role in Lemma~\ref{lem:WK} below.

A $j$-dependent norm on $\Vcal_h$ is defined, using the weights from the $\h=\ell$ case of
\eqref{e:elldef}--\eqref{e:hsigdef}, by
\begin{equation}
\label{e:Vnormdef}
\begin{aligned}
\|V\|_{\Vcal_h} &=
\max\Big\{
|g|, L^{2j}|\nu_j|, |z_j|, |y_j|,
\ell_j^p\ell_{\sigma,j}(|\lambda_a|\vee|\lambda_b|),\;
\\
& \qquad\qquad\qquad
\ell_{\sigma,j}(|t_a|\vee|t_b), \ell_{\sigma,j}^{2} (|q_a|\vee|q_b|), L^{4j}|u|
\Big\},
\end{aligned}
\end{equation}
where $x \vee y = \max\{x,y\}$.
We extend the domain in $\R^3$ appearing in \eqref{e:domRGbulk}
by including now the coupling constants $\lambda_a,\lambda_b$ (for $n=0$ these are
permitted to be complex), and define
\begin{equation}
\label{e:DVdef}
    \DV_j = \{U\in \Vcal_h^{(0)} :
    g> C_{\DV}^{-1} \ggen  , \;  \|U\|_{\Vcal_h} < C_{\DV} \ggen \}.
\end{equation}
The $\Wcal$ norm is built from the $T_\phi = T_{\phi,j}(\h_j)$ norm used in \cite{BS-rg-step}.
Concerning observables, the $T_\phi$ norm obeys
(recall \eqref{e:Fdecomp})
\begin{equation}
\label{e:Fnormsum}
    \|F\|_{T_\phi}
    =
    \|F_\varnothing\|_{T_\phi}
    +
    \h_\sigma \left( \|F_{a}\|_{T_\phi} + \|F_{b}\|_{T_\phi}\right)
    +
    \h_{\sigma}^2 \|F_{ab}\|_{T_\phi}
    .
\end{equation}
This is the same as what is used in \cite{BS-rg-step},
except we now define $\h_\sigma$ by \eqref{e:hsigdef}.

We also need the following mass intervals.
Given $\delta>0$, let
\begin{equation}
\label{e:massint}
    \Iint_j = \begin{cases}
    [0,\delta) & (j<N)
    \\
    [\delta L^{-2(N-1)},\delta) & (j=N),
    \end{cases}
\end{equation}
and, for $\mgen^2 \in \Iint_j$, let
\begin{equation}
\label{e:Itilint}
    \Igen_j = \Igen_j(\mgen^2) =
    \begin{cases}
    [\frac 12 \mgen^2, 2 \mgen^2] \cap \Iint_j & (\mgen^2 \neq 0)
    \\
    [0,L^{-2(j-1)}] \cap \Iint_j & (\mgen^2 =0).
    \end{cases}
\end{equation}
Let $\sgen_j = (\mgen^2, \ggen_j)$, and let
$\chigen_j$ be given by \eqref{e:chicCovdef} with $j_m$ determined by
mass $\mgen^2$ rather than $m^2$.
We extend the bulk domain of \eqref{e:domRGbulk} to a
domain $\domRG_j^\varnothing(\sgen)
\subset \Vcal_h^{(0)} \times \Kcal_j$,
(with the same convention when the parameter $\sgen$ is omitted),
defined by
\begin{equation} \label{e:domRG}
  \domRG_j(\sgen_j)
  = \DV_j
    \times B_{\Wcal_j}(\DVa\chigen_j\ggen^3).
\end{equation}
The domain $\domRG$ also depends on the vector $h$, but we regard $h$ as
fixed and do not include it in the notation.

\subsection{A single renormalisation group step including observables}
\label{sec:1rg}

The following theorem is the centrepiece of the proof of
Theorems~\ref{thm:mr-stars}--\ref{thm:mr-aa}.
For observables, it provides the non-perturbative counterpart to
the perturbative statement of Proposition~\ref{prop:pt}.
Its proof, which requires adjustments to some arguments in \cite{BS-rg-IE,BS-rg-step},
is deferred to Section~\ref{sec:step}.

One of its consequences is that if $h \in \R^n$ is chosen to lie in one
of the eigenspaces $E^\pm$, then $V_j\in \Vcal_h$ for all $j$.
In other words, the complete renormalisation group flow keeps the vector $h \in \R^n$
fixed for all $j$.
Proposition~\ref{prop:pt} gives the perturbative version of this fact.
The norms in Theorem~\ref{thm:step-mr-fv} depend on the choice of the eigenspace
$E^\pm$ via the appearance of $\gamma_{n,p}^\pm$ in the definition of $\h_{\sigma,j}$
in \eqref{e:hsigdef}, and thus the estimates it provides also depend on the choice
of eigenspace for $h$.  This is the source of the two distinct powers $\gamma_{n,p}^\pm$
for the
logarithms appearing in Theorems~\ref{thm:mr-stars}--\ref{thm:mr-ab}.

\begin{theorem} \label{thm:step-mr-fv}
Let $d = 4$.  Let $n = 0$ and $p \ge 1$, or $n \ge 1$ and $p=1,2$.
Let $C_\DV$ and $L$ be sufficiently large.
Let $h=h^\pm \in E^\pm$, and choose
$\gamma = \gamma_{n,p}^\pm$ in \eqref{e:gratdef} and \eqref{e:hsigdef}.
There exist $M>0$ and $\delta >0$ such that
for $\ggen \in (0,\delta)$ and $\mgen^2 \in \Iint_+$, and with the domain
$\domRG$ defined using any $\DVa> M$, the maps
\begin{equation}
\label{e:RKplusmaps}
R_+:\domRG(\sgen) \times \Igen_+(\mgen^2) \to \Vcal_h^{(1)},
\quad
K_+:\domRG(\sgen) \times \Igen_+(\mgen^2) \to \Wcal_{+}(\sgen_+)
\end{equation}
define $(U,K)\mapsto (V_+,K_+)$ as in \eqref{e:RGmap} and obeying \eqref{e:IcircKdu},
and satisfy the estimates
\begin{equation}
\label{e:RKplus}
\|R_+\|_{\Vcal_h}
\le
M\chigen_+\ggen_+^{3}
, \qquad
\|K_+\|_{\Wcal_+}
\le
M\chigen_+ \ggen_+^{3}
.
\end{equation}
In addition, $R_+,K_+$ are jointly continuous in all arguments $m^{2}, V,K$.
\end{theorem}

In particular, the bounds of \eqref{e:RKplus} hold when
$\mgen^2=m^2 \in \Iint_j$, and in this case $\chigen_+=\chi_{j+1}$.  Also, $\ggen_j$
can be replaced in estimates by $\gbar_j$, due to \eqref{e:gbarggen}.  This leads
to the replacement of the right-hand sides of \eqref{e:RKplus} by $\chi_{j+1}\gbar_{j+1}$,
which itself can be replaced by $\chi_j \gbar_j$.  Thus there is no need for distinction
between these various options.

More can be said about $R_+$, for which we have the exact formulas
\eqref{e:Vplusdef}--\eqref{e:Rdef}.
It follows exactly as in \cite[Proposition~\ref{step-prop:Rcc}]{BS-rg-step} that
\begin{equation} \label{e:Rrange}
  \pi_a R_+ = \pi_b R_+ =0 \text{ for } j\geq j_{\pp\qq}, \quad
  \pi_{ab} R_+ = 0 \text{ for } j< j_{\pp\qq}.
\end{equation}
We write $R^{\lambda_x}_{+}$ for the coupling constant corresponding to $\lambda_x$
in $R_+$, and similarly for $R^{q_x}_+$.
We write
\begin{equation}
    \text{$f_j \prec g_j$ to mean that $f_j \le c g_j$.}
\end{equation}
By definition of the $\Vcal_h$ norm, and with \eqref{e:Rrange},
the first bound of
\eqref{e:RKplus} implies that, for $(U,K) \in \domRG_j(\sgen_j)$,
\begin{align}
\label{e:vlam}
   |R^{\lambda_x}_+|
&
\prec
\ell_j^{-\nl}\ell_{\sigma,j}^{-1}\chi_j\gbar_j^{3} \1_{j< j_{\pp \qq}}
\prec
    \grat_j \chi_{j}  \gbar_{j}^{2} \1_{j< j_{\pp \qq}}
    ,
\\
\label{e:vq}
    |R^{q_x}_+|
&
\prec
\ell_{\sigma,j}^{-2}\chi_j\gbar_j^{3}
\prec
    \grat_{j_{ab}}^2 L^{-2 \nl j_{ab}}2^{-2 \nl (j-j_{ab})} \chi_{j}\gbar_{j}\1_{j \ge  j_{\pp \qq}}
.
\end{align}
As discussed below the statement of Proposition~\ref{prop:pt},
the first scale for which $\qpt$ of \eqref{e:qpt} can be nonzero
is $q_{\pt,j_{ab}+1}$.
The indicator function in \eqref{e:vq} shows that this
remains true on a non-perturbative level.

With observables, according to \cite[\eqref{step-e:plusindep}]{BS-rg-step},
the statement for the bulk flow in \eqref{e:bulk} is accompanied by
the statement that, for $x=a$ or $x=b$,
\begin{equation}
\label{e:plusindep}
\begin{aligned}
    &\text{if $\pi_x V=0$ and $\pi_x K(X)=0$ for all $X \in \Pcal$ then}
    \\
    &\text{$\pi_x R_+=\pi_{ab} R_+=0$ and $\pi_x K_+(U)
    =\pi_{ab} K_+(U)=0$ for all $U \in \Pcal_+$.}
\end{aligned}
\end{equation}
Moreover, as discussed below \cite[\eqref{step-e:plusindep}]{BS-rg-step},
$\lambda_{a,+}$ is independent of each of $\lambda_b$,
$\pi_b K$, and $\pi_{ab}K$, and a similar statement holds for
$\lambda_{b,+}$.

\subsection{Complete renormalisation group flow}
\label{sec:crg}

Given $(m^2,g_0) \in [0,\delta)^2$,
the initial conditions for the global existence of the
bulk renormalisation group flow
are given by
\begin{equation} \label{e:V0c}
  \pi_\varnothing U_{0} = U_0^c = (g_0, z_0^c(m^2,g_0), \nu_0^c(m^2,g_0)),
\end{equation}
and this gives rise via Theorem~\ref{thm:bulkflow} to the sequence $U_j(m^2,g_0)$.
The next three propositions show that
the flow with observables, and with initial condition $U_0 \in \Vcal_h^{(0)}$ defined by
\begin{equation} \label{e:V0ob}
  \pi_\varnothing U_0 = U_0^c, \quad \lambda_{x,0}\in \{0,1\},
  \quad u_0=q_{x,0}=t_{x,0}=0,
  \qquad (x=a,b),
\end{equation}
exists for all $j \leq N$, and they state properties of that flow.
The flow of $\lambda_x,q_x,t_x$ does depend on the choice of
the vector $h=h^\pm\in E^\pm$, and on the choice of initial condition
$\lambda_{a,0},\lambda_{b,0}$,
but we do not add labels to indicate this dependence.
When $\lambda_{a,0}=0$ or
$\lambda_{b,0}=0$, we define the coalescence scale $j_{ab}$ to
be $j_{ab}=\infty$ rather than via \eqref{e:Phi-def-jc}, since in this case
at least one of the observable fields $\sigma_a,\sigma_b$ is absent
and its point $a$ or $b$ no longer plays a special role.

\begin{prop}
\label{prop:qlamN}
Let $d =4$.  Let $n = 0$ and $p \ge 1$, or $n \ge 1$ and $p=1,2$.
Let $h=h^\pm \in E^\pm$, and choose
$\gamma = \gamma_{n,p}^\pm$ in \eqref{e:gratdef} and \eqref{e:hsigdef}.
Let $(\zeta_0 = 0, U_0)$ be given by \eqref{e:V0ob},
and let $K_0=\1_\varnothing$.
Let $N \in \N$ and $(m^2,g_0) \in [\delta L^{-2(N-1)},\delta) \times (0,\delta)$.
There exist $(\zeta_j,U_j,K_j)$ such that
$(U_j,K_j) \in \domRG_j$ and
\eqref{e:IcircKdu} hold for $0 \le j \le N$.
This choice is such that $\pi_\varnothing U_j = U^c_j(m^2,g_0)$.
For $x = a$ or $x=b$,
if
$\lambda_{x,0}=0$ then $\lambda_{x,j}=0$ for all $0 \le j \le N$,
whereas if $\lambda_{x,0}=1$ then
\begin{equation}
\label{e:lam}
    \lx{j}
    =
\begin{cases}
    \Pi_{j-1}^\pm
    \left(
    1
    + \sum_{k=0}^{j-1}  \vlx{k}^\pm
    \right)
    & (j \le j_{ab})
    \\
    \lx{j_{ab}-1} & (j > j_{ab})    ,
\end{cases}
\end{equation}
where $\vlx{k}^\pm \in \R$
obey, for some $c>0$,
\begin{equation}
\label{e:vkbds}
    |\vlx{k}^\pm|\le c\chi_k \gbar_k^2 .
\end{equation}
Also, with
$M$ given by Theorem~\ref{thm:step-mr-fv}, for all $j$,
\begin{equation}
\label{e:K}
\|K_j\|_{\Wcal_j} \le M \chi_j \gbar_j^3.
\end{equation}
\end{prop}

\begin{proof}
We first observe that if
$\lambda_{x,0}=0$ then $\lambda_{x,j}=0$ for all $0 \le j \le N$,
due to \eqref{e:lampt} and \eqref{e:plusindep}.  We therefore assume that $\lambda_{x,1}=1$.

The proof is by induction on $j$.
We make the induction hypothesis:
\begin{align*}
{\rm IH}_j:  \hspace{3mm} &
\text{for all $k \le j$, $(U_k,K_k) \in \domRG_k$,
\eqref{e:lam} and \eqref{e:K} hold with $j$ replaced by $k$;}
\\ & \text{and \eqref{e:vkbds} holds for all $k<j$.}
\end{align*}
By direct verification, ${\rm IH}_0$ holds since $\Pi^\pm_{-1}=1$
and $\|K_0\|_{\Wcal_0}=0$ by definition.
We now assume ${\rm IH}_j$ and show that it implies ${\rm IH}_{j+1}$.

We apply Theorem~\ref{thm:step-mr-fv} with $\grat_j=\grat_j^\pm$ in \eqref{e:hsigdef},
where $\grat_j^\pm = (g_j/g_0)^{\gamma_{n,p}^\pm}$ as in \eqref{e:gratdef}.
By the induction hypothesis and \eqref{e:RKplus},
$K_{j+1} \in B_{\Wcal_j}(\DVa \chigen_{j+1} \ggen_{j+1})$ and satisfies \eqref{e:K}.
According to Theorem~\ref{thm:bulkflow},
the sequence $U^c$ satisfies the bounds required for $\pi_\varnothing U$ in the definition of $\domRG$ and obeys \eqref{e:bulk}--\eqref{e:RGmapbulk},
so that $\pi_\varnothing U_{j} = U_{j}^c$ for all $j$.
Therefore, to verify $(U_{j+1},K_{j+1}) \in \domRG_{j+1}$,
it suffices to show that $|\lambda_{x,j+1}| < C_{\DV} \ell_{j+1}^{-p}\ell_{\sigma,j+1}^{-1}$.

Let $x$ denote $a$ or $b$.
Since $R_{j+1}(U_j,K_j)\in \Vcal_h$ by \eqref{e:RKplusmaps},
the inclusion of the non-perturbative remainder $R^{\lambda_x}_{j+1}$ in
the flow of $\lambda_x$ gives, by \eqref{e:Rdef} and Proposition~\ref{prop:pt},
\begin{equation} \label{e:lamRGflow}
\lambda_{x, j+1} =
\begin{cases}
f_j^\pm \lx{j} + R^{\lambda_x}_{j+1}  & (j + 1 < j_{ab}) \\
\lx{j_{ab} - 1} & (j + 1 \geq j_{ab}).
\end{cases}
\end{equation}
The flow of $\lambda_x$ stops at the coalescence scale,
so we restrict to  $j + 1 < j_{ab}$.
In this case, we insert \eqref{e:lam} into \eqref{e:lamRGflow} to obtain
\begin{equation}
\label{e:lambda-remainder-defn}
\lx{j+1}
= \Pi_{j}^\pm \left(1 + \sum_{k=0}^{j-1} \vlx{k}^\pm \right) + R^{\lambda_x}_{j+1}
= \Pi_{j}^\pm \left(1 + \sum_{k=0}^{j} \vlx{k}^\pm \right),
\end{equation}
with $\vlx{j}^\pm = (\Pi_j^{\pm})^{-1} R^{\lambda_x}_{j+1}$.
By \eqref{e:vlam}, this gives
\begin{equation}
\label{e:vbarbd}
|\vlx{j}^{\pm}|
= |\Pi_j^{\pm}|^{-1}|R_{j+1}^\lambda|
\le c' (\Pi_j^{\pm})^{-1}\grat_j^\pm \chi_j\gbar_j^2, \text{ for some } c' > 0.
\end{equation}
Then \eqref{e:vkbds} follows since
$(\Pi_j^{\pm})^{-1}$ and $\grat_j^\pm$ are comparable by Lemma~\ref{lem:Pi} and \eqref{e:gratdef}.

To complete the induction,
it remains to prove that $|\lx{j+1}|\ell_{j+1}^{p}\ell_{\sigma,j+1}< C_{\DV} \ggen_{j+1} $.
By \eqref{e:elldef}--\eqref{e:hsigdef}, it suffices to prove that
\begin{equation}
\label{e:lamsmall}
    |\lx{j+1}|<
    C_{\DV}  \grat_{j+1}
    \quad\quad (j+1<j_{ab})
\end{equation}
(the case of $j+1 \ge j_{ab}$ then also follows).
This follows from \eqref{e:lam}--\eqref{e:vkbds},  the
estimate $\sum_{k=0}^j |\vlx{k}^\pm| = O(g_0)$ (by \eqref{e:chisum}), and Lemma~\ref{lem:Pi},
since we may assume that $C_\DV >2$.
\end{proof}

\begin{prop}
\label{prop:qN}
Let $d = 4$.  Let $n = 0$ and $p \ge 1$, or $n \ge 1$ and $p = 1, 2$.
Let $h = h^\pm \in E^\pm$, and choose
$\gamma = \gamma_{n,p}^\pm$ in \eqref{e:gratdef} and \eqref{e:hsigdef}.
Let $(\zeta_0 = 0, U_0)$ be given by \eqref{e:V0ob} with $\lambda_{a,0}=\lambda_{b,0}=1$,
and let $K_0=\1_\varnothing$.
Let $N \in \N$ and  $(m^2,g_0) \in [\delta L^{-2(N-1)},\delta) \times (0,\delta)$.
For $j \le N$ and $x = a,b$, the entry $q_{x,j}$ in $\zeta_j$ produced by
Proposition~\ref{prop:qlamN} obeys
\begin{equation}
\label{e:q}
\qx{j}
= \nl! \lambda_{a, j_{ab}} \lambda_{b, j_{ab}}  w^\nl_{j; ab} + \sum_{i = j_{ab}}^{j - 1} \vqx{i},
\end{equation}
with $|\vqx{i}| \prec \grat_{j_{ab}}^2 L^{-2 \nl j_{ab}}2^{-2p(j-j_{ab})}
\chi_{j}\gbar_{j}\1_{j \ge  j_{\pp \qq}}$.
\end{prop}

\begin{proof}
By \eqref{e:qpt} and \eqref{e:Rdef},
\begin{equation} \label{e:dq}
\delta \qx{j+1}
= \delta \qx{\pt} + \vqx{j}
= \nl! \lambda_{a, j_{ab}} \lambda_{b, j_{ab}}  \delta[w^\nl_{j; ab}] + \vqx{j}.
\end{equation}
For all $j < j_{ab}$, both $\delta q_\pt$ and $\vqx{j}$ vanish,
and summation of $\delta[w^\nl_{j; ab}]$ produces a telescoping sum, so that
\begin{equation} \label{e:qind}
\qx{j} = \sum_{i=j_{ab}}^{j-1} \delta \qx{i}
= \nl! \lambda_{a, j_{ab}} \lambda_{b, j_{ab}}  w^\nl_{j; ab} + \sum_{i = j_{ab}}^{j - 1} \vqx{i}.
\end{equation}
The desired bound on $\vqx{j}$ is provided by \eqref{e:vq}, and the proof is complete.
\end{proof}

\begin{prop}
\label{prop:conty}
For $x=a,b$ and $j \le N$,
each of $\lambda_{x,j}, q_{x,j}$ is independent of $N$, meaning that, e.g.,
the finite sequence $\{\lambda_{x,1},\ldots,\lambda_{x,N}\}$ takes the same values
on the torus $\Lambda_N$ as on a larger torus $\Lambda_{N'}$ with $N' > N$.
Also, each  of $\lambda_{x,j}, q_{x,j}$
is defined as a continuous function of $(m^2,g_0) \in [0,\delta)^2$,
$\lambda_{a,j}$ is independent of $\lambda_{b,0}$,
and $\lambda_{b,j}$ is independent of $\lambda_{a,0}$.
\end{prop}

\begin{proof}
The proof is identical to the proof of \cite[Proposition~\ref{saw4-prop:qlamN}(ii)]{BBS-saw4},
which provides the $(n,p)=(0,1)$ version of the statement and extends without modification
to our more general context here.
Note that by definition of $V_+$ in \eqref{e:Vplusdef},
$\lambda_{x,N}$ and $q_{x,N}$ are constructed from $K_{N-1}$ and $I_{N-1}$,
so they are independent of whether the torus has scale $N$, or a larger scale.
\end{proof}

\subsection{Inductive limit of observable flow}
\label{sec:indlim}

Proposition~\ref{prop:conty} permits the observable coupling constants
to be defined
as \emph{infinite} sequences, not stopped at $j=N$, via an inductive limit $N \to \infty$.
Indeed, since $\lx{j}$, $\qx{j}$
are independent of $N>j$,
we obtain sequences defined for any given $j \in \N_0$ by choosing any $N>j$.
For the case of initial condition $\lambda_{b,0}=0$, we write $\lambda_{a,j}^*$
for the inductive limit of the sequence $\lambda_{a,j}$, and define $\lambda_{b,j}^*$
similarly.
By \eqref{e:lam},
\begin{equation}\label{e:lam-star-flow}
\lambda_{x,j}^*
= \Pi_{j-1}^\pm \left(1 + \sum_{k=0}^{j-1}  \vlx{k}^\pm\right)
    \quad \text{for $x=a,b$ and $j \in \N_0$}.
\end{equation}
By Proposition~\ref{prop:conty} and by definition,
\begin{equation}
\label{e:lamlamstar}
    \lx{j} = \lambda_{x,j \wedge (j_{ab} - 1)}^*
\end{equation}
for any choice of initial
conditions $\lambda_{a,0},\lambda_{b,0}\in\{0,1\}$.
The following two lemmas analyse the sequences defined by inductive limits.
The constants $v_{x}^\pm$ in the first lemma ostensibly depend on $x$,
but they are shown below in Proposition~\ref{prop:infty}
to be independent of $x=a,b$.
The function $g_\infty(m^2)$ in its statement is given by Lemma~\ref{lem:ginfty}.

\begin{lemma}
\label{lem:lamstar}
Fix $h \in E^\pm$ and make the corresponding choice of $\gamma = \gamma^\pm$.
Let $(m^2, g_0) \in (0,\delta)^2$.  For $x = a,b$,
there exist constants $v_x^\pm = 1 + O(g_0)$, such that for all $j \in \N_0$,
\begin{equation}
\label{e:lam-star}
\lx{j}^*
= v_x^\pm \grat_j^\pm \big(1 + O(\chi_j \gbar_j)\big).
\end{equation}
The limit $\lambda_{x, \infty}^* (m^2)
= \lim_{j \to \infty} \lx{j}^*$ exists, and
\begin{equation}\label{e:lambda-star-infty}
    \lambda_{x, \infty}^* (m^2)
    = v_x^\pm \left( \frac{g_\infty(m^2)}{g_0} \right)^{\gamma^\pm}
    .
\end{equation}
On the other hand, if $\lambda_{a,0}=\lambda_{b,0}=1$ and if $j_{ab}<j_m$ then,
as $|a-b|\to\infty$ with $j_{ab} < j_m < N$,
\begin{equation}
\label{e:lambda-ab}
    \lambda_{x,j_{ab}} = v_x^\pm
    \left( \frac{1}{{\sf b} g_0 \log|a-b|} \right)^\gamma
     ( 1+\Ecal_{ab}^{(p)}).
\end{equation}
\end{lemma}

\begin{proof}
Let $r_x^\pm = \sum_{k=0}^\infty r_{x,k}^\pm$.  By \eqref{e:vkbds} and \eqref{e:chisum},
the sum converges and is $O(g_0)$, and in addition
$r_x^\pm- \sum_{k=0}^{j-1} r_{x,k}^\pm= O(\chi_j\gbar_j)$.  Let $u_x^\pm=1+r_x^\pm$.
Then, by \eqref{e:lam-star-flow},
\begin{equation}
\label{e:lamstar1}
    \lambda_{x,j}^*
    = \Pi_{j-1}^\pm \Big(1+r_x^\pm +   O(\chi_j\gbar_j) \Big)
    = u_x^\pm \Pi_{j-1}^\pm (1 + O(\chi_j\gbar_j)).
\end{equation}
With \eqref{e:gratdef} and Lemma~\ref{lem:Pi}, this implies that there
exists $\alpha_0=1+O(g_0)$ such that
\begin{equation}
\label{e:lamstar2}
    \lambda_{x,j}^*
    = \alpha_0 u_x^\pm \grat_{j}^\pm (1 + O(\chi_j\gbar_j)).
\end{equation}
This proves \eqref{e:lam-star} with $v_x^\pm=\alpha_0u_x^\pm$,
and then \eqref{e:lambda-star-infty} follows
immediately from the definition \eqref{e:gratdef}, Lemma~\ref{lem:ginfty}, and \eqref{e:chisum}.

The proof of \eqref{e:lambda-ab} follows similarly, using
\eqref{e:lam} and Lemma~\ref{lem:grat}, with Lemma~\ref{lem:ggenasy}
(and \eqref{e:ggbar}) to bound
the error term $ O(\chi_{j_{ab}}\gbar_{j_{ab}})$.
\end{proof}

For $m^2 \ge 0$, we write
\begin{equation}
\label{e:Cabmdef}
    G_{ab}(m^2) = (-\Delta_{\Z^4}+m^2)^{-1}_{ab},
    \quad\quad
    G_{ab}=G_{ab}(0).
\end{equation}

\begin{lemma}
\label{lem:qflow}
Fix $h \in E^\pm$, $|h| = 1$, and make the corresponding choice of $\gamma = \gamma^\pm$.
Let $(m^2,g_0) \in [0,\delta)^2$.
For both $x = a,b$,
the limit $\qx{\infty}(m^2,g_0) = \lim_{j \to \infty} \qx{j}(m^2,g_0)$ exists, is continuous,
and is given by
\begin{equation}
\label{e:qinfty}
\qx{\infty}(m^2)
= \nl!\, \lambda_{a, j_{ab}} \lambda_{b, j_{ab}} G_{ab}^{p}(m^2)
+ \sum_{i=j_{ab}}^{\infty} \vqx{i}.
\end{equation}
Under the restriction that $j_{ab}<j_m$ if $p>1$, and for all $a,b$ if $p=1$,
\begin{equation}\label{e:R-bound}
\sum_{i=j_{ab}}^{\infty} \vqx{i} =
\frac{G_{ab}^\nl}{(g_0\log|a-b|)^{2\gamma^\pm}} O(\chi_{j_{ab}}\gbar_{j_{ab}}),
\end{equation}
As  $|a-b|\to\infty$,
\begin{equation}
\label{e:qinf0}
    \qx{\infty}(0)
    =
    \nl! v_a^\pm v_b^\pm
     \left( \frac{1}{{\sf b} g_0 \log|a-b|} \right)^{2\gamma^\pm}
    G_{ab}^\nl
    ( 1 + \Ecal_{ab}^{(p)}).
\end{equation}
\end{lemma}

\begin{proof}
The formula \eqref{e:qinfty} follows from
Proposition~\ref{prop:qN} and the fact that $\lim_{j\to \infty}w_{j;ab}= G_{ab}(m^2)$
by definition.
The sum on the right-hand side of \eqref{e:qinfty}
converges uniformly in $(m^2,g_0)$ by
Proposition~\ref{prop:qN} and is therefore continuous by Proposition~\ref{prop:conty}.
By Proposition~\ref{prop:qN} and
the fact that $\chi_{j}\gbar_{j} \le O(\chi_{j_{ab}}\gbar_{j_{ab}})$
(see \cite[Lemma~\ref{flow-lem:elementary-recursion}(i)]{BBS-rg-flow}),
we obtain
\begin{equation}
\label{e:sum-Rq-bds}
\sum_{i=j_{ab}}^{\infty} |\vqx{i}|
\prec \grat_{j_{ab}}^2 L^{-2 \nl j_{ab}}
\sum_{i=j_{ab}}^{\infty}
2^{-2p(i-j_{ab})} \chi_{i}\gbar_{i}
= \grat_{j_{ab}}^2 L^{-2 \nl j_{ab}}O \left( \chi_{j_{ab}} \gbar_{j_{ab}}\right).
\end{equation}
Note that the exponential factor in the second sum is needed for convergence, which
is not otherwise guaranteed by \eqref{e:chisum}.
Then \eqref{e:R-bound}
follows from \eqref{e:gratasy} (using $j_{ab} < j_{m}$ when $p > 1$) and \eqref{e:jabbds}, together with the fact that
$G_{ab} = \frac{1}{4\pi^2} |a-b|^{-2}(1 + O(|a-b|^{-2}) )$ by \eqref{e:Casym}.
Finally,
since \eqref{e:R-bound} is true for all $a,b$ when $m = 0$,
\eqref{e:qinf0} follows from Lemma~\ref{lem:lamstar}.
\end{proof}

\section{Analysis of renormalisation group flow}
\label{sec:rg-flow-analysis}

We now complete the proofs of our main results
Theorems~\ref{thm:mr-stars}--\ref{thm:mr-aa}.
As a first step, in Section~\ref{sec:cfpf}, we rewrite the correlation
functions of interest in terms of derivatives of
the partition function $Z_N$ of \eqref{e:ZN}.
This rewrite permits us to prove, in Proposition~\ref{prop:infty}, that the
constants $v_a^\pm,v_b^\pm$ in \eqref{e:qinf0} are actually independent of $a,b$,
and hence the asymptotic behaviour as $|a-b|\to\infty$ is given by the logarithmic
and $G_{ab}$ factors in \eqref{e:qinf0}.
The derivatives of $Z_N$ naturally lead us to study derivatives of $W_N$ and $K_N$,
and estimates for these are given in
Section~\ref{sec:WKbds}.
In Section~\ref{sec:pf-stars}, we identify the correlation functions
of Theorem~\ref{thm:mr-stars} in terms of the limiting values $\lambda_{x,\infty}^*$
of Lemma~\ref{lem:lamstar} and prove Theorem~\ref{thm:mr-stars}.
Finally, in Section~\ref{sec:pfmr}, we prove Theorems~\ref{thm:mr-ab}--\ref{thm:mr-aa}.

\subsection{Correlation functions and the partition function}
\label{sec:cfpf}

Recall the definition of the partition function $Z_N$ from \eqref{e:ZN}.
For $n \ge 1$,
we write $Z_N(\varphi)$ to emphasise its dependence on the field $\varphi$.
For $n=0$, we write $Z_N^0(\phi,\phib)$ for the degree-zero part of the form $Z_N$.
For $n \ge 1$, we define the (un-normalised) \emph{pressure}
\begin{equation}
\label{e:pressuredef}
    P_N(\varphi) = \log Z_N(\varphi).
\end{equation}
We use the notation used in Section~\ref{sec:ef}
for derivatives with respect to external and observable fields.
We also write $D_{\phib} Z_N^0$ for the directional derivative of $Z_N$ with
respect to $\phib$ in the direction of the constant field 1,  evaluated at
$\phi=\phib=0$.

\begin{lemma}\label{lem:corr-to-pressure}
Fix $m^2>0$ and $z_0>-1$.
For $n \ge 1$ and $p=1,2$,
\begin{align}
\label{e:DaPN}
\left\langle \varphi_a^\nl \cdot h \right\rangle_{g,\nu,N}
&= \left(1 + z_0\right)^{\nl/2}
D_{\sigma_a} P_N(0),
\\
\label{e:DaDbPN}
\inprod{ \varphi_a^p \cdot h\,}{\varphi_b^p \cdot h}_{g,\nu,N}
&= (1+z_0)^p D_{\sigma_a, \sigma_b}^2 P_N(0),
\\
\label{e:DaDphiPN}
\inprod{ (\varphi, H)^\nl }{ \varphi_a^\nl \cdot h }_{g,\nu, N}
&= \frac{(1 + z_0)^\nl}{m^{2\nl}}
D_\varphi^\nl(H) D_{\sigma_a} P_N
.
\end{align}
For $n=0$ and $p \ge 1$,
\begin{align}
\label{e:DaDbPNn0}
\wm^{(\nl)}_{ab,N}(g,\nu)
&= (1+z_0)^p
D_{\sigma_a\sigma_b}^2
Z_N^0(0),
\\
\label{e:DaDphiZN}
\starpol^{(p)}_N (g,\nu)
&= \frac{(1+z_0)^p}{m^{2\nl}}
D_{\bar \phi}^\nl D_{\sigma_a} Z_N^0 .
\end{align}
\end{lemma}

\begin{proof}
We first prove \eqref{e:DaPN}--\eqref{e:DaDphiPN}.
The identity \eqref{e:DaDbPN} is the same as \eqref{e:corrdiff}, and \eqref{e:DaPN}
also follows similarly from explicit differentiation.
For \eqref{e:DaDphiPN},
we let $\Sigma (J) = \Ex_C e^{- V_{0} (\Lambda) + (\varphi, J)}$, and
\eqref{e:stardiff} becomes
\begin{equation}
\inprod{ (\varphi, H)^p }{ \varphi_a^p \cdot h }_{g,\nu,N}
= (1 + z_0)^p
D_J^p(H)
D_{\sigma_a}
\log \Sigma (J).
\end{equation}
As in \cite[\eqref{phi4-log-e:LTC}]{BBS-phi4-log},
we obtain
\begin{equation} \label{e:SigmaaZN0azzz}
  \Sigma(J) = \Ex_C\left(e^{-V_{0}(\Lambda)+(J,\varphi)}\right)
  = e^{\frac{1}{2}(J,C J)} Z_N(CJ)
\end{equation}
by the translation $\varphi \mapsto \varphi + CJ$ to complete the square in
the middle member of \eqref{e:SigmaaZN0azzz}.
This gives $\log \Sigma(J) = \frac{1}{2} (J, CJ) + \log Z_N(CJ)$.
Since $(J,CJ)$ is independent of the observable field $\sigma_a$,
\begin{equation}
D_J^p(H) D_{\sigma_a}  \log \Sigma(J)
= D_J^p(H) D_{\sigma_a} \log Z_N(CJ)
.
\end{equation}
Since $H$ is a constant field, $CH = m^{-2}H$.  The chain rule then gives
\begin{equation}
D_J^p(H)  D_{\sigma_a} \log \Sigma(J)
= m^{-2p} D_\varphi^p(H) D_{\sigma_a} \log Z_N ,
\end{equation}
and the proof of \eqref{e:DaDphiPN} is complete.

The case  $n=0$ is similar, except that the logarithm is superfluous
due to the self-normalisation property of the Gaussian super-expectation.
The identity \eqref{e:DaDbPNn0} is a restatement of \eqref{e:intobs}.
For \eqref{e:DaDphiZN}, we define
$\Sigma (J,\bar J) = \Ex_C e^{-V_{0}(\Lambda) + (J,\phib) + (\bar J,\phi)}$,
and rewrite \eqref{e:star-to-integral} as
\begin{equation}
\starpol^{(p)}_N (g,\nu)
= (1+z_0)^p
D_{\bar J}^\nl D_{\sigma_a} \Sigma (J,\bar J).
\end{equation}
Now completion of the square (as in \cite[\eqref{log-e:GamZ0}]{BBS-saw4-log}) gives
\begin{equation} \label{e:SigmaaZN0a}
\Sigma(J,\bar J)
= e^{(J,C\bar J)} Z_N^0(CJ, C\bar J),
\end{equation}
and \eqref{e:DaDphiZN} again follows by differentiation and the chain rule.
\end{proof}

\subsection{Non-perturbative estimates}
\label{sec:WKbds}

The following lemma allows us to control the non-perturbative quantities in
the proofs of our main theorems.
We write $D^k_\sigma$ to mean
no derivative for $k=0$, the derivative with respect to $\sigma_a$
for $k=1$, and derivatives with respect to $\sigma_a$ and $\sigma_b$ for $k=2$.

\begin{lemma}
\label{lem:WK}
Let $h=h^\pm \in E^\pm$, and let $\gamma=\gamma^\pm$.
For $n=0$, $p \ge 1$,
the following estimates (all at zero field)
hold uniformly in $g\in(0,\delta)$ and
$m^2 \in [\delta L^{-2(N-1)},\delta)$.
For initial conditions $\lambda_{a,0}=\lambda_{b,0}=1$ and for $l=0,1,2$,
\begin{align}
\label{e:Kg1}
    |D_{\sigma}^l K_{N}^0(\Lambda)|
    &
    \prec
    \chi_N  \gbar_N^{3-l}
        \left(\frac{1}{2^{p(N-j_{ab})_+}}
    \frac{1}{|a-b|^p} \frac{1}{(g \log |a-b|)^{\gamma}}\right)^l
    .
\end{align}
For initial conditions $\lambda_{a,0}=1$, $\lambda_{b,0}=0$, and for $k=0,1,\ldots,p$
and $l=0,1$,
\begin{align}
\label{e:DKbd}
  |D_{\phib}^k D_{\sigma}^l K_{N}^{0}(\Lambda)|
  &\prec \chi_N \gbar_N^{3-l}
  \frac{L^{N (k-l\nl)}}{(g_0\log m^{-2})^{l\gamma}}
  ,
\\
\label{e:WNbdz}
  |D_{\phib}^k D_{\sigma}^l W_{N}^0(\Lambda)|
  &\prec \chi_N \bar g_N^{2-l}
  \frac{L^{N (k-l\nl)}}{(g_0\log m^{-2})^{l\gamma}}
  .
\end{align}
The bounds \eqref{e:Kg1}--\eqref{e:WNbdz} also
hold for $n \ge 1$ and $p=1,2$, after
changing $K_N^0$ to $K_N$ and making
directional
derivatives with respect to $\varphi$ in the direction of a constant field $1$.
\end{lemma}

\begin{proof}
We give the proof for $n=0$.  The proof for $n \ge 1$ involves only slight
changes in notation.
By \eqref{e:gbarggen}, $\ggen_j$ and $\gbar_j$ are interchangeable
in estimates.

Recall the definitions of the $T_{0,j}(\ell_j)$ and $\Phi_j(\ell_j)$ norms
from \cite[Section~\ref{log-sec:flow-norms}]{BBS-saw4-log}.
In \eqref{e:Fnormsum}, in the $T_{0,j}(\ell_j)$ norm
each occurrence of $\sigma$ or $\sigmab$ produces the weight
\begin{equation}
\label{e:ellsigdef}
    \ell_{\sigma ,j}
=
    \ell_0^{-p} \grat_{j\wedge j_{\pp\qq}}^{-1} 2^{p(j-j_{ab})_+}L^{p(j\wedge j_{\pp\qq})} \ggen_{j}
\end{equation}
defined in \eqref{e:hsigdef}.
We apply \cite[\eqref{step-e:KWbd}]{BS-rg-step} which
uses this fact, together with \eqref{e:K}, to see that for $l=0,1,2$ the bound
\begin{equation}
\begin{aligned}
\label{e:KWbd}
    |D^l_{\sigma}K_{N}^0(\Lambda;0,0)|
    &\le
    \ell_{\sigma , N}^{-l}
    \|K_N(\Lambda)\|_{T_{0,N}(\ell_N)}
    \le
    \ell_{\sigma , N}^{-l}
    \|K_N\|_{\Wcal_N}
    \\&
    \prec
    \grat_{j_{\pp\qq}}^{l}
    2^{-lp(N-j_{ab})_+} L^{-lpj_{\pp\qq}} \chi_N \gbar_{N}^{3-l}
\end{aligned}
\end{equation}
holds uniformly in $m^2 \in [\delta L^{-2(N-1)},\delta)$.  By \eqref{e:jabbds},
$L^{-j_{ab}} \prec |a-b|^{-1}$.  The logarithmic behaviour of  $\grat_{j_{\pp\qq}}$
is given by \eqref{e:gratasy}, and \eqref{e:Kg1} is proved.

For any $k \le p_\Ncal$, $l=0,1$, $F\in \Ncal$, and test functions $J_i:\Lambda \to \C$
($i=1,\ldots,k$),
it follows from the
definition of the $T_{0,N}(\ell_N)$ norm that
\begin{equation}
\label{e:DphibF}
  |D_{\phib}^k D_{\sigma}^l F^0(0, 0; J_1,\ldots,J_p)|
  \leq  \ell_{\sigma,N}^{-l}
  \|F\|_{T_{0,N}(\ell_N)} \|J_1\|_{\Phi_N(\ell_N)} \cdots \|J_k\|_{\Phi_N(\ell_N)}.
\end{equation}
By definition, $\|1\|_{\Phi_N(\ell_N)}=\ell_N^{-1}$ (as in
\cite[\eqref{log-e:1norm}]{BBS-saw4-log}).  As in \eqref{e:KWbd}, this
gives
\begin{equation}
\label{e:DphibK}
\begin{aligned}
  |D_{\phib}^k D_{\sigma}^l K_{N}^{0}(\Lambda;0,0;1,\ldots,1)|
  &\leq
  \ell_{\sigma,N}^{-l} \|K_N(\Lambda)\|_{T_{0,N}(\ell_N)} \|1\|_{\Phi_N(\ell_N)}^k
  \le
  \ell_{\sigma,N}^{-l}\ell_N^{-k} \|K_N\|_{\Wcal_N}
  .
\end{aligned}
\end{equation}
With the initial conditions assumed for \eqref{e:DKbd}--\eqref{e:WNbdz},
we have $j_{ab}=\infty$.
By \eqref{e:ellsigdef}, \eqref{e:elldef}, \eqref{e:gratdef}, \eqref{e:ginfty}, and \eqref{e:newbubble},
\begin{equation}
    \ell_{\sigma,N}^{-l} \ell_N^{-k}
    =\ggen_N^{-l} \grat_{N}^l \left(\ell_0 L^{-N}\right)^{l\nl - k}
    \prec
    \gbar_N^{-l} (g_0\log m^{-2})^{-l\gamma} L^{N (k-l\nl )}.
\end{equation}
With \eqref{e:K}, this proves \eqref{e:DKbd}.
Finally, for the bound on $W_N$, we recall from
\cite[Proposition~\ref{IE-prop:Wnorms}]{BS-rg-IE} that
\begin{equation} \label{e:Wbilinbd}
  \|W_{N}(\Lambda)\|_{T_{0,N}} \prec \chi_N \gbar_N^2,
\end{equation}
and \eqref{e:WNbdz} then follows exactly as in \eqref{e:DphibK}.
\end{proof}

\subsection{Proof of Theorem~\ref{thm:mr-stars}}
\label{sec:pf-stars}
Let $n \ge 0$.
For small $g,\varepsilon > 0$, set $\nu = \nu_c + \varepsilon$, and
let $(m^2,g_0,\nu_0,z_0) = (\tilde m^2,\tilde g_0, \tilde \nu_0, \tilde z_0)$
be the functions of $(g,\varepsilon)$ given by Proposition~\ref{prop:changevariables}.
This choice is consistent with the initial condition \eqref{e:V0ob}
that guarantees the existence of the global flow with observables.
By
\cite[\eqref{log-e:chi-mtil0}]{BBS-saw4-log}
for $n=0$,
and \cite[\eqref{phi4-log-e:chim2}]{BBS-phi4-log} for $n \ge 1$, it provides the
\emph{identity}
\begin{equation}\label{e:chim2}
    \chi =
    \chi(g,\nu) = \frac{1+\tilde z_0}{\tilde m^2} = \frac{1 + z_0}{ m^2}.
\end{equation}

\begin{prop}
\label{prop:infty}
Let $h=h^\pm\in E^\pm$ and $\gamma=\gamma^\pm$.
Let $n=0$ and $p \ge 1$, or  $n\ge 1$ and $p = 1, 2$.
For $n \ge 1$, let $H$ be a constant field with value $H_0$.
For $(g, \varepsilon)\in (0,\delta)^2$, and for $x = a,b$,
\begin{align}
\label{e:lamainftyn0}
\frac{1}{\chi^\nl} \starpol^{(p)} (g,\nu)
&= \nl !\, \lambda_{x,\infty}^*
& (n=0)
\\
\label{e:lamainftyn}
\frac{1}{\chi^\nl} \inprod{ (\varphi, H)^\nl }{ \varphi_a^\nl \cdot h }_{g,\nu}
& =
\nl !
(H_0^\nl \cdot h)
\lambda_{x,\infty}^*
& (n \geq 1).
\end{align}
In particular, the infinite volume limit on the left-hand side of \eqref{e:lamainftyn} exists.
\end{prop}

\begin{proof}
We use initial conditions $\lambda_{a,0}=1$ and $\lambda_{b,0}=0$.
We start with \eqref{e:ZNfinal}, namely
\begin{equation}\label{e:ZNfinal(phi)}
Z_N
= e^{\zeta_N}\left(I_N  + K_N \right),
\end{equation}
where $Z_N,I_N,K_N$ depend on $(\phi,\phib)$ for $n=0$, and on $\varphi$ for $n \ge 1$.

We first prove \eqref{e:lamainftyn0}.
In this case, $\zeta_N=\frac 12 (q_{a,N}+q_{b,N})\sigma_a\sigma_b$.
By \eqref{e:Idef} (since $\Lambda$ is a single
block at scale $N$),
\begin{equation}
  Z_N^0 = e^{\zeta_N}(I_N^0 + K_N^0)
  = e^{\zeta_N}(e^{-V_N^0} (1+W_N^0) + K_N^0).
\end{equation}
Since $\pi_\varnothing e^{-V_N^0} = e^{-U_N^0}$
and $D_{\sigma_a} e^{-V_N^0} = \lambda_{a,N}^* \bar\phi_a^\nl$,
\begin{equation}
    D_{\sigma_a} Z_{N}^0
    = \lambda_{a,N}^*\bar\phi_a^\nl   e^{-U_N^0} \left(1 + \pi_\varnothing W_N^{0}\right)
    + e^{-U_N^0} D_{\sigma_a} W_{N}^{0} + D_{\sigma_a} K_{N}^{0}.
\end{equation}
We differentiate with respect to $\phib$ in direction $1$,
$\nl$ times, and set $\phi = \phib = 0$ to obtain
\begin{equation}
D_{\bar \phi}^\nl D_{\sigma_a} Z_N^{0}
= \nl! \lambda_{a,N}^*
+ D_{\phib}^\nl D_{\sigma_a} W_{N}^{0}
+ D_{\phib}^\nl D_{\sigma_a} K_{N}^{0} ,
\end{equation}
where we used the facts that $e^{-U_N^0} = 1$ and
$W_N^{0,\varnothing} = 0$ when $\phi=\phib=0$.
By Lemma~\ref{lem:WK} and \eqref{e:DaDphiZN},
\begin{equation}
\label{e:TZ}
\left(\frac{1+z_0}{m^{2}}\right)^{-\nl}
\starpol^{(p)}_N (g,\nu)
= D_{\bar \phi}^\nl D_{\sigma_a} Z_N^{0}
= \nl! \lambda_{a,N}^*
+ O \left(\frac{\chi_N \bar g_N}{(g\log m^{-2})^{\gamma^+}}\right)
\end{equation}
(since $\gamma^-=\gamma^+$ for $n=0$).
We let $N\to\infty$ in \eqref{e:TZ}, using the facts that
$\chi_N \bar g_N \to 0$ by \eqref{e:chisum},
$\starpol_N^{(p)} \to \starpol^{(p)}$ by Proposition~\ref{prop:infinite-volume-limit},
and $\lambda_{a,N}^* \to \lambda_{a,\infty}^*$ by Lemma~\ref{lem:lamstar}.
We also use \eqref{e:chim2} to identify the factor $\chi^{-\nl}$ on the left-hand side.
This proves \eqref{e:lamainftyn0}.

To prove \eqref{e:lamainftyn},
we apply both $D_\varphi^\nl D_{\sigma_a}$ to the logarithm of
the right-hand side of \eqref{e:ZNfinal(phi)}.  Since $\zeta_N$ is independent of $\varphi$,
in terms of the pressure \eqref{e:pressuredef} this gives
\begin{equation}
D_\varphi^\nl D_{\sigma_a} P_N = D_\varphi^\nl D_{\sigma_a} \log \left(I_N + K_N\right).
\end{equation}
By definition, $I_N = e^{-U_N} (1 + W_N)$, and we write
$I_N + K_N = e^{-U_N} (1 + W_N + e^{U_N} K_N)$.
Since $D_\varphi^\nl D_{\sigma_a}(-U_N) = \nl! \, \lambda_{a,N}^* (H_0^\nl \cdot h)$,
\begin{equation}
D_\varphi^\nl D_{\sigma_a} P_N
= \nl! \, \lambda_{a,N}^* (H_0^\nl \cdot h)
+ D_\varphi^\nl D_{\sigma_a} \log (1 + W_N + e^{U_N} K_N).
\end{equation}
It is now an exercise in calculus to apply Lemma~\ref{lem:WK}
(and the fact that $U_N$ lies in the domain $\DV_N$ of \eqref{e:DVdef}) to conclude that
\begin{equation}\label{e:DphiDaLog}
D_\varphi^\nl D_{\sigma_a} \log (1 + W_N + e^{U_N} K_N)
= O \left(\frac{\chi_N \bar g_N}{(g\log m^{-2})^{\gamma}}\right).
\end{equation}
Then, by \eqref{e:DaDphiPN},
\begin{equation}
\left(\frac{1 + z_0}{m^{2}}\right)^{-\nl}
\inprod{ (\varphi, H)^\nl }{ \varphi_a^\nl \cdot h }_{g,\nu, N}
= \nl! \, \lambda_{a,N}^* (H_0^\nl \cdot h)
+ O \left(\frac{\chi_N \bar g_N}{(g\log m^{-2})^{\gamma}}\right).
\end{equation}
Again we use $\chi_N \bar g_N \to 0$ and
$\lambda_{a,N}^* \to \lambda_{a,\infty}^*$ to see that
the limit as $N \to \infty$ of the right-hand side exists and equals
the right-hand side of \eqref{e:lamainftyn}.
Therefore, the limit of the left-hand side must also exist and so
\begin{equation}
\inprod{ (\varphi, H)^\nl }{ \varphi_a^\nl \cdot h }_{g,\nu}
= \lim_{N\to \infty} \inprod{ (\varphi, H)^\nl }{ \varphi_a^\nl \cdot h }_{g,\nu, N}
\end{equation}
exists in the sense of \eqref{e:cfsumlim}.
We complete the proof of \eqref{e:lamainftyn} by appealing to \eqref{e:chim2}.
\end{proof}

\begin{cor}\label{cor:vx}
The constant $v_x^\pm$ in \eqref{e:lam-star} is, in fact, independent of $x$,
and moreover, for $\nl = 1$ and for all $n \geq 0$, $\lambda_{x,\infty}^* = 1$.
\end{cor}

\begin{proof}
Since the left-hand sides of \eqref{e:lamainftyn0} and \eqref{e:lamainftyn} are independent of $x$,
$\lambda_{x,\infty}^*$ and $v_x^\pm$ must also be independent of $x$.

Let $\nl = 1$.
By definition, $\starpol^{(1)}$ is the susceptibility $\chi$,
so \eqref{e:lamainftyn0} yields $\lambda_{x,\infty}^* = 1$
(as was proved in \cite[Lemma~\ref{saw4-lem:lamlim}]{BBS-saw4}).
For $n \geq 1$, since $(\varphi, H) = \sum_x \varphi_x \cdot H_0$,
we take $H_0 = \hat e_1$, the first standard basis vector.
Using $\varphi_x^i \mapsto - \varphi_x^i$ symmetry and \eqref{e:susceptdef},
\begin{equation}
\inprod{ (\varphi, H) }{ \varphi_a \cdot h }_{g,\nu}
= \lim_{N\to\infty}
\sum_{x\in\Lambda_N} \inprod{ \varphi_x^1 }{ \varphi_a \cdot h }_{g,\nu,N}
= \lim_{N\to\infty}
\sum_{x\in\Lambda_N} h^1  \left\langle \varphi_x^1 \varphi_a^1 \right\rangle_{g,\nu,N}
= (H_0 \cdot h) \; \chi.
\end{equation}
Thus \eqref{e:lamainftyn} simplifies to $\lambda_{x,\infty}^* = 1$.
\end{proof}

\begin{proof}[Proof of Theorem~\ref{thm:mr-stars}]
(i) By \eqref{e:chin} and \eqref{e:chi0}, \eqref{e:chim2} gives
\begin{equation}
  m^2
  \sim
  (1+z_0) A_g^{-1} \varepsilon (\log \varepsilon^{-1})^{-\gamma^+}
  \quad \text{as $\varepsilon \downarrow 0$},
\end{equation}
and hence $\log m^{-2} \sim \log \varepsilon^{-1}$.
Using \eqref{e:lambda-star-infty} and Lemma~\ref{lem:ginfty}, and since $g_0 = g \big(1 + O(g)\big)$,
\begin{equation}\label{e:lam-indep-infty}
\lambda_{\infty}^* (m^2)
\sim \frac{\tilde v^\pm}{\left(\log \varepsilon^{-1}\right)^{\gamma^\pm}}, \qquad
\tilde v^\pm = \frac{v^\pm}{\left(g_0 {\sf b}\right)^{\gamma ^\pm}}
=
\frac{1}{\left(g {\sf b}\right)^{\gamma ^\pm}}
 \big(1 + O(g)\big),
\end{equation}
where,
in view of Corollary~\ref{cor:vx},
we have dropped the labels $x$ on $\lambda_\infty^*$ and $v^\pm$.
For $n = 0$, we use \eqref{e:lamainftyn0}
(recall that $\gamma^+ = \gamma^-$ for $n = 0$) to obtain
\begin{equation}
\frac{1}{\chi^\nl} \starpol^{(p)} (g,\nu)
= \nl! \lambda_{\infty}^*(m^2)
\sim \nl! \frac{\tilde v^+}{\left(\log \varepsilon^{-1}\right)^{\gamma^+}}.
\end{equation}
This proves \eqref{e:starasy}.

\smallskip \noindent (ii)
Let
$n \ge 1$ and $\nl = 2$.  Now we use
\eqref{e:lamainftyn} to obtain
\begin{equation}
\label{e:H01}
\frac{1}{\chi^2}
\inprod{ (\varphi, H)^2 }{ \varphi_a^2 \cdot h }_{g,\nu}
=
2 !
(H_0^2 \cdot h)
\lambda_{x,\infty}^*
\sim  (H_0^2 \cdot h)
\frac{2 \tilde v^\pm}{\left(\log \varepsilon^{-1}\right)^{\gamma^\pm}},
\end{equation}
where $H_0$ is the constant value of the field $H$, and $H_0^2\in\R^n$ is the vector whose
components are the squares of the components of $H_0$.
For the choice $h = n^{-1/2}e^+ \in E^+$,
we have $\varphi_a^2 \cdot h = n^{-1/2} |\varphi_a|^2$
and $H_0^2 \cdot h = n^{-1/2} |H_0|^2$.
We cancel the $n^{-1/2}$ factor on both sides of \eqref{e:H01} and obtain
\begin{equation}
\frac{1}{\chi^2}
\inprod{ (\varphi, H)^2 }{ \varphi_a^2 \cdot h }_{g,\nu}
\sim |H_0|^2  \frac{2 \tilde v^+}{\left(\log \varepsilon^{-1}\right)^{\gamma^+}} .
\end{equation}
We take $H_0 = \hat e_k$ to be the $k^{\rm th}$ standard basis vector, and then sum over $k$,
to obtain
\begin{equation}
\label{e:Hek}
\frac{1}{\chi^2} \sum_{x,y} \inprod{ \varphi_x \cdot \varphi_y }{ |\varphi_a|^2 }_{g,\nu}
= \frac{1}{\chi^2} \sum_{x,y} \sum_{k = 1}^n \inprod{ \varphi_x^{k} \varphi_y^{k} }{ |\varphi_a|^2 }_{g,\nu}
\sim  \frac{2n \tilde v^+}{\left(\log \varepsilon^{-1}\right)^{\gamma^+}} .
\end{equation}
This proves \eqref{e:corrdot}.
Suppose now that $n \geq 2$.
By symmetry, \eqref{e:Hek} gives
\begin{equation}
\label{e:Hh1}
\frac{1}{\chi^2}
\sum_{x, y}
\inprod{ \varphi_{x}^1 \varphi_{y}^1 }{ (\varphi_a^1)^2 }_{g,\nu}
+ (n - 1) \frac{1}{\chi^2} \sum_{x,y} \inprod{ \varphi_{x}^1 \varphi_{y}^1 }{ (\varphi_a^2)^2 }_{g,\nu}
\sim \frac{2 \tilde v^+}{\left(\log \varepsilon^{-1}\right)^{\gamma^+}}.
\end{equation}
In \eqref{e:H01} we take $H_0=\hat e_1$ and
$h = 2^{-1/2}(1,-1,0,\ldots,0) \in E^-$. Since $h \in E^-$,
now $\gamma = \gamma^-$. We obtain
\begin{equation}
\label{e:Hh2}
\frac{1}{\chi^2}
\sum_{x, y}
\inprod{ \varphi_{x}^1 \varphi_{y}^1 }{ (\varphi_a^1)^2 }_{g,\nu}
- \frac{1}{\chi^2} \sum_{x,y} \inprod{ \varphi_{x}^1 \varphi_{y}^1 }{ (\varphi_a^2)^2 }_{g,\nu}
\sim \frac{2 \tilde v^-}{\left(\log \varepsilon^{-1}\right)^{\gamma^-}}.
\end{equation}
Since $\gamma^-<\gamma^+$, the combination of \eqref{e:Hh1}--\eqref{e:Hh2} gives
\begin{align}
\frac{1}{\chi^2}
\sum_{x, y}
\inprod{ \varphi_{x}^1 \varphi_{y}^1 }{ (\varphi_a^1)^2 }_{g,\nu}
&\sim \frac{n-1}{n}
\frac{2 \tilde v^-}{\left(\log \varepsilon^{-1}\right)^{\gamma^-}}
,
\\
\frac{1}{\chi^2}
\sum_{x, y}
\inprod{ \varphi_{x}^1 \varphi_{y}^1 }{ (\varphi_a^2)^2 }_{g,\nu}
&\sim -\frac{1}{n}
 \frac{2\tilde v^-}{\left(\log \varepsilon^{-1}\right)^{\gamma^-}}
 ,
\end{align}
which proves \eqref{e:corr1111}--\eqref{e:corr1122}.

\smallskip \noindent (iii)
The asymptotic formula \eqref{e:mrcasy1} follows from \eqref{e:lam-indep-infty},
and  the proof is complete.
\end{proof}

\subsection{Proof of Theorems~\ref{thm:mr-ab}--\ref{thm:mr-aa}}
\label{sec:pfmr}

\begin{proof}[Proof of Theorem~\ref{thm:mr-ab}] (i-ii)
We denote the parameters $(n,p)$ by superscripts.
The infinite volume limit of the watermelon network can be computed as
a limit using Proposition~\ref{prop:infinite-volume-limit}, and for $n \ge 1$
we have defined the critical infinite volume limits of correlation
functions as in \eqref{e:cflim}.
For $n \geq 1$, let
$(\corrfcn_c)_{ij} = \langle (\varphi_a^i)^p ; (\varphi_b^j)^p \rangle_{\nu_c(n)}$
denote the matrix of critical correlation functions.
According to \eqref{e:DaDbPN} and \eqref{e:DaDbPNn0}
(we drop the notation for evaluation at zero as all fields are evaluated at zero here),
\begin{equation}\label{e:pressure-lims}
\big(1 + \tilde z_0(g,0) \big)^p
\lim_{\varepsilon \downarrow 0} \lim_{N\to\infty}
D^2_{\sigma_a\sigma_b} P_N^{(n,p)}
= \begin{cases}
\wm^{(\nl)}_{ab}(\nu_c(0)) & (n = 0, \,\nl \geq 1), \\
h \cdot \corrfcn_c h & (n \geq 1, \,\nl = 1,2).
\end{cases}
\end{equation}
For the prefactor on the left-hand side, we have used
Proposition~\ref{prop:changevariables} for existence of the limit
$\tilde z_0(g,\varepsilon) \to \tilde z_0(g,0)$ as $\varepsilon \downarrow 0$.
It also follows from Theorem~\ref{thm:bulkflow}  that  $\tilde z_0 = O(g)$.

By \eqref{e:ZNfinal},
\begin{equation}
P_N^{(n,p)} = \log Z_N^{(n,p)} = \zeta_N + \log(1 + K_N(\Lambda)),
\end{equation}
with
$\zeta_N = \tfrac{1}{2}(q_{a,N} + q_{b,N})\sigma_a\sigma_b + t_{a,N}\sigma_a + t_{b,N}\sigma_b + u_N|\Lambda|$
(if $n=0$ then $t_{a,N} = t_{b,N} = u_N = 0$).  Differentiation gives
\begin{equation}
\label{e:DaDbPN-exact-result}
D^2_{\sigma_a\sigma_b} P_{N}^{(n,p)}
= \frac{1}{2}(q_{a,N} + q_{b,N})
+ \frac{D^2_{\sigma_a\sigma_b}K_{N}}{1 + \pi_\varnothing K_{N}}
- \frac{\left(D_{\sigma_a}K_{N}\right)
\left(D_{\sigma_b}K_{N}\right)}{(1 + \pi_\varnothing K_{N})^2}.
\end{equation}
According to \eqref{e:Kg1}, the last two terms vanish in the $N \to \infty$ limit,
so that
\begin{equation}\label{e:infinite-volume-limit}
\lim_{N\to\infty} D^2_{\sigma_a\sigma_b} P_{N}^{(n,p)}
= \frac{1}{2} \big( q_{a,\infty}(m^2) + q_{b,\infty}(m^2) \big).
\end{equation}
We write $v^\pm$ for the common value of $v_a^\pm$ and $v_b^\pm$
(recall Corollary~\ref{cor:vx}).
Since $m^2 \downarrow 0$ as $\varepsilon \downarrow 0$,
by Lemma~\ref{lem:qflow},
\begin{equation} \label{e:mainlim}
\begin{aligned}
\lim_{\varepsilon \downarrow 0}
\lim_{N\to\infty}
 D^2_{\sigma_a\sigma_b} P_{N}^{(n,p)}
&= \frac{1}{2} \big(q_{a,\infty}(0) + q_{b,\infty}(0)\big) \\
&=
\nl! (v^\pm)^2
     \left( \frac{1}{{\sf b} g_0 \log|a-b|} \right)^{2\gamma^\pm}
  G_{ab}^\nl \left( 1 + \Ecal_{ab}^{(p)}\right).
\end{aligned}
\end{equation}
When we make the choice $h=h^+=n^{-1/2}e^+\in E^+$ and carry out the renormalisation group analysis,
it is the exponent $\gamma^+$ that occurs in \eqref{e:mainlim},
and we conclude \eqref{e:Wasy-0}--\eqref{e:Wasy-1b}
(for \eqref{e:Wasy-1a} we use $\gamma^+=0$ when $p=1$).

\smallskip \noindent (iii)
Next we prove \eqref{e:Wasy-2}--\eqref{e:phi12p-new}.
Let $n \ge 2$ and $p = 2$.
We make the two choices $h^+ = n^{-1/2} e^+ \in E^+$ and
$h^- =  2^{-1/2}(1,-1,0,\ldots,0) \in E^-$,
which obey $|h^\pm| = 1$.
By symmetry,
\begin{align}
\inprod{\varphi_a^2 \cdot h^+}{\varphi_b^2 \cdot h^+ }_{\nu_c}
& = \inprod{(\varphi_a^1)^2}{(\varphi_b^1)^2}_{\nu_c}
  + (n-1)\inprod{(\varphi_a^1)^2}{(\varphi_b^2)^2}_{\nu_c}, \\
\inprod{\varphi_a^2 \cdot h^-}{\varphi_b^2 \cdot h^-}_{\nu_c}
& = \inprod{(\varphi_a^1)^2}{(\varphi_b^1)^2}_{\nu_c}
  - \inprod{(\varphi_a^1)^2}{(\varphi_b^2)^2}_{\nu_c},
\end{align}
and hence
\begin{align}
n \inprod{(\varphi_a^1)^2}{(\varphi_b^1)^2}_{\nu_c}
& = \inprod{\varphi_a^2 \cdot h^+}{\varphi_b^2 \cdot h^+ }_{\nu_c}
  + (n - 1)\inprod{\varphi_a^2 \cdot h^-}{\varphi_b^2 \cdot h^-}_{\nu_c}, \\
\label{e:12pf}
n \inprod{(\varphi_a^1)^2}{(\varphi_b^2)^2}_{\nu_c}
& = \inprod{\varphi_a^2 \cdot h^+}{\varphi_b^2 \cdot h^+ }_{\nu_c}
  - \inprod{\varphi_a^2 \cdot h^-}{\varphi_b^2 \cdot h^-}_{\nu_c}
     .
\end{align}
The first term on the right-hand sides has been computed already in the proof of \eqref{e:Wasy-1b}.
For the second term, we instead
use $h=h^-$, and now obtain \eqref{e:mainlim} with $\gamma = \gamma^-$.
This leads to \eqref{e:Wasy-2}--\eqref{e:phi12p-new}.

\smallskip \noindent (iv)
The asymptotic formula \eqref{e:mrcasy} for the amplitudes $\mrc_{n,p,\pm}'$
follows directly, using
the amplitude $\frac{1}{(2\pi)^2}$ for $G_{ab}$ in \eqref{e:Casym} and \eqref{e:lam-indep-infty}.
\end{proof}

Our results in Theorem~\ref{thm:mr-ab} concern the decay of correlation functions
exactly at the critical point.  On the other hand, the results of Theorem~\ref{thm:mr-stars}
concern behaviour of correlation functions summed over the lattice, as the critical
point is approached.  These results do not reveal the nature of the control we have
for correlation functions (not summed) near but not at the critical point.
In the following remark,
we indicate the nature of the control our proof provides in this respect,
for the special case of the two-point function ($p=1$) where the absence of a logarithmic
correction to the leading behaviour simplifies matters.

\begin{rk} \label{rmk:massive-Gab}
Let
\begin{equation}
\label{e:Gab-interacting-defn}
G_{ab}(g, \nu; n) =
\begin{cases}
\wm_{ab}^{(1)}(g, \nu) & (n = 0) \\
\inprod{\varphi^1_a}{\varphi^1_b}_{g, \nu} & (n \geq 1).
\end{cases}
\end{equation}
The quantities defined in \eqref{e:Cabmdef} are then given by
$G_{ab}(0, m^2) = (-\Delta_{\Z^4}+m^2)^{-1}_{ab}$ (independent of $n$).
From \eqref{e:infinite-volume-limit} we obtain
\begin{equation}
\label{e:Gab-to-qinfty-massive}
G_{ab}(g, \nu; n)
= \frac{1 + z_0}{2} \big(q_{a,\infty}(m^2) + q_{b,\infty} (m^2) \big).
\end{equation}
By Lemma~\ref{lem:lamstar} with $p=1$, $\lambda_{x,j}^* = v_x^+(1+O(\chi_j \gbar_j)$.
From Corollary~\ref{cor:vx}, we see that $v_x^+=1$ and therefore
$\lambda_{x,j} = 1 + O(\chi_j\gbar_{j})$.
We insert this into \eqref{e:qinfty} and
apply \eqref{e:R-bound}
to get
\begin{equation}
\label{e:Gab-to-Gm0}
G_{ab}(g, \nu; n)
= (1 + z_0)
\bigg[\big(1 + O(\chi_{j_{ab}}\gbar_{j_{ab}}) \big) G_{ab}(0, m^2) + O(\chi_{j_{ab}}\gbar_{j_{ab}}) G_{ab}(0,0) \bigg].
\end{equation}
Since $G_{ab}(0, m^2)$ is monotone decreasing in $m^2$ (e.g., by \refeq{wmdef} with $p=1$,
or by \eqref{e:srw}),
\begin{equation}
\label{e:Gm0}
G_{ab}(g, \nu; n)
= (1 + z_0)
\Big[G_{ab}(0, m^2) + O(\chi_{j_{ab}}\gbar_{j_{ab}}) G_{ab}(0,0) \Big].
\end{equation}
For the critical case, where $m^2=0$, the last term on the above right-hand side
is smaller than the leading term by a factor $(\log|a-b|)^{-1}$, and
in this case \eqref{e:Gm0} shows that the interacting two-point function is well approximated
by the non-interacting two-point function with renormalised parameters.
Away from the critical point, if $|a-b|$ is sufficiently large
that $j_{ab}>j_m$, then, roughly speaking, $G_{ab}(0, m^2)$ will have exponential
decay $e^{-m|a-b|}$.
On the other hand, taking $\Omega =2$ in \eqref{e:chicCovdef}, and using the fact
that the flow of $\gbar_j$ essentially stops at the mass scale $j_m$
with $\gbar_{j_m} \approx (\log m^{-1})^{-1}$, we find
that now the error term  decays as
\begin{equation}
\label{e:Gm0-error-decay-massive}
    2^{-(j_{ab}-j_m)}\gbar_{j_m}|a-b|^{-2} \approx
    \frac{1}{(\log m^{-1})^{1-t}} \frac{1}{|a-b|^{2+t}} \quad \text{with $t=1/\log_2 L$}
    .
\end{equation}
Since $L$ is large, $t$ is small, and the error term is larger than the leading term
once $|a-b|$ is large enough that the coalescence scale exceeds the mass scale.
A new idea would be needed to obtain control uniform in $|a-b|$, and such control would
be required for example to study the asymptotic behaviour of the correlation length
as $\nu \downarrow \nu_c$.
\end{rk}

\begin{proof}[Proof of Theorem~\ref{thm:mr-aa}.]
We must show that
\begin{align}
\label{e:Wasy-0-aa-pf}
    \wm_{aa}^{(p)}(\nu_c(0)) & =
    G_{aa}^p
    (p! +O(g)) & (p\ge 1),
\\
\label{e:Wasy-1a-aa-pf}
    \inprod{\varphi_a^1}{\varphi_a^1}_{\nu_c(n)}
    & = G_{aa}
    (1+O(g)) & (n\ge 1),
\\
\label{e:Wasy-1b-aa-pf}
    \inprod{|\varphi_a|^2}{|\varphi_a|^2}_{\nu_c(n)}
    & = G_{aa}^2
    (2!n +O(g)) & (n\ge 1),
\\
\label{e:phi12p-new-aa-pf}
     \inprod{(\varphi_a^1)^2}{(\varphi_a^2)^2}_{\nu_c(n)}
     &=
     O(g) & (n\ge 2).
\end{align}
Now the coalescence scale is $j_{aa}=0$, and hence $\lambda_j = 1$ for all $j$.
Also,
\eqref{e:hsigdef} now gives $\ell_{\sigma,j}= 2^{pj}\ggen_j$, and
\eqref{e:Kg1} is replaced by
\begin{align}
\label{e:Kg1-aa}
    | D^k_\sigma K_{N}(\Lambda)|
    &
    \prec
    \chi_N  \gbar_N^{3-k} 2^{-kpN}
     .
\end{align}
Minor changes to the proof of Lemma~\ref{lem:qflow} show that for the case of $a=b$
we obtain
\begin{equation}
    \frac 12 (q_{a,\infty}(0) + q_{b,\infty}(0)) = p! G_{aa}^p + O(g_0),
\end{equation}
and using this in place of \eqref{e:mainlim} leads to the desired results.
In particular, the main terms cancel now in \eqref{e:12pf}, leading to
\eqref{e:phi12p-new-aa-pf}.
\end{proof}

\section{Proof of Theorem~\ref{thm:step-mr-fv}}
\label{sec:step}

Theorem~\ref{thm:step-mr-fv} is
an adaptation of \cite[Theorems~\ref{step-thm:mr-R}--\ref{step-thm:mr}]{BS-rg-step} to include
more general observables.
Its proof
requires modification
to some aspects of \cite{BS-rg-IE,BS-rg-step}, which focus
specifically on the case of $p=1$ and WSAW,
to handle arbitrary $p\ge 1$ for WSAW, and $p=1,2$ for $|\varphi|^4$.
These modifications can be sorted into three categories:
\begin{enumerate}[(i)]
\item
Different choices of parameters and changes to stability estimates
are small details, which
are provided in Section~\ref{sec:choices-params}.
\item
Modification needed in one aspect of the renormalisation group map \eqref{e:RGmap}
is also a small detail, which is discussed in Section~\ref{sec:chages-Map6}.
\item
For $n \ge 2$ and $h\in E^\pm$, we use new ideas to
prove that the full non-perturbative flow of the
coupling constants remains in the space $\Vcal_h$.  This is seen perturbatively in
Proposition~\ref{prop:pt}, and non-perturbatively from the fact that $R_+$ maps
into $\Vcal_h$ in Theorem~\ref{thm:step-mr-fv}.
The new ingredient is the requirement of
$h$-factorisability in Definition~\ref{def:Kspace}, and the fact that this
property is preserved by the renormalisation group map.  We discuss this
 in Section~\ref{sec:hfac}.
\end{enumerate}

\subsection{Choices of parameters, stability and regularity estimates}\label{sec:choices-params}

\subsubsection{Restriction to real coupling constants}
\label{sec:realcc}

Complex coupling constants are used in \cite{BS-rg-step}
only to enable Cauchy estimates
in the proof of Theorems~\ref{thm:bulkflow}
and \ref{thm:step-mr-fv},
but otherwise complex coupling constants are not used.
In \cite{BS-rg-step}, real $(V,K)$ does
indeed yield real $(R_+,K_+)$ for $|\varphi|^4$,
as the vector space $\Kcal$ is a real
vector space, and when $V$ is real there is no way to produce an imaginary part
in $R_+$ or $K_+$.
For WSAW, the complex field can be reexpressed in terms of a real field, and
the bulk coupling constants $g,\nu,z,y$ can be seen to remain real.
For the observable coupling constants $\lambda,q$, the complexity plays a more
prominent role and we have not ruled out the possibility that $\lambda,q$ become
complex.  We  permit them to be complex here, and this
creates no difficulties.

\subsubsection{Choice of \texorpdfstring{$h_{\sigma}$}{h-sigma}}
\label{sec:choice-h-sigma}

Our choices of $\ggen_j$ and $\h_j$ in \eqref{e:ggendef} and
\eqref{e:elldef} are identical to those used in \cite{BS-rg-IE,BS-rg-step},
but the choice of $\h_{\sigma,j}$ in \eqref{e:hsigdef} differs by the appearance of
$\grat_j$ (and thus $\gamma=\gamma^\pm_{n,p}$) and by allowing all $p \ge 1$.
By \cite[\eqref{pt-e:gbarmono}]{BBS-rg-pt},
$\frac 12 \ggen_{j+1} \le \ggen_j \le 2 \ggen_{j+1}$.
Therefore, by \eqref{e:hsigdef},
\begin{equation}
\label{e:h-assumptions-p}
    \frac{\h_{\sigma,j+1}}{\h_{\sigma,j}}
    \le
    {\rm const}\,
    \begin{cases}
     L^p & j < j_{ab}
     \\
     1 & j \ge j_{ab},
     \end{cases}
\end{equation}
where the improved bound occurs for $j \ge j_{ab}$ since the power of $L$ in
\eqref{e:hsigdef} stops changing at the coalescence scale.
On the other hand,
it is indicated in \cite[\eqref{IE-e:h-assumptions}]{BS-rg-IE} that
what is required in \cite{BS-rg-IE,BS-rg-step} is that
\eqref{e:h-assumptions-p} should hold with $L^p$ replaced by $L^1$,
which is a stronger requirement than \eqref{e:h-assumptions-p}.

The $L^p$ growth in \eqref{e:h-assumptions-p} can be accommodated because now
we take $d_+(a)=d_+(b) = p\1_{j<j_{ab}}$ (see Section~\ref{sec:loc}), rather than the choice
$\1_{j<j_{ab}}$ used in \cite[Section~\ref{IE-sec:cl}]{BS-rg-IE}.
Because of this, in the proof of \cite[Propositions~\ref{IE-prop:cl},\,\ref{IE-prop:1-LTdefXY}]{BS-rg-IE},
the computation of the small parameter $\cgam_{\alpha,\beta}(Y)$
(not to be confused with $\gamma^\pm_{n,p}$
despite its similar name) gives exactly the same value
$\cgam_{\alpha,\beta}(Y)=L^{-d -1} +
L^{-1}\1_{Y \cap \{a,b\} \not = \varnothing}$
present in \cite[Proposition~\ref{IE-prop:cl}]{BS-rg-IE}, and the
analysis of \cite{BS-rg-IE,BS-rg-step}
can continue to be based on the crucial contraction
\cite[Proposition~\ref{IE-prop:cl}]{BS-rg-IE} which remains unchanged.

The $L^p$ growth in \eqref{e:h-assumptions-p} also violates the hypotheses
of \cite[Lemma~\ref{IE-lem:Imono}]{BS-rg-IE}, whose
conclusion is used in several places
in \cite{BS-rg-IE} (e.g., in the proofs of the important results
\cite[Proposition~\ref{IE-prop:Istab}, \ref{IE-prop:h}, \ref{IE-prop:ip}]{BS-rg-IE}).
However, the conclusion of \cite[Lemma~\ref{IE-lem:Imono}]{BS-rg-IE} continues
to hold if its hypotheses are modified to use our definition of gauge invariance
in Definition~\ref{def:fieldsym}, and to use the bounds
\eqref{e:h-assumptions-p},
$\h_{\phi,j+1}' \prec L^{-1}\h_{\phi,j}$, and
$\h_{\sigma,j+1}'(\h_{\phi,j+1}')^p \prec \h_{\sigma,j}\h_{\phi,j}^p$
that hold in our present context.

Thus the consequences of \cite[Lemma~\ref{IE-lem:Imono}]{BS-rg-IE} continue
to hold in our present setting of general values of $p$.

\subsubsection{Choice of \texorpdfstring{$p_\Ncal$}{regularity parameter}}
\label{sec:pNcal}

By the definition of $\h_{\sigma,j}$ in \eqref{e:hsigdef},
\begin{equation}
\label{e:hsigratio}
\frac{\ell_{\sigma,j}}{h_{\sigma,j}} = \ggen_j^{1 - \nl/4},
\end{equation}
and this grows for $p>4$.
This plays a role in \cite[Lemma~\ref{step-lem:KKK}]{BS-rg-step}, which is the
place that determines the choice $p_\Ncal =10$ used in \cite{BBS-phi4-log,BBS-saw4-log,
BBS-saw4}.
We continue to use $p_\Ncal=10$ when $p < 4$.
For $p\ge 4$ (which we consider only for WSAW), we take a larger choice, as follows.

First, \cite[Lemma~\ref{step-lem:KKK}]{BS-rg-step} is proved using
\cite[Proposition~\ref{norm-prop:KKK}]{BS-rg-norm},
which in turn relies on \cite[Proposition~\ref{norm-prop:Tphi-bound}]{BS-rg-step}.
We must choose $p_\Ncal \ge A+1$, where $A$ appears in the proof of
\cite[Proposition~\ref{norm-prop:Tphi-bound}]{BS-rg-step}.
In the factor $\rho^{(A+1)}$ in
\cite[Proposition~\ref{norm-prop:Tphi-bound}]{BS-rg-step},
there can appear at most two bad ratios
\eqref{e:hsigratio}, since the worst case contains two observable fields, together with at least $A-1$ good ratios $\ell_j/h_j$ which each
yield a factor $\ggen_j^{1/4}$ by \eqref{e:elldef}.  Thus, at worst, $\rho^{(A+1)}$ gives
\begin{equation}
    \ggen_j^{(A-1)/4} \ggen_j^{2(1-p/4)},
\end{equation}
and we require in \cite[Lemma~\ref{step-lem:KKK}]{BS-rg-step} that this is at most
$\ggen_j^{10/4}$.  Therefore, the minimal $p_\Ncal$ we can permit is
\begin{equation}
    p_\Ncal = A+1 \quad \text{where} \quad
    \tfrac 14 (A-1) + 2 - \tfrac p2 \ge \tfrac{10}{4}, \quad \text{i.e.,\; $A=2p+3$}.
\end{equation}
Thus we can take any fixed $p_\Ncal \ge \max\{10,2p+4\}$.

\subsubsection{Stability estimate: value of \texorpdfstring{$\epV$}{epsilon-V}}

The term $(|\lambda_a|+|\lambda_b|)\h_j\h_{\sigma,j}$ appears in the definition
of $\epsilon_{V,j}$  in \cite[\eqref{IE-e:monobd}]{BS-rg-IE},
for the estimates of \cite[Proposition~\ref{IE-prop:monobd}]{BS-rg-IE}.
This term arises as the $T_0$ norm of
$\lambda_a \sigma_a \phib_a + \lambda_b \bar\sigma_b \phi_b$, and is suitable
for $p=1$.  For general $p \ge 1$, it needs
replacement by $(|\lambda_a|+|\lambda_b|)\h_j^p\h_{\sigma,j}$.  This replacement has been
incorporated into
the definition \eqref{e:DVdef} of $\DV_j$,
so that membership in $\DV_j$ implies that
$(|\lambda_a|\vee|\lambda_b|)\ell_j^p\ell_{\sigma,j} \le C_\DV \ggen_j$.
Also, by \eqref{e:elldef}--\eqref{e:hsigdef}, and since $\ell_0\ge 1$ and $k_0 \le 1$
(as chosen in \cite[Section~\ref{step-sec:2np}]{BS-rg-step}),
\begin{equation}
\begin{aligned}
|\lambda_x|h_j^ph_{\sigma,j}
&=|\lambda_x| \ell_j^p \ell_{\sigma,j}^p(h_j/\ell_j)^p (h_{\sigma,j}/\ell_{\sigma,j}) \\
&= |\lambda_x| \ell_j^p \ell_{\sigma,j}^p (k_0/\ell_0)^p \ggen_j^{-p/4} \ggen_j^{p/4-1} \\
&\le C_\DV \ggen_j(k_0/\ell_0)^p  \ggen_j^{-1}\le C_{\DV}k_0^p.
\end{aligned}
\end{equation}
This fulfills the required bound on $\epsilon_{V,j}$
of \cite[Proposition~\ref{IE-prop:monobd}]{BS-rg-IE}.

\subsubsection{Stability estimate: case of \texorpdfstring{$p\ge 4$}{p greater than 4}}
\label{sec:bigp}

For $\nl > 2$, the proof of
\cite[Proposition~\ref{IE-prop:Iupperzz}]{BS-rg-IE}
must be modified.
In particular, for $p \ge 4$, we must justify placing such a large power in the exponent,
as this appears to make the expectation of $e^{-V}$ divergent since the measure provides
only exponentially quadratic decay.
Justification is possible because functions of $\sigma_a$ and $\sigma_b$ are equivalent
to second-order polynomials, by definition of the quotient space in
\eqref{e:Ncaldecomp4}--\eqref{e:Ncaldecomp}.  Because of this, the placement of the observables
in the exponent is an option that superficially appears worse than
it actually is.

In more detail,
by definition of $\Ncal$, we have $e^{\lambda_a\sigma_a \phib_a^p}=1+\lambda_a\sigma_a \phib_a^p$.
Therefore,
\begin{equation}
\begin{aligned}
\|e^{\lambda_a\sigma_a \phib_a}\|_{T_\phi}
&\le 1 + |\lambda_a|\h_\sigma \|\phib_a^p\|_{T_\phi}
\le 1+|\lambda_x|\h_\sigma \h^p (1+\|\phi\|_\Phi)^{2p} \\
&\le e^{2p(|\lambda_x|\h_\sigma \h^p )^{1/p} (1+\|\phi\|_\Phi^{2})},
\end{aligned}
\end{equation}
where in the
second inequality we used \cite[Proposition~\ref{norm-prop:T0K}]{BS-rg-norm}, and
in the third we used the elementary fact
(see \cite[Lemma~\ref{IE-lem:exp-bounds}]{BS-rg-IE})
that $1 + u^{p} (1+x)^{2p} \le e^{2pu (1+ x^{2})}$
for any $x,u>0$ and $p\ge \max\{1,u\}$, with
the choice $u=(|\lambda_x|\h_\sigma \h^p )^{1/p}$.
This modification permits the proof of \cite[Proposition~\ref{IE-prop:Iupperzz}]{BS-rg-IE}
to proceed as it is otherwise written.

\subsection{Modification to \texorpdfstring{\cite[Map~6]{BS-rg-step}}{the Map 6 of RG construction}}
\label{sec:chages-Map6}

For the analysis of Map~6 in
\cite[Section~\ref{step-sec:int-by-parts2}]{BS-rg-step},
we must estimate the increments
$\delta q_a$, $\delta q_b$, $\delta t_a$, $\delta t_b$, and $\delta u$ that arise
in $R_+$.
The discussion of $\delta u$ provided there holds without change here.
There is a small modification to
the treatment of $\delta q_a$, $\delta q_b$, which we discuss
first, and $\delta t_a$, $\delta t_b$ are new here.
We use the notation of \cite[Section~\ref{step-sec:int-by-parts2}]{BS-rg-step}.

Let $x = a,b$. It suffices to show that $\|\delta q_x \sigma_a\sigma_b\|_{T_0} \prec 1$,
and for this we may assume that $j \ge j_{ab}$.
In this case, $\lambda_x =\lambda_{j_{ab},x}$
and $\lambda_x$ is not updated by $Q$.
By \cite[Proposition~\ref{pt-prop:Cdecomp}]{BBS-rg-pt},
for $m^2 \in \Iint_j$,
$|C_{j;xy}|
  \leq c
  L^{-2(j-1)}$.
From this we conclude that $\delta [w_{ab}^p] \prec L^{-2pj} \prec \ell_j^{2p}$.
Therefore,
\begin{equation}
\delta q_x
= p! \lambda_a\lambda_b  \delta[w_{ab}^p]
\prec \lambda_a\lambda_b  \ell_j^{2p}.
\end{equation}
Since $V \in \DV_j$,
we have $|\lambda_x|  \le C_\DV \ggen_j \ell_j^{-p} \ell_{\sigma,j}^{-1}$.
Therefore,
\begin{equation}
    \|\delta q_x \sigma_a\sigma_b\|_{T_0}
    =
    |\delta q_x| \h_{\sigma,j}^2
    \prec
    \ggen_j^2 (\h_{\sigma,j}/\ell_{\sigma,j})^2.
\end{equation}
Since the right-hand side is $\ggen_j^2$ for $\h=\ell$, and is $\ggen_j^{2p/4}$
for $\h=h$, this is sufficient.

Finally,  $\delta t_x$ only arises for $n\ge 1$ and $p=2$, which we assume in the following
sketch.
It suffices to show that $|\delta t_x| \h_\sigma \prec 1$.
By \eqref{e:tpt},
\begin{equation}
\label{e:tpt-bis}
    \delta t_x = t_{\pt,x}(V-Q) - t_x
    =
    \1_{n \ge 1}  \hat\lambda_x (e^+ \cdot h) \hat\varsigma,
\end{equation}
where
\begin{equation}
\label{e:varsighat}
\begin{aligned}
\hat\varsigma
= \Big( C_{0,0}( &1 - \1_{j + 1 < j_{ab}}2\hat\nu w^{(1)})
+ \1_{j + 1 < j_{ab}} \hat\nu^+\delta[ w^{(2)}]
+ \1_{j + 1 \ge j_{ab}}\delta[\hat\nu w^{(2)}]
\Big),
\end{aligned}
\end{equation}
with $\hat\lambda_x,\hat\nu$ the relevant coupling constants of $V-Q$.  Thus,
$\hat\lambda_x = \lambda_x - \lambda_{x,Q}$ and $\hat\nu = \nu-\nu_Q$, with $\lambda_{x,Q}$ and
$\nu_Q$ from $Q$.
As above, we have $C_{00}\prec \ell_j^2$
and
$|\lambda_x|  \le C_\DV \ggen_j \ell_j^{-2} \ell_{\sigma,j}^{-1}$.
As in \cite[\eqref{step-e:rhoFcaldef}]{BS-rg-step}, we define
\begin{equation}
    \label{e:rhoFcaldef}
    \epdV_{j}
    =
    \begin{cases}
    \chicCov_{j}^{1/2} \ggen_{j} & (\h =\ell )
    \\
    \chicCov_{j}^{1/2} \ggen_{j}^{1/4}   & (\h = h ).
    \end{cases}
\end{equation}
In the setting of Map~6, we have $|\lambda_{x,Q}| \h^2 \h_{\sigma} \prec \epdV$.
The largest term on the right-hand side of \eqref{e:varsighat}
is the first one, and
its contribution to $|\delta t_x| \h_\sigma$ is bounded by a multiple of
\begin{equation}
    |\hat \lambda_x| \ell_j^2 \h_{\sigma,j}
    \prec
    (\ggen_j \ell_j^{-2} \ell_{\sigma,j}^{-1}+ \epdV_j \h_j^{-2}\h_{\sigma,j}^{-1})
    \ell_j^2 \h_{\sigma,j}
    \prec
    \begin{cases}
     \ggen_{j} & (\h =\ell )
    \\
    \ggen_{j}^{1/2}   & (\h = h )
    \end{cases}
\end{equation}
(recall \eqref{e:elldef}--\eqref{e:hsigdef}).
This is sufficient.

\subsection{Renormalisation and reduced symmetry}
\label{sec:hfac}

As discussed in Section~\ref{sec:sym}, for $n \ge 2$ the $O(n)$ symmetry can be
reduced by choice of $h$.  To handle this, we replaced the definition
of the space $\Kcal$ in \cite[Definition~\ref{step-def:Kspace}]{BS-rg-step}
by the adapted version in
Definition~\ref{def:Kspace}.  With Definition~\ref{def:Kspace},
we can prove that
if $h \in E^\pm$, and if $U \in \Vcal_h$ and $K \in \Kcal(h)$ obey appropriate
estimates, then under the renormalisation group map
it is also the case that $V_+\in \Vcal_h$ and $K_+ \in \Kcal_+(h)$.
This is the content of the following proposition, in which we place more prominence
than usual on $h$ in the notation.

\begin{prop}\label{prop:range-RKplus}
The renormalisation group map of \cite[Section~\ref{step-sec:mr}]{BS-rg-step} obeys
$(R_+, K_+):\domRG(\sgen,h) \times \Igen_+(\mgen^2) \to \Vcal_h^{(1)}
\times \Wcal_{+}(\sgen_+,h)$.
\end{prop}

The proof of Proposition~\ref{prop:range-RKplus} is organised as follows.
In Section~\ref{sec:hfac-properties}, we
prove elementary properties of $h$-factorisability. In Section~\ref{sec:range-Rplus},
we prove that $R_+$ maps into $\Vcal_h^{(1)}$.
In Section~\ref{sec:hfac-Kplus}, we prove that $K_+$ maps into $\Wcal_{+}(\sgen_+)$.

\subsubsection{Elementary properties of \texorpdfstring{$h$}{h}-factorisability}
\label{sec:hfac-properties}

\begin{lemma}
\label{lem:efac-product-rule}
Let $n \ge 1$ and $h \in \R^n$.
If $F,K \in \Ncalefac$,
and if $\pi_\varnothing F$ and $\pi_\varnothing K$ are $S(n)$-invariant,
then $FK\in\Ncalefac$ with
$\fac{FK}_\alpha = (\pi_\varnothing F) K^*_\alpha +  F^*_\alpha(\pi_\varnothing K)$  for $\alpha
=a,b$.
\end{lemma}

\begin{proof}
We write $F_\varnothing = \pi_\varnothing F$ and   $K_\varnothing = \pi_\varnothing K$.
Since we work in a quotient space with $\sigma_\alpha^2 = 0$,
\begin{equation}
    \pi_\alpha FK =
    F_\varnothing  (\pi_\alpha K) +  (\pi_\alpha F)K_\varnothing
    =
    \sigma_\alpha \left( [F_\varnothing  K^* + F^* K_\varnothing ]\cdot h  \right),
\end{equation}
so the first requirement of Definition~\ref{def:efac}
holds with $(FK)^*$ as indicated.
Secondly, by the hypotheses on $F_\varnothing$ and $K$, for $P \in S(n)$,
\begin{equation}
\begin{aligned}
(P (F_\varnothing K_\alpha^*))(\varphi)
&= (F_\varnothing (PK^*_\alpha))(\varphi)
= F_\varnothing(\varphi)(PK^*_\alpha)(\varphi) \\
&= F_\varnothing (P\varphi) K_\alpha^*(P \varphi)
= (F_\varnothing K_\alpha^*)(P \varphi).
\end{aligned}
\end{equation}
The $F^* K_\varnothing$ term is similar, and the proof is complete.
\end{proof}

\begin{lemma}
\label{lem:efac-operations-invariance}
Let $n \ge 1$, $X \subset \Lambda$, and $F \in \Ncalefac$.
Then $\LT_X F \in \Ncalefac$ with $\fac{\Loc_X F}_\alpha = \Loc_X F_\alpha^*$.
Also, $\Ex \theta F \in \Ncalefac$ with $\fac{\Ex \theta F}_\alpha = \Ex \theta F_\alpha^*$.
Here $\Loc_X F_\alpha^*$ and $\Ex \theta F_\alpha^*$ are defined component-wise.
\end{lemma}

\begin{proof}
The statement has content only for $n \ge 2$, so we write the proof for this case.
Since $\Loc_X$ commutes with $\pi_\alpha$ and is linear,
\begin{equation}
    \pi_\alpha \Loc_X F = \Loc_X \pi_\alpha F = \Loc_X \sigma_\alpha (F_\alpha^*\cdot h)
    = \sigma_\alpha (\Loc_X F_\alpha^*\cdot h).
\end{equation}
The invariance under permutations follows easily.

Again by linearity,
$\pi_\alpha \Ex \theta F = \Ex \theta \pi_\alpha F
= \Ex \theta \sigma_\alpha ( F_\alpha^* \cdot h)
= \sigma_\alpha (\Ex\theta F_\alpha^*\cdot h)$.
For the invariance under permutations $P \in S(n)$ of the fields, we use
\begin{equation}
\begin{aligned}
(P (\Ex \theta F_\alpha^*))(\varphi)
&= \Ex \theta (P F_\alpha^*)(\varphi)
= \Ex (P F_\alpha^*)(\varphi + \zeta)
= \Ex F_\alpha^*(P(\varphi + \zeta)) \\
&= \Ex F_\alpha^*(P\varphi + \zeta)
= (\Ex \theta F_\alpha^*) (P\varphi),
\end{aligned}
\end{equation}
where $\zeta$ is the integration variable,
and where the fourth equality follows by making the change of variables $\zeta \mapsto P \zeta$
(with Jacobian equal to 1) in the integral.
\end{proof}

The following lemma shows that
$I_j$ and related quantities are Euclidean covariant (recall Definition~\ref{def:latticesym})
and inherit $h$-factorisability from $V$.

\begin{lemma}
\label{lem:efac-functions-of-V}
Let $V \in \Vcal_h$, $X \in \Pcal_j$, and $x \in \Lambda$.  Each of
$W_j(V,X)$, $I_j(V,X)$, $P_{j,x}(V)$ and $V_{\pt,x} (V)$
is in $\Ncalefac$.
Each of
$\pi_\varnothing W_j(V)$, $\pi_\varnothing I_j(V)$
(as functions of $X\in \Pcal_j$), and
$\pi_\varnothing P_{j}(V)$ and $\pi_\varnothing V_\pt (V)$
(as functions of $x\in \Lambda$)
is Euclidean covariant.
\end{lemma}

\begin{proof}
Let  $A\in \Ncalefac$ be a polynomial in the fields, and let $\alpha =a,b$.
Then $\pi_\alpha A = \sigma_\alpha (A_\alpha^* \cdot h)$,
and we can assume that every component of $A_\alpha^*$ is a polynomial.
Recall the definition of $\Lcal_C$ in \eqref{e:LapC}.  Note that
$\pi_\alpha \Lcal_C A = \sigma_\alpha (\Lcal_C A_\alpha^* \cdot h)$.
Let $P \in S(n)$ be a permutation matrix.
Since $\Lcal_C$ acts component-wise,
$P\Lcal_C A_\alpha^* =\Lcal_C PA_\alpha^*$, and hence, since $A \in \Ncalefac$,
$(P\Lcal_C A_\alpha^*)(\varphi) =(\Lcal_C PA_\alpha^*)(\varphi)= (\Lcal_C A_\alpha^*)(P\varphi)$.
This shows that $\Lcal_C A \in \Ncalefac$.
Consequently,
\begin{equation}
e^{\pm \Lcal_C} A = \sum_{k = 0}^{\deg(A)} \frac{(\pm 1)^k}{k!} \Lcal_C^k A
\in \Ncalefac.
\end{equation}

Let $V \in \Vcal_h$ and $X \in \Pcal_j$.  Then $V \in \Ncalefac$ by definition and
every component of $V_\alpha^*$ is a polynomial.
Using Lemmas~\ref{lem:efac-product-rule}--\ref{lem:efac-operations-invariance}
and the above observations concerning $\Lcal_C$, we see from \eqref{e:FCAB}--\eqref{e:Fpi}
that $F_{\pi,C}(V,V)$ is \efac{}, as are
$W_j(V,X)$, $I_j(V,X)$, $P_{j}(V,X)$ and $V_\pt (V,X)$.

The Euclidean covariance is a consequence of the definitions,
the
Euclidean invariance of $w_j$, and the
Euclidean covariance property $A(\pi_\varnothing \LT_X K)= \pi_\varnothing \LT_{AX}(AK)$ of
\cite[Proposition~\ref{loc-prop:9LTdef}]{BS-rg-loc}.
\end{proof}

\subsubsection{Range of \texorpdfstring{$R_+$}{Rplus}}
\label{sec:range-Rplus}

The following proposition gives the $R_+$ part of Proposition~\ref{prop:range-RKplus}.

\begin{prop}\label{prop:range-Rplus}
Let $h \in E^\pm$.
If $(U,K) \in \domRG(\sgen)$ and $m^2 \in \Igen_+(\mgen^2)$, then
$R_+(U,K) \in \Vcal_h^{(1)}$.
\end{prop}

The main step in the proof of Proposition~\ref{prop:range-Rplus}
is provided by Lemma~\ref{lem:Locsym} below, which in turn relies on
Lemma~\ref{lem:Sprime}.
For the latter, we observe that
the linear span of the permutation subgroup $S(n)$ consists of the set $\bar S(n)$ of
$n\times n$ matrices whose row and column sums are
all equal.
Given a set $Z$ of matrices,
we write $Z' = \{B : AB=BA \; \text{for all $A\in Z$}\}$ for its commutant.
Recall the set of matrices $M_2(n)$ from Definition~\ref{def:M2}.
The following lemma states that $\bar S(n)$ and $M_2(n)$ are each other's commutant;
we omit the elementary proof.

\begin{lemma}
\label{lem:Sprime}
For $n \ge 1$, $M_2'(n)= \bar S(n)$ and $\bar S'(n)=M_2(n)$.
\end{lemma}

The  proof of the following lemma uses the fact that
\begin{equation}
\label{e:TLoc}
    T(\LT_X F) = \LT_X (TF)
    \quad
    \text{for \emph{any} $n\times n$ matrix $T$ and $F \in \Ncal$}.
\end{equation}
A proof of \eqref{e:TLoc} is given in
\cite[Proposition~\ref{phi4-log-prop:9LTdef}]{BBS-phi4-log}
for the case $F \in \Ncal_\varnothing$, and the same proof holds for $F\in \Ncal$.
Also, it is shown in \cite[Sections~\ref{loc-sec:LocDefinition}, \ref{loc-sec:ssloc}]{BS-rg-loc}
that $\LT$ preserves Euclidean invariance,
gauge ($U(1)$) invariance, and supersymmetry.

\begin{lemma}
\label{lem:Locsym}
Let $X \subset \Lambda$ and $\alpha \in \{a,b,ab\}$.
For $n = 0$, $p\ge 1$, and for $n \ge 1$ and $p=1,2$,
\begin{equation}
\label{e:Locsym}
\Loc_X (\pi_\alpha \Ncal_{h}) \subset \bigcup_{m \in M_2(n)} \pi_\alpha \Vcal_{m h}(X) .
\end{equation}
In particular, if $h \in E^\pm$, the right-hand side of \eqref{e:Locsym} becomes simply
$\pi_\alpha \Vcal_h (X)$.
\end{lemma}

\begin{proof}
We use properties of $\LT$ from \cite{BS-rg-loc}.
By \eqref{e:Vmulticomponent}, an element of $\pi_{ab}\Vcal_h$
can be written as $-\sigma_a\sigma_b \frac 12 (q_a\1_{x=a} + q_b\1_{x=b})$ (independent of $h$).
Thus \eqref{e:Locsym} follows from our choice $d_+(ab) = 0$.
Similarly, for $\alpha =a,b$ and
$j \geq j_{ab}$, elements of $\pi_\alpha\Loc_X (\Ncal_h)$ are constant multiples of $\sigma_\alpha$.
Thus we assume henceforth that $j < j_{ab}$ and consider $\alpha =a,b$.
In this case, $d_+(\alpha)=p$.

For $n=0$ and $p \ge 1$, $h$ plays no role, and in this case $\sigma_a=\sigma^p$ and
$\sigma_b=\bar\sigma^p$
(recall Definition~\ref{def:fieldsym}).
The only $U(1)$-invariant monomials containing $\sigma_a=\sigma^p$ or $\sigma_b=\bar\sigma^p$,
and with dimension at most $p$, are
$\{\sigma^p\bar\phi^p,  \bar\sigma^p\phi^p\}$.
Since $\Loc_X$ preserves $U(1)$ invariance, this implies that, as required,
\begin{equation}
\label{e:LocNcalU1}
\Loc_X (\pi_a\Ncal_{h}) =  \1_{a\in X}\sigma^p \Span \left\{\phib_a^p \right\},
\qquad
\Loc_X (\pi_b\Ncal_{h}) =  \1_{b\in X}\sigmab^p \Span \left\{\phi_b^p \right\}.
\end{equation}
The appearance of the indicator functions on the right-hand sides
of \eqref{e:LocNcalU1} follows directly
from the definition of $\LT$ in \cite[Definition~\ref{loc-def:LTXYsym}]{BS-rg-loc}.

For $n \ge 1$ and $p=1,2$, since $d_+(\alpha) = p$,
\begin{align}
\label{e:LocNcal-a-p1}
\Loc_X (\pi_\alpha\Ncal)
&\subset
\1_{\alpha\in X}
    \sigma_\alpha \Span \left\{1, \varphi^{i}_\alpha \mid
1 \leq i \leq n \right\}
& &(p = 1),  \\
\label{e:LocNcal-a-p2}
\Loc_X (\pi_\alpha\Ncal)
&\subset
\1_{\alpha\in X}
    \sigma_\alpha \Span \left\{1, \varphi^{i}_\alpha, \varphi^i_\alpha \varphi^k_\alpha, \nabla \varphi^{i}_\alpha  \mid
1 \leq i,k \leq n \right\}
& &(p = 2),
\end{align}
where the superscripts on $\varphi_\alpha$ indicate components.

For the case $p = 2$, it
follows from \eqref{e:TLoc} that
the $R(n)$-invariance of $\Ncal_h$ is preserved by $\Loc_X$.
The linear, mixed quadratic, and gradient monomials from \eqref{e:LocNcal-a-p2} are not invariant
under replacement of one component of $\varphi_\alpha$ by its negative, and thus are not
in $\Loc_X (\pi_\alpha\Ncal_h)$ when $p = 2$.  Therefore, for both $p=1$ and $p=2$,
\begin{equation}
\label{e:LocNcal-a-p12}
    \Loc_X (\pi_\alpha\Ncal_h) \subset
    \1_{\alpha\in X}
    \sigma_\alpha \Span
    \left\{1, (\varphi_\alpha^{1})^p, \ldots, (\varphi_\alpha^{n})^p \right\} .
\end{equation}
By Definition~\ref{def:Ncalh}, if $F \in \Ncal_h$ then
$\pi_\alpha F = \sigma_\alpha (F_\alpha^*\cdot h)$, with
$F^* \in \left(\pi_\varnothing \Ncal\right)^n$ such that $(PF_\alpha^*)(\varphi)
= F_\alpha^*(P\varphi)$ for all $P \in S(n)$.
By \eqref{e:LocNcal-a-p12}, each component of
$\Loc_X F_\alpha^*$ lies in
$\Span \left\{1, (\varphi_\alpha^{1})^p, \ldots, (\varphi_\alpha^{n})^p \right\}$.
Therefore, there exist an $n\times n$ matrix $m_\alpha$ and a vector $v_\alpha\in \R^n$ such that
$\Loc_X F_\alpha^* = m_\alpha \varphi_\alpha^p + v_\alpha$.
With Lemma~\ref{lem:efac-operations-invariance}, this implies that
\begin{equation}
    P(m_\alpha \varphi_\alpha^p + v_\alpha) = m_\alpha P\varphi_\alpha^p + v_\alpha
\end{equation}
for every $P \in S(n)$, from which we conclude that $Pv_\alpha=v_\alpha$ and $Pm_\alpha=m_\alpha P$ for every
$P \in S(n)$.  The first of these conclusions implies that $v_\alpha=s_\alpha e^+$
for some $s_\alpha \in \R$
(with the vector $e^+$ of \eqref{e:epm}), and by Lemma~\ref{lem:Sprime}
the second implies that $m_\alpha\in M_2(n)$.
Since $m^T=m$ for $m \in M_2(n)$,
\begin{equation}
\begin{aligned}
\Loc_X \pi_\alpha F
&= \1_{\alpha\in X}
\sigma_\alpha \big( m_\alpha \varphi_\alpha^p \cdot h + s_\alpha e^+ \cdot h \big)
\\
&= \1_{\alpha\in X}
\sigma_\alpha \big( \varphi_\alpha^p \cdot (m_\alpha h) + s_\alpha e^+\cdot h \big)
.
\end{aligned}
\end{equation}
The right-hand side lies in $\pi_\alpha \Vcal_{m_\alpha h} (X)$
(with $t_\alpha = s_\alpha e^+\cdot h$), and this completes the proof.
\end{proof}

\begin{proof}[Proof of Proposition~\ref{prop:range-Rplus}]
Let $h$ be in one of the eigenspaces $E^\pm$ of the matrices in $M_2(n)$
(and $h=1$ if $n=0$).
The definition of $R_+$ is given in \eqref{e:Vplusdef}--\eqref{e:Rdef}.
It is already established in \cite[Section~\ref{step-sec:Iconstruction}]{BS-rg-step}
that $\pi_\varnothing R_+ \in \pi_\varnothing \Vcal_h^{(1)}$ (for this $h$ plays
no role).  Thus we concentrate on $\pi_\alpha R_+$ for $\alpha \in \{a,b,ab\}$.
Note that the superscript in $\Vcal_h^{(1)}$ plays no role in these observable subspaces.

By assumption, $U \in \Vcal_h^{(0)} \subset \Vcal_h$, and
by Proposition~\ref{prop:pt},
$V_\pt : \pi_\alpha \Vcal_h \to \pi_\alpha \Vcal_{h_\pt}=\pi_\alpha \Vcal_h$ (with the last equality
due to $h \in E^\pm$).  Thus, by definition of $R_+$,
it suffices to show that the polynomial $Q$ defined by \eqref{e:Vplusdef} obeys
$\pi_\alpha Q \in \pi_\alpha \Vcal_h$.
By definition of $Q$, to prove that $\pi_\alpha Q \in \pi_\alpha \Vcal_h$ it suffices
to prove that $\LT_X : \pi_\alpha \Ncal_h \to \pi_\alpha \Vcal_h$, because
$K(Y)I(Y,V)^{-1} \in \Ncal_h$ by Lemma~\ref{lem:efac-functions-of-V}
and because $K(Y)\in \Ncal_h$
since $K \in \Kcal(h)$ (recall Definition~\ref{def:Kspace}).
This last requirement is provided by Lemma~\ref{lem:Locsym}, and the proof is complete.
\end{proof}

\subsubsection{Range of \texorpdfstring{$K_+$}{Kplus}}
\label{sec:hfac-Kplus}

We now complete the proof of Proposition~\ref{prop:range-RKplus},
by proving its $K_+$ part.

We extend the notion of $h$-factorisation in Definition~\ref{def:efac}
to maps $F:\Pcal_j \to \Ncal$, as follows.
We say that $F$ is $h$-\emph{factorisable} if
$F(X) \in \Ncalefac$ for all
$X \in \Pcal_j$.
By Lemma~\ref{lem:efac-product-rule}, if $F,G \in \Kcal$ are \efac{}, then
$F \circ G$ is \efac{} as well since the $O(n)$-invariance of $\pi_\varnothing F$ and $\pi_\varnothing G$
is guaranteed by the definition of $\Kcal$ in Definition~\ref{def:Kspace}.

\begin{proof}[Proof of Proposition~\ref{prop:range-RKplus}]
By Proposition~\ref{prop:range-Rplus}, $R_+(U,K) \in \Vcal_h^{(1)}$,
so it remains to check that $K_+ \in \Wcal_{+}(\sgen_+)$.
This statement is provided by
\cite[Theorem~\ref{step-thm:mr}]{BS-rg-step},
apart from the requirement that $K_+$ is \efac{}, and, if $p=2$, that $K_+$ is
$R(n)$-invariant.
To check that the map $K_+$ constructed in \cite{BS-rg-step}
is \efac{}, we recall that the construction is a composition of six maps
which produce $K^{(1)}, \ldots, K^{(6)}=K_+$.
We examine these one by one and show that $K^{(i)}\in \Ncalefac$ implies
$K^{(i + 1)} \in \Ncalefac$.
We omit the simpler proof that $K_+$ is $R(n)$-invariant when $p=2$.
\begin{enumerate}
\item
According to its construction in \cite[Lemma~\ref{step-lem:K1}]{BS-rg-step},
$K^{(1)}$ is a polynomial in $I$, $K$, and $J$
(see \cite[\eqref{step-e:Kout-poly}]{BS-rg-step}).
Since $J$ is given by localised products of $I$ and $K$
(see \cite[\eqref{step-e:Map1JXB}--\eqref{step-e:Map1JBB}]{BS-rg-step}),
it is \efac{}, and hence so is $K^{(1)}$, by Lemma~\ref{lem:efac-product-rule}.

\item By \cite[Lemma~\ref{step-lem:K2}]{BS-rg-step},
$K^{(2)}$ is a circle product of $\delta I^{(2)}$ and $K^{(1)}$. Both of these are \efac{},
and hence so is $K^{(2)}$.

\item The definition of $K^{(3)}$ is given in
\cite[\eqref{step-e:K3van}]{BS-rg-step}.
All of the quantities on the right-hand side of \cite[\eqref{step-e:K3van}]{BS-rg-step}
are \efac{}, and hence so is $K^{(3)}$.

\item
According to its construction in \cite[Lemma~\ref{step-lem:K5a}]{BS-rg-step},
$K^{(4)}$ is a polynomial in $\Ipttil$, $K^{(3)}$, and $\hldg$
(see \cite[\eqref{step-e:Kout-poly}]{BS-rg-step}).
By \cite[\eqref{step-e:hptdefqq}]{BS-rg-step}, $\hldg$ is a truncated expectation of $V$'s,
so it is \efac{} by Lemma~\ref{lem:efac-operations-invariance}, and hence so is $K^{(4)}$.

\item
Map~5 replaces $W(V_\pt)$ by $W(V_+)$.  Since
both $V_\pt(V)$ and $V_+$ are \efac{}, so is $K^{(5)}$.

\item
The role of Map~6 is to perform summation by parts and
to move constant fields out of the circle product.  Only the second aspect is
different in our present setting, in which \cite[\eqref{step-e:removeu}]{BS-rg-step}
becomes replaced by
\begin{equation}
    ((e^{\delta \zeta}\Ipt^+)\circ K^{(5)})(\Lambda)
    =
    e^{\delta \zeta (\Lambda)}(\Ipt^+ \circ (e^{-\delta \zeta}K^{(5)})(\Lambda),
\end{equation}
where
\begin{equation}
\delta \zeta (X) = \sum_{x\in X} V_{\pt,x}(V-Q)|_{\varphi=0}.
\end{equation}
We have shown above that $V - Q$ and $\Vpt(V-Q)$ are \efac{}, and hence so
is $\delta\zeta$.  It can then be seen from its definition
in \cite[\eqref{step-e:K7-def}]{BS-rg-step} that $K^{(6)}$ is \efac{}.
\end{enumerate}
Since $K_+=K^{(6)}$ by definition, this completes the proof.
\end{proof}

\setcounter{section}{0}
\renewcommand{\thesection}{\Alph{section}}

\section{Proof of Proposition~\ref{prop:Integral-Representation}}\label{app:intrep}

In this appendix, we prove Proposition~\ref{prop:Integral-Representation}
using ideas from \cite{BIS09}, but organise the proof
in a more direct manner for our current goal.
The particular approach we present here arose in \cite{BEI92}, but these
ideas have a long history
going back to \cite{Syma69} and including \cite{McKa80,PS80,BIS09,BFS82,LeJa88,Dynk83,FFS92}.
Proposition~\ref{prop:Integral-Representation} can be equivalently stated as the
identity
\begin{equation}
\label{e:intrep1a}
\begin{aligned}
    \int e^{-\sum_{x\in\Lambda} \big(\tau_{\Delta ,x} + g\tau_x^2 + \nu\tau_x \big)}
    \bar{\phi}_{a_1}\cdots &\phib_{a_p} \phi_{b_1} \cdots \phi_{b_p} \\
    &=
    \sum_{\sigma\in S_p}
    \int_{\R_+^\nl} E_A^N \big[e^{- I_p(T)} \1_{X(T)= \sigma(B)}\big] e^{-\nu \|T\|_1 } dT
    ,
\end{aligned}
\end{equation}
where now the $i^{\rm th}$ walk $X^i$ begins at $a_i$ and ends at $\sigma(b_i)$.
The proof of \eqref{e:intrep1a} is based on three different formulas for the Green function
$(-\Delta +V)^{-1}$, where
$V$ is a complex diagonal matrix whose diagonal entries $v_x$ obey ${\rm Re}(v_x) > 0$.
The three formulas are presented in the following three lemmas.

\begin{lemma}
\label{lem:srw}
Let $\mathcal{W}_{ab}^n$ denote the set of nearest-neighbour $n$-step paths
from $a$ to $b$.  Then
\begin{equation}
\label{e:srw}
(-\Delta_\Lambda + V)^{-1}_{ab}
= \sum_{n=0}^\infty \sum_{Y \in \mathcal{W}_{ab}^n} \prod_{j=0}^{n}\frac{1}{2d + v_{Y_j}}.
\end{equation}
\end{lemma}

\begin{proof}
We write $-\Delta = 2d \1 - J$ and let $U = 2d \1 + V$.
Then $(-\Delta + V)^{-1}$ is given by the Neumann series
\begin{equation}
(-\Delta_\Lambda + V)^{-1}
= (U - J)^{-1} = \bigg(U(\1 - U^{-1}J)\bigg)^{-1}
= \sum_{n=0}^{\infty} \big(U^{-1}J\big)^n U^{-1},
\end{equation}
which converges since ${\rm Re} (V) >0$.
The $ab$ matrix element of the right-hand side is the right-hand side of \eqref{e:srw},
and the proof is complete.
\end{proof}

\begin{lemma}
\label{lem:intrepLHS}
Let $X(T)$ be a continuous time simple random walk on $\Lambda$ with local time $L_T(x)$.
Let $V$ be a complex diagonal matrix with entries $v_x$ such that ${\rm Re}(v_x) > 0$, then
\begin{equation}
    (-\Delta_\Lambda + V)^{-1}_{ab}
    =
    \int_{\R_+} E_a^N \big[e^{-\sum_{x\in\Lambda} v_x L_T(x) } \1_{X(T)= b}\big]  dT
    .
\end{equation}
\end{lemma}

\begin{proof}
We think of $X$ as a discrete time simple random walk $Y$ with
independent and identically distributed  $\operatorname{Exp}(2d)$ holding times $(\sigma_i)_{i\geq 0}$.
We set $\gamma_j = \sum_{i = 0}^j \sigma_i$, and condition on $Y$ to obtain
\begin{equation}
\begin{aligned}
\int E_a &\left[e^{-v \cdot L_T} \1_{X(T) = b}\right] dT
\\
&= \sum_{n=0}^\infty \sum_{Y \in \mathcal{W}_{ab}^n} \left(\frac{1}{2d}\right)^n
E \left[e^{-\sum_{j=0}^{n-1} v_{Y_j} \sigma_j} \int_{\gamma_{n-1}}^{\gamma_n} e^{-v_{Y_n} (T - \gamma_{n-1}) } dT \right]
\\
&= \sum_{n=0}^\infty \sum_{Y \in \mathcal{W}_{ab}^n} \left(\frac{1}{2d}\right)^n
E \left[\left(e^{-\sum_{j=0}^{n-1} v_{Y_j} \sigma_j}\right) \frac{-1}{v_{Y_n}} \left(e^{-v_{Y_n} \sigma_{n}} - 1\right)\right].
\end{aligned}
\end{equation}
Since the $\sigma_i$ are i.i.d., the expectation factors
into a product of $n+1$ expectations that can each be evaluated explicitly,
with the result that
\begin{equation}
\begin{aligned}
\int E_a &\left[e^{-v \cdot L_T} \1_{X(T) = b}\right] dT \\
&= \sum_{n=0}^\infty \sum_{Y \in \mathcal{W}_{ab}^n} \left(\frac{1}{2d}\right)^n
\left(\prod_{j=0}^{n-1}\frac{2d}{2d + v_{Y_j}}\right)  \left(\frac{2d}{2d + v_{Y_n}} - 1\right) \left(\frac{-1}{v_{Y_n}}\right) \\
&= \sum_{n=0}^\infty \sum_{Y \in \mathcal{W}_{ab}^n} \prod_{j=0}^{n}\frac{1}{2d + v_{Y_j}}.
\end{aligned}
\end{equation}
By Lemma~\ref{lem:srw}, this completes the proof.
\end{proof}

The next lemma uses the complex Gaussian probability
measure on $\C^\Lambda$ with covariance $C$,
defined by
\begin{equation}
    d\mu_C =
    \frac{\det A}{(2\pi i)^M}  e^{-\phi A\bar\phi} d\bar\phi d\phi,
\end{equation}
with $A=C^{-1}$ and $d\bar\phi d\phi$ is the Lebesgue measure
$d\bar\phi_1 d\phi_1 \cdots d\bar\phi_\Lambda d\phi_\Lambda$
(see, e.g., \cite[Lemma~2.1]{BIS09} for a proof that this measure is properly normalised).
The statement that $d\mu_C$ has covariance $C$ means that
$\int \bar\phi_a\phi_b d\mu_C = C_{ab}$.  Integration by parts
(see, e.g., \cite[Lemma~2.2]{BIS09}) gives the formula
\begin{equation}
\label{e:ibp}
    \int_{\C^{\Lambda}} \bar\phi_a F e^{-\phi A\bar\phi} \;d\bar\phi d\phi
    =
    \sum_{x} C_{ax} \int_{\C^{\Lambda}} \frac{\partial F}{\partial \phi_x} e^{-\phi A\bar\phi} \;d\bar\phi d\phi
    .
\end{equation}

\begin{lemma}
\label{lem:intrepRHS}
Let $V$ be a complex diagonal matrix with entries $v_x$ such that ${\rm Re}(v_x) > 0$.
Let $A=-\Delta_\Lambda + V$ and
set $C = A^{-1}= (-\Delta_\Lambda + V)^{-1}$.
Then
\begin{equation}
\label{e:intrefRHS}
    \sum_{\sigma \in S_p} \prod_{i=1}^p (-\Delta_\Lambda + V)^{-1}_{a_ib_{\sigma(i)}}
    =
    \int e^{-\phi A \bar\phi - \psi A \bar\psi}
    \phib_{a_1}\cdots \phib_{a_p} \phi_{b_1}\cdots \phi_{b_p}
    .
\end{equation}
\end{lemma}
\begin{proof}
By definition,
\begin{equation}
e^{-\psi A\bar\psi}
= \sum_{n=0}^M \frac{(-1)^n}{n!} \left(\psi A\bar\psi\right)^n
= \frac{(-1)^M}{M!} \left(\psi A\bar\psi\right)^M + (\text{forms of deg} < 2M),
\end{equation}
and only the first (top degree) form on the right-hand side can contribute to the integral.
Using $\psi A\bar\psi
= \sum_{x,y} A_{xy} \psi_x\bar\psi_y$ and anti-symmetry, we obtain
\begin{equation}
\begin{aligned}
\left(\psi A \bar\psi\right)^M
&= \sum_{x_1,y_1} \cdots \sum_{x_M,y_M} A_{x_1 y_1} \cdots A_{x_M y_M} \psi_{x_1}\bar\psi_{y_1} \cdots \psi_{x_M} \bar\psi_{y_M}
\\
&= \sum_{\eta \in S_M} \sum_{\sigma \in S_M} A_{\eta(1) \sigma(1)} \cdots A_{\eta(M) \sigma(M)} \psi_{\eta(1)}\bar\psi_{\sigma(1)} \cdots \psi_{\eta(M)} \bar\psi_{\sigma(M)}
\\
&= M! \sum_{\sigma \in S_M} A_{1 \sigma(1)} \cdots A_{M \sigma(M)} \psi_{1}\bar\psi_{\sigma(1)} \cdots \psi_{M} \bar\psi_{\sigma(M)}
\\
&= M! \sum_{\sigma\in S_M} \sgn (\sigma) A_{1 \sigma(1)}  \cdots A_{M \sigma(M)} \psi_{1}\bar\psi_{1} \cdots \psi_{M} \bar\psi_{M}
\\
&= (-1)^M M! \left(\det A\right) \bar\psi_{1}\psi_{1} \cdots \bar\psi_{M} \psi_{M}
,
\end{aligned}
\end{equation}
so the top degree part of $e^{-\psi A\bar\psi}$ is
$\left(\det A\right) \bar\psi_{1}\psi_{1} \cdots \bar\psi_{M} \psi_{M}$.
Since $\bar\psi_x\psi_x = \frac{1}{2\pi i} d\bar\phi_x d\phi_x$, this gives
\begin{equation}
\label{e:pre-ibp}
    \int e^{-\phi A\bar\phi - \psi A\bar\psi}
    \phib_{a_1}\cdots \phib_{a_p} \phi_{b_1}\cdots \phi_{b_p}
    =
    \int_{\C^\Lambda}
    \phib_{a_1}\cdots \phib_{a_p} \phi_{b_1}\cdots \phi_{b_p}
    d\mu_C.
\end{equation}
We apply the integration by parts formula \eqref{e:ibp} $p$ times to see that the
right-hand is equal to
the left-hand side of \eqref{e:intrefRHS},
and the proof is complete.
(The last step is an instance of Wick's Theorem \cite{GJ87}.)
\end{proof}

\begin{proof}[Proof of Proposition~\ref{prop:Integral-Representation}]
We prove \eqref{e:intrep1a}.
First,
we define $F : \R^{\Lambda_N} \to \R$ by
\begin{equation}
\label{e:intrep1}
    F(S) = e^{-\sum_{x\in\Lambda_N} \big( gS_x^2 + (\nu-1)S_x \big)}
    \quad \quad
    (S \in \R^{\Lambda_N}).
\end{equation}
Then, by the definition given in \eqref{e:wmNdef} and the fact that $\sum_{x}L_T(x)=\|T\|_1$,
the summand on the right-hand side of \eqref{e:intrep1a} is equal to
\begin{equation}
    \int_{\R_+^\nl} E_A^N \big[e^{- I_p(T)} \1_{X(T)= \sigma(B)}\big] e^{-\nu \|T\|_1 } dT
    =
    \int_{\R_+^\nl} E_A^N \big[F(L_T) \1_{X(T)= \sigma(B)}\big] e^{-\|T\|_1 } dT.
\end{equation}
Also,
\begin{equation}
    \int
    e^{-\sum_{x\in\Lambda}
    \big(\tau_{\Delta ,x} + g\tau_x^2 + \nu\tau_x \big)} \bar{\phi}_a^{\nl} \phi_b^\nl
    =
    \int
    F(\tau)
    e^{-\sum_{x\in\Lambda} \big(\tau_{\Delta ,x} +\tau_x \big)}
    \bar{\phi}_a^{\nl} \phi_b^\nl.
\end{equation}
We write $F$ in terms of its Fourier transform $\hat F$ as
\begin{equation}
    F(S) =
    \int e^{-i \sum_{x\in\Lambda} r_x S_x} \hat F(r) dr.
\end{equation}
With an appropriate argument to justify interchanges of integration
(done carefully in \cite{BIS09}), it therefore
suffices to show that for all $s_x \in \C$ with ${\rm Re}(s_x) >0$,
\begin{equation}
\label{e:intrepF}
    \int
    e^{-\sum_{x\in\Lambda} \big(\tau_{\Delta ,x} + s_x\tau_x \big)}
    \phib_{a_1}\cdots\phib_{a_p} \phi_{b_1}\cdots \phi_{b_p}
    =
    \sum_{\sigma \in S_p}
    \int_{\R_+^\nl} E_A^N \big[e^{-\sum_{x\in\Lambda} s_x L_T(x) } \1_{X(T)= \sigma(B)}\big]  dT
    .
\end{equation}
Let $V$ be the diagonal matrix with entries $s_x$.
Since the components of $X$ are independent and identically distributed,
the integral on the right-hand side of \eqref{e:intrepF} factors with each factor being
$(-\Delta_\Lambda + V)^{-1}_{a_i \sigma(b_i)}$ by Lemma~\ref{lem:intrepLHS}.
By Lemma~\ref{lem:intrepRHS}, the left-hand side of \eqref{e:intrepF} is therefore
equal to the right-hand side, and the proof is complete.
\end{proof}

\section*{Acknowledgements}

We thank Roland Bauerschmidt for
invaluable advice, discussions, and input;
David Brydges for countless influential conversations;
and John Cardy for correspondence which helped clarify our understanding
of the role of reduction of $O(n)$ symmetry for $n \ge 2$.
We gratefully acknowledge the hospitality and support of
the Mathematical Institute of Leiden University (GS and AT),
and of RIMS at Kyoto University (GS),
where part of this work was done.
This work was supported in part by NSERC of Canada.

\bibliography{../../bibdef/bib}
\bibliographystyle{plain}

\end{document}